\documentclass[11pt]{amsart}
\usepackage{latexsym,amssymb,calc,amsmath,amsthm,mathrsfs,amssymb,color,graphicx,graphics,psfrag}
\usepackage{bbold, epsfig, euscript,enumitem}
\usepackage[bf]{caption2}
\usepackage{fancyhdr,wrapfig}
\usepackage{aliphat}
\usepackage[windows]{umlaute}

\usepackage[T1]{fontenc}
\usepackage{fancyhdr}

\newcommand{\divS}{{\rm div}_\Sigma \,}

\renewcommand{\div}{{\rm div \,}}

\newcommand{\esssup}[1]{\mathop{\rm ess\ sup}}
\newcommand{\essinf}[1]{\mathop{\rm ess\ inf}}
\newcommand{\N}{{\rm I\kern - 2.5pt N}}
\newcommand{\Z}{{\rm Z\kern - 5.5pt Z}}
\newcommand{\Q}{{\rm I\kern - 5.25pt Q}}
\newcommand{\C}{{\rm I\kern - 6.25pt C}}
\newcommand{\R}{{\rm I\kern - 2.5pt R}}

\newcommand{\bbf}{\mathbf{b}}
\newcommand{\dbf}{\mathbf{d}}
\newcommand{\Dbf}{\mathbf{D}}
\newcommand{\fbf}{\mathbf{f}}
\newcommand{\gbf}{\mathbf{g}}
\newcommand{\Ibf}{\mathbf{I}}
\newcommand{\jbf}{\mathbf{j}}
\newcommand{\Jbf}{\mathbf{J}}
\newcommand{\nbf}{\mathbf{n}}
\newcommand{\Nbf}{\mathbf{N}}

\newcommand{\qbf}{\mathbf{q}}

\newcommand{\Sbf}{\mathbf{S}}
\newcommand{\ubf}{\mathbf{u}}
\newcommand{\vbf}{\mathbf{v}}
\newcommand{\xbf}{\mathbf{x}}
\newcommand{\Phibf}{\mathbf{\Phi}}
\newcommand{\na}{\nabla}
\newcommand{\pa}{\partial}
\newcommand{\stress}{\mathbf{S}}
\begin{document}
\title[Sharp-Interface multicomponent continuum thermodynamics]{Sharp-Interface Continuum Thermodynamics of multicomponent fluid systems with interfacial mass}
\author[Dieter Bothe]{Dieter Bothe \vspace{0.1in}}
\address{Department of Mathematics and Profile Topic \emph{Thermofluids \&  Interfacial Phenomena}\\
Technical University of Darmstadt\\
Alarich-Weiss-Str.~10\\
D-64287 Darmstadt, Germany}
\email{bothe@mma.tu-darmstadt.de}
\date{July 1, 2022}
\begin{abstract}
\noindent
We revisit the sharp-interface continuum thermodynamics of two-phase multicomponent fluid systems with interfacial mass.
Since the published work is not fully consistent, we provide a rigorous derivation of the local balance equations and
the entropy production rates, including all relevant steps and mathematical tools.
Special emphasis is put on an axiomatic form of the entropy principle which, at the same time, allows for an efficient closure process. The latter builds on an appropriate representation of the entropy production rates, which is based on structural
information being obtained from a more refined model which includes partial momenta and is given in an appendix.
The closure relations obtained for the one-sided bulk-interface species transfer rates are further investigated,
which leads to a novel model for mass transfer at fluid interfaces, influenced by the presence of adsorbed surface
active agents. We finally compare our results in detail with the existing literature on continuum thermodynamics of two-phase multicomponent fluid systems with interfacial mass.
\end{abstract}
\maketitle
\noindent
{\bf Keywords:}
Multicomponent two-phase fluid systems, adsorbed species, mass transfer resistance, soluble surfactant, interface chemical potentials, interfacial entropy production, interfacial jump conditions, surface tension effects, energy barrier model.\vspace{0.2in}\\
\section{Introduction}
Multicomponent two-phase fluid systems are ubiquitous in nature, science and engineering.
Prominent examples are droplets in the atmosphere, forming clouds or fog,  the spray of water droplets in breaking waves, gas bubbles rising through water in rivers, natural lakes or oceans, or bubbles being dispersed in other liquids within technical contact apparatuses such as bubble columns or extraction columns.
In all these examples, two bulk fluid phases are in contact at a deformable interface which is free to move.
The appearance of such an interface is the result of either no/partial miscibility on the molecular scale of two different fluids in contact, e.g.\ oil and water, or the coexistence of a liquid and its own vapour.
In the vast majority of such two-phase fluid systems, at least two different chemical constituents (species) are present; the only exception would be a liquid/vapour system without any impurities.
In fact, the prototypical case is that of several chemical species being mixed on the molecular scale. For instance, in case of an air bubble in water, the list of involved constituents includes water, nitrogen, oxygen and carbon dioxide,
among others. Consequently, the generic two-phase fluid system is composed of two multicomponent mixtures which form the two bulk phases, being in contact at their common interface.

Two-phase fluid systems which are out of equilibrium exchange in particular mass, momentum and energy.
In multicomponent systems away from chemical equilibrium, matter will be exchanged across the interface,
i.e.\ a transfer of chemical constituents takes place.
Any such process in which a certain species is exchanged is termed ''mass transfer'' in Chemical Engineering, a notion which we also follow here. Mass transfer occurs in all of the examples given above, such as ocean-atmospheric exchange of gaseous components (including CO$_2$ as a most relevant topic)
\cite{Veron, Hall, deike2022}, cloud physicochemistry \cite{tilgner2021},
gas scrubbing processes \cite{charpentier1981},
aeration for oxygen supply in bio-reactors, e.g.\ for waste water treatment \cite{Tramper, Rosso},
reactive bubble column processes \cite{Deckwer, Schlüter2021} etc.

In particular in two-phase chemical reactors, the transfer of chemical species is the necessary prerequisite
in order for the desired chemical reactions to occur. In fact, this mass transfer is often the limiting step of the overall process. This is even more true for extraction processes, in which mass transfer is the core process step to be performed.
Therefore, detailed and fundamental quantitative knowledge on the local mass transfer across fluid interfaces is of utmost importance for process control and intensification.
Besides external operating conditions which can be used to influence, say, the size and shape of bubbles or droplets, the flow conditions, pressure and temperature, small-scale interfacial phenomena can have a significant impact on the mass transfer rates. In particular,
it is observed in many applications and experiments that surface active substances (so-called surfactants)
are present, as impurities in contaminated systems or as additives, brought intentionally into the fluid system in order
to change certain properties (see, e.g., \cite{Rosen, Chowdhury, Singh, Palmer}) or being an educt or product in a chemical reaction network. Let us note in passing that even certain fluorescence tracers used for local concentration measurements turn out to be surface active, thus influencing the process which is to be monitored \cite{Weiner2019}.
Surfactants adsorb to the fluid interface, changing its interfacial free energy and interfacial tension.
Even at very small surfactant concentrations in the range of a few ppm or less within the bulk, their interfacial
concentration will typically be large and lead to significant changes of the macroscopic system behavior; see \cite{Takagi} for a review.
A classical example is the strong impact of surface active substances on the rise velocity of bubbles in aqueous systems, slowing them down considerably.

Adsorption of surfactant strongly influences the transfer rates of, e.g., gaseous components into a liquid phase.
It is commonly accepted that there are two different ways, how this influence is mediated.
First, since surfactant is inhomogeneously distributed along the interface, the interfacial tension displays non-zero
surface gradients. This induces so-called Marangoni stresses which act as tangential forces at the interface, thus changing the flow field locally, resulting for instance in the above mentioned deceleration of a rising bubble.
This in turn alters the ratio between diffusive and convective transport time scales, changing the diffusive mass transfer fluxes at the interface. Experimental results on the influence of Marangoni stress on mass transfer are reported, e.g., in \cite{Alves2005, Garcia2010, Jimenez2014, Huang2017, Nekoeian2019}.
Second, even in the absence of fluid flow, the partial coverage of the interface with surfactant molecules constitutes a barrier against the passage of other molecules across the interface. This leads to an additional hindrance effect, also referred to as mass transfer resistance, sometimes also called ''steric'' effect. Experimental results on this hindrance effect can be found, e.g., in
\cite{Sardeing2006, Hebrard2009, Aoki2015, Hori2019, Hori2020}.
Let us note that, similarly to mass transfer, the rate of evaporation is influenced by the presence of surfactants \cite{Langmuir, Barnes1986}.
Already in \cite{Langmuir}, Langmuir introduced an energy barrier model to explain
this phenomenon, at least qualitatively. Despite of this classical work, how to derive and incorporate a local barrier/hindrance effect in a thermodynamically consistent way within a rigorous continuum physical framework is an open question and one main aim of the present paper is to close this gap.
For comparison, let us briefly recall the current sharp-interface continuum mechanical model for (species) mass transfer, which we shall call the ''standard model''.

The standard model for the continuum physical description of mass transfer across fluid interfaces represent the interface as a surface of zero thickness. Assuming small Mach number flows and isothermal conditions, the model is based on the incompressible two-phase Navier-Stokes equations. Inside the fluid phases the governing equations are
\begin{align}
	&\na \cdot \vbf=0,\label{E1}\\
	&\pa_t (\rho\vbf)+ \na \cdot (\rho \vbf \otimes \vbf)+\na p=\na \cdot \Sbf^{\rm visc}+\rho\gbf \label{E2}
\end{align}
with the viscous stress tensor
\begin{equation}\label{E3}
\Sbf^{\rm visc}=\eta(\na\vbf+\na \vbf^{\sf T}),
\end{equation}
where the material parameters depend on the respective phase.
These bulk equations need to be complemented by so-called jump conditions which apply at the interface.
Without phase change, as is assumed, the normal component $V_\Sigma=\vbf^\Sigma \cdot \nbf_\Sigma$
of the interfacial velocity coincides with the normal component of both the adjacent fluid velocities, resulting
in the kinematic boundary condition
\begin{equation}\label{KineticBC}
V_\Sigma = \vbf \cdot \nbf_\Sigma,
\end{equation}
where $V_\Sigma$ denotes the speed of normal displacement of the interface and $\nbf_\Sigma$ is the interfacial unit normal.
Moreover, no-slip between the two phases at the interface is usually imposed. In this situation, the interfacial jump conditions for total mass and momentum are
\begin{equation}\label{E4}
[\![\vbf]\!]=0, \quad [\![p \Ibf - \Sbf^{\rm visc} ]\!] \cdot \nbf_\Sigma=\sigma \kappa_\Sigma \nbf_\Sigma + \nabla_\Sigma \sigma,
\end{equation}
where $\sigma$ denotes the interfacial tension, $\kappa_\Sigma=-\na \cdot \nbf_\Sigma$ is twice the mean curvature of $\Sigma$ and $[\![\phi]\!]$
stands for the jump of a quantity $\phi$ across the interface.

The molar concentration $c_i$ of a chemical species $A_i$ is governed by the balance equation
\begin{equation}\label{E6}
\pa_t c_i+\na \cdot (c_i \vbf+\Jbf_i)=r_i,
\end{equation}
where the diffusive flux $\Jbf_i$ is typically modeled according to Fick's law as
\begin{equation}\label{E7}
\Jbf_i=-D_i \na c_i
\end{equation}
with diffusivity $D_i$. The source term on the right-hand side in (\ref{E6}) accounts for chemical reactions. At the interface, the diffusive fluxes in normal direction are commonly supposed to be continuous, i.e.\ \begin{equation}\label{E8}
[\![-D_i \na c_i]\!] \cdot \nbf_\Sigma=0.
\end{equation}
Furthermore, continuity of the chemical potentials at the interface is assumed, i.e.\ \begin{equation}\label{E9}
[\![\mu_i]\!] =0.
\end{equation}
Using standard relations for the $\mu_i$, this leads to Henry's law. For molar concentrations, the latter states that
the one-sided limits $c_k^\pm$ of the bulk concentrations satisfy
\begin{equation}\label{E10}
c^-_k=c^+_k/H_k
\end{equation}
at the interface, with the Henry coefficient $H_k$ often assumed to be constant. Equations (\ref{E1})--(\ref{E10}) comprise what is called the ``standard model'' throughout this paper.

This standard model, sometimes with further simplifications like constant surface tension, constant diffusivities or  homogeneous gas phase concentrations is the basis of almost all detailed numerical simulations of mass transfer across fluid interfaces up to now; cf.\ \cite{1, Trygg2013, 13, Dirk2016, Weiner2017, Oliva2019, Weiner2019, Claassen2020}.
If surfactants are included in this  model, their effect is a change of the local surface tension, mediated via a surface equation of state.
This leads to surface gradients which directly influence the interfacial force balance. While these so-called
Marangoni effects strongly impact the local hydrodynamics and, hence, the mass transfer rates due to a modified interplay
of diffusive and convective transport processes, the direct hindrance effect of surfactant on mass transfer processes is not included in this standard model, but accounted for via global mass transfer correlations; see, e.g., \cite{Sardeing2006}.

The standard model has several shortcomings, from which we list those being most relevant in the context
of mass transfer phenomena:
\begin{enumerate}
\item[(i)]
the diffusion fluxes modeled with Fick's law are inconsistent to mass conservation and to the second law of thermodynamics (cf.\ \cite{BD-MCD});
\item[(ii)]
the transmission condition \eqref{E8} does not account for the relative (normal) motion between the interface and the adjacent bulk phases, thus being inappropriate to describe, e.g., the dissolution of a pure gas bubble (cf.\ \cite{1});
\item[(iii)]
the condition \eqref{E9} of continuous chemical potentials is an idealization in several respects: even for systems without adsorbed species, additional kinetic and viscous terms appear (cf.\ below); for systems with adsorbed species, condition \eqref{E9} provides no means to couple surfactant concentrations to transfer rates.
\end{enumerate}
The main focus of the present paper is closely related to the last item.
\begin{figure}
\includegraphics[width=5.4in]{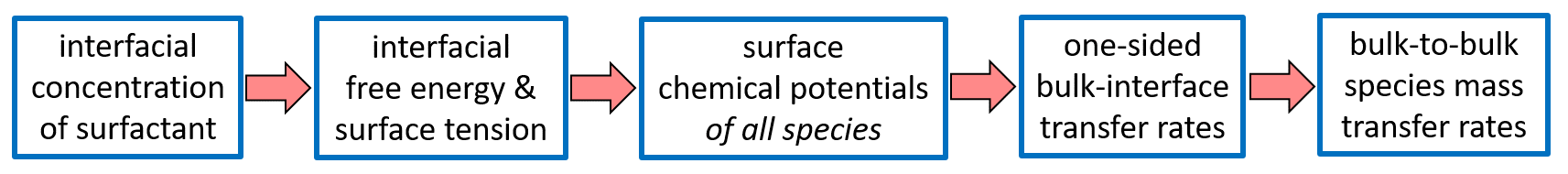}
\caption{Schematic illustration of how surfactant influences local mass transfer rates.}\label{causal-chain}
\end{figure}
In a nutshell, Figure~\ref{causal-chain} illustrates how a surfactant, say constituent $k$, can influence the
mass transfer rates of any transfer species: The presence of surfactant adsorbed at the interface changes the interfacial tension. Since the interface chemical potentials of all species depend strongly on the interfacial tension,
they change accordingly. Now, in a consistent and sufficiently general model, the mass transfer rates are not only determined by the one-sided limits of the bulk chemical potentials but of their mutual difference to the interfacial chemical potential. This way, the transfer rates become dependent on surfactant concentration,
independently of any flow velocity in the fluid system.
What is therefore missing for a predictive, fundamental modeling is a coupling of interfacial
properties--in particular the interfacial free energy--with mass transfer
rates across the interface. This coupling requires that the transfer species, in addition to surfactants, are also a part of the interfacial thermodynamics.
Therefore, to introduce a cross-effect between the surface coverage by a certain constituent
and the mass transfer rate of another constituent, both have to interact on the interface, which requires both constituents to be represented in the thermodynamical state of the interface
via non-vanishing interfacial (i.e., area-specific) partial mass densities.

To conclude this introduction, the above mentioned shortcomings of the standard model to describe relevant
mass transfer phenomena stress the need for more general (sharp-interface) models, accounting for interfacial
mass densities of \emph{all involved constituents} together with a complete thermodynamical description
of the bulk and interfacial mixtures, undergoing irreversible processes. The latter include transport and transfer processes, in particular the local exchange of chemical constituents between the interface and the adjacent bulk phases.
We will refer to this exchange processes as adsorption (bulk to interface) and desorption (interface to bulk), also
for constituents which do not accumulate on the interface, but are still present there with a small but positive area-specific concentration.

In this generality, i.e.\ for the multicomponent two-phase fluid systems with interfacial mass for all constituents,
only very few papers have been published, which consider the interfacial entropy production rate and resulting restrictions to closure relations.
Building on prior sharp-interface modeling of irreversible processes given in \cite{Waldmann} and \cite{BAM},
in the seminal paper \cite{Bedeaux}, Dick Bedeaux derived such a sharp-interface model with entropy inequality
in the spirit of T.I.P.\ (Theory of Irreversible Processes cf.\ De Groot, Mazur \cite{dGM}).
A few years later, the monograph \cite{Slattery-Interfaces} by Slattery appeared, which contains a similar
sharp-interface model with interfacial mass densities, deriving and evaluating bulk and interfacial entropy production rates in the spirit of Truesdell and Noll; cf.\ \cite{TN2003}.
In these fundamental works, as well as in all other papers on this topic to be discussed below, an interfacial entropy production rate is given, but actually never with the same representation.
More severe, most of the given rates are not even equivalent and those
which are correct are often arranged in a disadvantageous way such that a resulting closure, which is not always given, would not result in a physically sound model.

The present paper revisits the sharp-interface continuum thermodynamical modeling
of multicomponent two-phase fluid systems with (partial) interfacial mass densities and adsorption of species at the interface.
We derive the correct entropy production rate and give a most natural arrangement of terms such that a closure without cross-effects nevertheless results in model equations which lead to physically sensible descriptions of the underlying interfacial phenomena.
In order to be able to derive this particular structure of the entropy production, a more refined, so-called class-II model for interfacial processes is obtained in the appendix.
Here, ''class-II'' refers to a model in which partial masses and \emph{partial momenta} of the
mixture are individually balances. Note that the standard multicomponent modeling with partial masses but a common single momentum
balance belongs to ''class I'', while for certain applications like plasma physics, a class-III modeling with partial internal
energy, hence individual, species-specific temperatures is required.
This notion of class-I, -II and -III models goes back to Hutter, cf.\ \cite{Hutter-book}.
Let us note in passing that such mixture models with individual velocity fields for different constituents require appropriate boundary conditions for all species.
This is an intrinsic difficulty since, while one might know the total traction or the velocity of the mixture at the boundary, one typically does not know the individual contributions. This issue is discussed in some detail in the book on mixture theory by Rajagopal and Tao \cite{Raj-Tao}.

Special emphasis is then placed on the modeling of mass transfer rates and with the above mentioned view of mass transfer
as a series of two one-sided bulk-interface transfer processes, a new model is derived which can describe the hindrance
effect of surfactants on the mass transfer of different constituents, like dissolved gases,
according to the causal chain as illustrated in Figure~\ref{causal-chain}.
This is relevant, for instance, to
describe the transfer of CO$_2$ between the atmosphere and rivers, seas or the oceans, where the fluid interface typically is
covered by organic molecules which are surface active.
Finally, we also survey the representations of the interfacial entropy production as given in the literature together with
a detailed comparison and critical assessment.

This work profited a lot from \cite{BD2015}, where many new ideas on the continuum thermodynamics of single-phase multicomponent chemically reacting fluid mixtures have been developed,
and from \cite{6}, where two-phase multicomponent systems without interfacial mass are investigated.
For fundamentals on continuum mechanical modeling of two-phase fluid systems we refer in particular to \cite{Hutter-book, Ishii, SK-DB-book, Slattery-two-phase, Slattery-etal-Interfaces}.
\section{Continuum Physical Interface Concepts}
We recall different concepts on how to model interfaces within continuum physics as well as some basic facts about the sharp interface modeling approach.

Josiah Willard Gibbs laid down the fundament of the modeling of fluid interfaces \cite{Gibbs-collection}.
The core point in his theory is the concept of a {\em dividing interface} and, as a result, the appearance
of so-called {\em surface excess quantities}.
\begin{figure}
\includegraphics[height=1.8in]{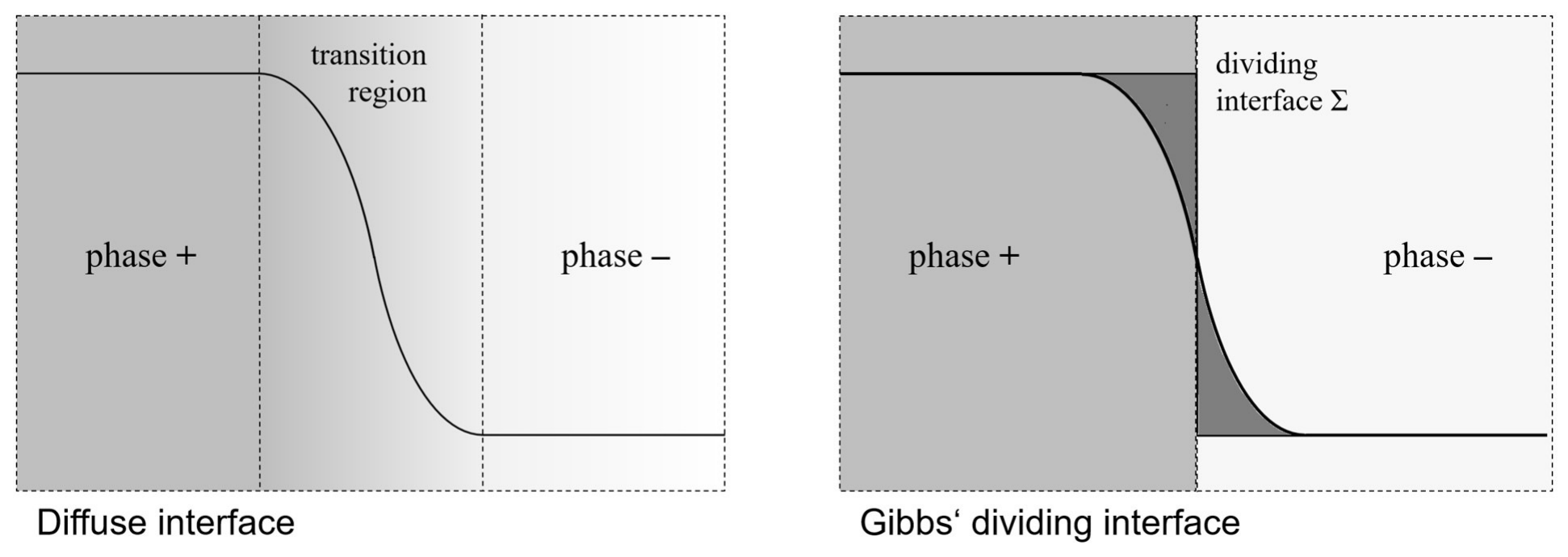}
\caption{Diffuse interface and Gibbs' definition of a dividing interface.}\label{dividing-interface}
\end{figure}
Figure~\ref{dividing-interface} explains this concept, which consists in a clever mapping of continuous densities which happen to
have a steep gradient in a thin transition layer within a so-called {\em diffuse interface} model (Figure~\ref{dividing-interface}, left)
to simpler, but discontinuous profiles, complemented by excess densities.
For this purpose, the (continuous) bulk densities of a certain extensive quantity are extrapolated from a region
adjacent to the transition layer of the diffuse interface picture to the interface region.
Within the interface region, this introduces a difference in the amount of the corresponding extensive quantity between the original distribution and the new, extrapolated one. This difference is attributed to the sharp interface position as a separate
entity, called the excess of the considered quantity; see (Figure~\ref{dividing-interface}, right).
The resulting idealized description is a useful simplification, since it avoids the necessity of knowing the exact transition profile. The resulting bulk densities are no longer continuous across the interface,
but admit one-sided limit values (coming from the different bulk phases) at every interface position. The excess quantity has an interfacial, i.e.\ area-specific
density which is assumed continuous along the sharp interface.

The concrete values of the interface excess density depend on the position, where the dividing interface is placed.
In the two-phase {\em single component} case, this position is usually chosen in such a way that the excess density vanishes.
If another species, say a surface active agent, is present, it is not possible to let both excess densities vanish  simultaneusly,
which is what actually makes this approach interesting, since one can place the dividing interface in such a way that the
''solvent'' still has vanishing interfacial density, but the surfactant has a positive interfacial concentration.
For multicomponent systems with $N>2$ chemical constituents, as considered here,
such a construction of a dividing interface can, however, result in
counter-intuitive outcomes like negative surface excess concentrations. Accordingly, there have been several attempts to find an optimal definition of the interface position. One approach due to Lucassen-Rynders and van den Tempel \cite{LR-vdT}
places the dividing interface in such a way that the total molar concentration of all matter which is put to the dividing surface as an excess quantity is constant. The point here is that such a position can always be found, while it is rather
unclear whether this is intuitive in any respect. But, at least, it leads to non-negative partial surface mass densities.
For this reason, this is the preferred choice in several works, like in \cite{Slattery-etal-Interfaces}.
There are other approaches which show that, while the excess densities themselves depend heavily on the position
of the dividing interface, certain derived quantities, which still carry all relevant information to build a sensible
interfacial thermodynamics on them, can be defined in a unique way.
In \cite{Savin}, this is called the principle of gauge invariance.
Rather recently, Bedeaux and Kjelstrup revisited this topic in \cite{BK-2018}
and showed that, despite the apparent difficulties, the
interfacial thermodynamics is not negatively affected thereby.

\begin{figure}
\includegraphics[height=1.8in]{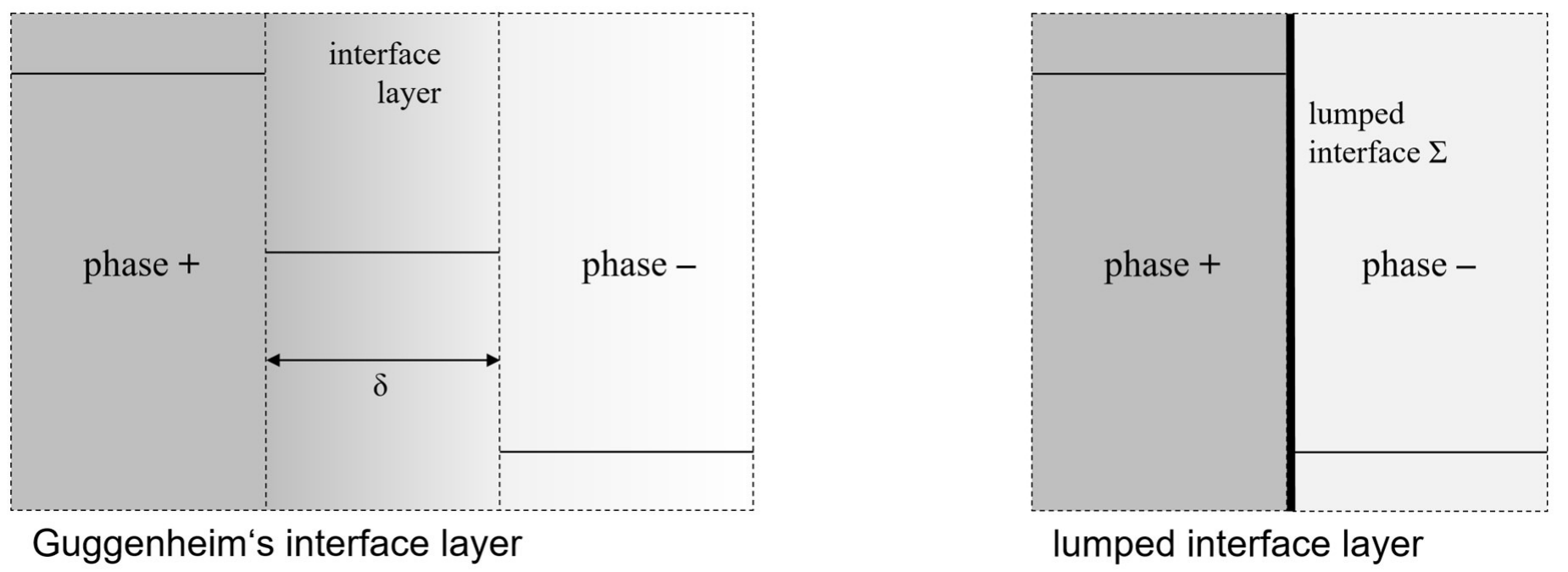}
\caption{Guggenheim's construction of a sharp interface
and the lumped interface layer construction employed here.}\label{Guggenheim}
\end{figure}

There are also alternative concepts like the one going back to Guggenheim \cite{Guggenheim}
in which the continuous profiles of the diffuse interface are replaced
by constant--in normal direction--profiles in a thin interfacial layer. This is illustrated by the left part
of Figure~\ref{Guggenheim}; cf.\ also \cite{Lang}.
This concept has later been employed in \cite{Guggenheim1940}.
We follow a related, but slightly different approach which is illustrated by the right picture in Figure~\ref{Guggenheim}.
The only difference to Guggenheim's construction is that we lump the (thin) transition layer into a sharp
interface, i.e.\ one with zero thickness.
The reason for this choice as the basis for the present paper is that the above Gibbs' viewpoint misses the fact that there are
always molecules or atoms inside the transition layer, where they experience different force fields, independently of
whether or not there is a non-zero surface excess of these species.
In particular, if a species is present on both sides of the interface,
in which case mass transfer of this species is possible and will occur unless the system is in equilibrium,
the transfer of molecules (or atoms) from one bulk phase to the other necessarily includes the occurrence of this constituent
in the transition layer.
Hence, this constituent is present in that part of the system,
which should be modeled separately due to the different force fields there.

In principle, however,
the same sharp-interface model as derived in the present paper can be obtained with positioning of a dividing interface according
to Lucassen-Rynders/van den Tempel \cite{LR-vdT}
or employing the gauge invariant quantities as, e.g., defined in Bedeaux/Kjelstrup \cite{BK-2018}.
The more important point, here, is that the simple concept of a lumped transition layer makes it clear that
all constituents should be included also on the interface with a (possibly) non-zero interfacial (molar) mass density.
In fact, surface concentrations in this model are necessarily non-negative, while excess mass densities can be negative
for the dividing interface due to Gibbs. This adds a further qualitative property to solutions of the lumped interface model, which is necessarily reflected by certain structural properties of the transfer coefficients,
an advantage from a mathematical viewpoint; cf.\ \cite{BD-MCD} for the implications of positivity requirements
on (cross-)diffusion coefficients.

This picture also makes clear that mass transfer necessarily has to be considered
as the succession of two bulk-interface transfer processes taking place in series.
This key conception is illustrated by Figure~3. Due to this series of two one-sided (and bi-directional) mass transfer processes,
the interfacial thermodynamical state automatically influences the one-sided mass transfer rates, hence also
the resulting total rate of the transfer series.
\begin{figure}
\includegraphics[height=2in]{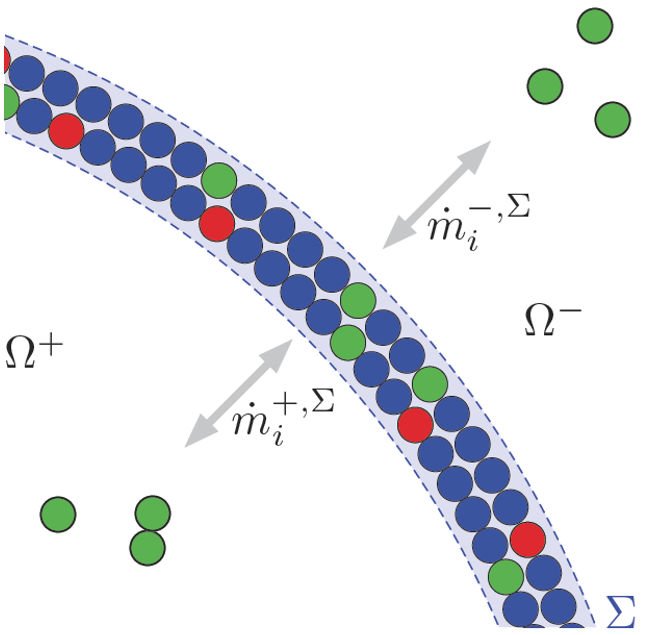}
\caption{Mass transfer between two bulk phases as a series of two one-sided bulk-interface transfer processes.}\label{transfer-series}
\end{figure}
Note that, if the transition layer between the two bulk-phases is lumped into an interface of zero thickness,
all extensive quantities inside this layer are rewritten using area-specific densities instead of volume-specific ones.
Hence, if $\Phi (V)$ stands for the amount of a certain extensive quantity inside a patch
$V=\{{\bf x}+s\, \nbf_\Sigma : \xbf \in A, |s|\leq \delta /2 \}$ of the interfacial layer with thickness $\delta$,
with $A$ the surface patch lying ''in the middle'' of $V$, then
\[
\Phi (V)= \int_A \int_{-\delta /2}^{\delta /2} \phi (\xbf + s \, \nbf_\Sigma (\xbf))\,{\rm d}s\, {\rm d}o (\xbf)
=: \int_A \phi^\Sigma (\xbf) \, {\rm d}o = \Phi^\Sigma (A)
\]
redefines the volume density into an area density, where $\phi$, $\phi^\Sigma$ are the volumetric, respectively
the area-specific densities of the extensive quantity $\Phi$ (bulk), respectively $\Phi^\Sigma$ (interface).
This allows to balance the extensive properties of the matter inside
the transition zone separately, which is sensible since the matter there experiences different force fields as compared to the bulk material.

The spatial region which is lumped into a surface and, hence, would actually be lost in this process, is ignored.
In other words, while there would be
a two-sided void region adjacent to the interface, this is still considered as bulk volume with the bulk densities. This introduces
an error, but this is negligible unless the volume of the transition zone becomes a visible fraction of the total volume. The latter is true if the thermodynamic state of the system approaches the critical state at which the surface starts to thicken and the two-phase character of the systems is finally replaced by that of a supercritical one-phase fluid system. We do not consider such near-critical cases.

Note that both the Gibbs as well as the Guggenheim concept are introduced in order to approximate a continuous (diffuse) transition zone by a sharp interface in a simple, intuitive way.
More refined studies have been undertaken for rigorously establishing a precise relationship between
the two different approaches; see, e.g., \cite{Grauel1988, 7, Lamorgese}.
Let us note that one can of course also keep the diffuse interface setting and employ gradient models rather than models with interface quantities.
In this case, gradients of the densities of the balanced quantities need to be included
in the list of independent variables and thermodynamic states, thus leading to the realm of phase field models in the spirit of Van der Waals and Korteweg, Cahn and Hilliard.
We refer to \cite{Lamorgese} for a brief historical perspective and, in particular, a comparison between phase field and sharp-interface models based on Gibb's dividing interface.

While possibly helpful for keeping some intuition about sharp-interface models fueled by molecular understanding, a connection to a corresponding diffuse interface model is
not a necessary prerequisite of the modeling.
In other words, the sharp-interface approach can also be seen as an independent
phenomenological approach, the value of which has to be judged on the basis of appropriate experiments.
Despite of the advantages of gradient theories, counter-balanced by the need to resolve the diffuse interface in numerical simulations, the accounting for even just point information about the matter
at the transition position in a sharp-interface model provides a valuable enrichment of the description.
This can for instance be seen in the novel mass transfer model derived below.
Let us finally stress, once again, that the essential implication of the above reasoning is that
a sharp-interface model for a multicomponent two-phasse fluid system with adsorbed species
should account for interfacial mass densities of {\em all} constituents being present in the overall fluid system.
%
%
%
%
\section{Sharp-Interface mathematical modeling}
We assume the two-phase multicomponent fluid system to fill the considered domain $\Omega \subset \R^3$
in such a way that two bulk phases
$\Omega^\pm (t)$ are separated by a sharp interface, i.e.\ by a hypersurface $\Sigma (t)$ in $\R^3$:
\[
\Omega = \Omega^+ (t) \cup \Omega^- (t) \cup \Sigma (t),
\]
a disjoint decomposition of $\Omega$. For technical simplicity, we assume the $\Sigma (t)$ to be closed surfaces without
contact to the domain boundary $\partial \Omega$, thus avoiding the appearance of contact lines; cf.\ \cite{FKB-2019}.
The family of moving hypersurfaces $\{\Sigma (t) \}_{t\in I}$, with $I=(a,b)\subset \R$ an open time interval, is assumed to form a
$\mathcal{C}^1$-hypersurface in space-time $\R^4$ such that the instantaneous surfaces $\Sigma (t)$ are $\mathcal{C}^2$-hypersurfaces
in $\R^3$ so that curvature is well-defined on $\Sigma (t)$. In particular, the total curvature (twice the mean curvature)
$\kappa_\Sigma =\divS (- n_\Sigma )$ is a well-defined continuous quantity. In addition, we assume that the normal field
$n_\Sigma (t,\cdot)$ of $\Sigma (t)\subset \R^3$ is continuously differentiable, jointly in both variables $(t,x)$.
Under these assumptions, we call $\{\Sigma (t) \}_{t\in I}$ a $\mathcal{C}^{1,2}$-family of moving hypersurfaces.
We also denote by $\mathcal M ={\rm gr}\,( \Sigma )$ the graph of the family $\{\Sigma(t)\}_{t \in I}$, defined as
\begin{equation}
\mathcal M := \{ (t,x)\in \R^4 : t\in I, \, x\in \Sigma (t)\} = \bigcup_{t\in I} \big( \{t\}\times \Sigma (t) \big).
\end{equation}
These definitions are also employed in \cite{Kimura.2008}, \cite{PrSi15}, \cite{Garcke-Handbook}
and in a similar form in \cite{Giga.2006}.

For a $\mathcal{C}^{1,2}$-family $\{\Sigma(t)\}_{t \in I}$ of moving hypersurfaces, $V_\Sigma$ denotes
the \emph{speed of normal displacement} of $\Sigma(\cdot)$ and is defined via the relation
\begin{equation}\label{VSigma0}
V_\Sigma (t,x) = \langle \gamma' (t) , n_\Sigma (t, \gamma (t)) \rangle
\end{equation}
for any $\mathcal{C}^1$-curve $\gamma$ with $\gamma (t)=x$ and ${\rm gr} \,(\gamma)\subset \mathcal M$,
where $\langle \cdot , \cdot \rangle$ denotes the inner product in $\R^3$.
It is not difficult to show that the definition via \eqref{VSigma0} is equivalent to
\begin{equation}\label{VSigma}
\lim_{h\to 0} \frac 1 h {\rm dist} (x+h V_\Sigma (t,x) n_\Sigma (t,x) , \Sigma (t+h)) =0 \quad \mbox{ for } t\in I, x\in \Sigma(t).
\end{equation}
The characterization via \eqref{VSigma} shows that $V_\Sigma$ is a purely kinematic quantity, determined only by the family $\{\Sigma(t)\}_{t \in I}$ of moving interfaces. Hence, the value of $V_\Sigma$ from \eqref{VSigma0}
does not depend on the choice of the specific curve; cf.\ Chapter~2.5 in \cite{PrSi15}.
The computation of $V_\Sigma$ is especially simple
if $\{\Sigma(t)\}_{t \in I}$ is given by a level set description, i.e.\
\begin{equation}\label{levelset}
\Sigma(t)=\{x \in \R^3: \phi(t,x)=0\}
\end{equation}
with $\phi \in \mathcal{C}^{1,2}(\mathcal N)$ for some open neighborhood $\mathcal N \subset \R \times \R^3$ of
$\mathcal M ={\rm gr}\,( \Sigma )$ such that $\nabla \phi \not= 0$ on $\mathcal{M}$. Then
\begin{equation}\label{E11}
V_\Sigma(t,x)=- \, \frac{\partial_t \phi(t,x)}{\|\nabla \phi(t,x)\|}\quad\text{ for } t \in J, \, x \in \Sigma(t).
\end{equation}
In the literature, $V_\Sigma$ is often called normal velocity of $\Sigma (\cdot)$, but we prefer to call it the speed
of normal displacement since $V_\Sigma$ is a scalar quantity.
In fact, the speed of normal displacement gives rise to a natural interface velocity field,
given as $w_\Sigma:=V_\Sigma n_\Sigma$, which therefore is intrinsic to any $\mathcal{C}^{1,2}$-family $\{\Sigma(t)\}_{t \in I}$ of moving hypersurfaces.
According to Corollary~1 together with Theorem~1 in \cite{Bo-2PH-ODE}, the associated initial value problem
\begin{equation}\label{2PH-IVP}
\dot x (t) = w_\Sigma (t,x(t)) \quad \mbox{ for } t\in I,\;\; x(t_0)=x_0
\end{equation}
is uniquely solvable (at least locally in time) for every $t_0\in I$ and $x_0\in \Sigma (t_0)$.
The unique (local) solution $x(\cdot ; t_0,x_0)$ of \eqref{2PH-IVP}, starting in $(t_0,x_0)$,
follows the moving hypersurface along a path which always runs normal to $\Sigma(t)$.
Consequently, the intrinsic interface velocity $w_\Sigma$ (hence $V_\Sigma$)
induces a Lagrangian-type derivative, the so-called
Thomas derivative $\partial^\Sigma_t$ (cf.\ \cite{Thomas}), defined as
\begin{equation}\label{Thomas-derivative}
\partial^\Sigma_t \phi^\Sigma (t_0,x_0) := \frac{d}{dt} \phi^\Sigma (t,x(t;t_0,x_0))_{|t=t_0},
\end{equation}
where $\phi^\Sigma$ denotes a quantity which is defined on $\mathcal{M}$.
According to Lemma~2 together with Theorem~1 in \cite{Bo-2PH-ODE}, unique solvability of the associated
initial value problems also holds true for any velocity field of the type
\begin{equation}\label{full-velocity}
{\bf v}^\Sigma = V_\Sigma \, {\bf n}_\Sigma + {\bf v}^\Sigma_{||},
\end{equation}
where the additional tangential part\footnote{We use $\Sigma$ as a subscript for intrinsic, geometrically defined quantities, while
we use $\Sigma$ as superscript for all other interface quantities.} ${\bf v}^\Sigma_{||}$ is continuous on $\mathcal{M}$ and
the maps ${\bf v}^\Sigma_{||} (t,\cdot )$ are locally Lipschitz continuous on $\Sigma (t)$ for every $t\in I$.
This gives rise to a Lagrangian derivative $\frac{D^\Sigma}{Dt}$
associated with ${\bf v}^\Sigma$, defined in analogy to the Thomas derivative above, for which one can show that
\begin{equation}\label{Lagrange-Thomas}
\frac{D^\Sigma \phi^\Sigma}{Dt} = \partial^\Sigma_t \phi^\Sigma + {\bf v}^\Sigma_{||} \cdot \nabla_\Sigma \phi^\Sigma
\end{equation}
with $\nabla_\Sigma$ denoting the surface gradient (w.r.\ to $\Sigma (t)$, where $t$ is understood); cf.\ \cite{Bo-2PH-ODE}.
Note also that ${\bf v}^\Sigma_{||} \cdot \nabla_\Sigma \phi^\Sigma = {\bf v}^\Sigma \cdot \nabla_\Sigma \phi^\Sigma$,
since $\nabla_\Sigma \phi^\Sigma$ is tangential to $\Sigma$.

The Thomas derivative allows to formulate a general surface transport theorem.
For this purpose, let $V\subset \Omega$ be a fixed control volume with outer normal field
${\bf n}$ and $\Sigma_V (t):=\Sigma (t) \cap V$.
Then the identity
\begin{equation}\label{2PH-trans-thm0}
\frac{d}{dt} \int_{\Sigma_V (t)} \!\! \phi^\Sigma \,do = \int_{\Sigma_V (t)}\!\!  \big( \partial_t^\Sigma \phi^\Sigma
- \phi^\Sigma \kappa_\Sigma V_\Sigma \big)\,do
 - \int_{\partial \Sigma_V (t)}\!\!  \phi^\Sigma V_\Sigma \frac{{\bf n} \cdot {\bf n}_\Sigma}{\sqrt{1-({\bf n}\cdot \mathbf{n}_{\Sigma})^2}}\,ds
\end{equation}
holds, see \cite{Romano, Gurtin, Alke}.
For a particular control volume $V$ such that ${\bf n}$ is tangential to $\Sigma_V (t)$ on $\partial \Sigma_V (t)$,
i.e.\ ${\bf n} \cdot {\bf n}_\Sigma =0$ on $\partial \Sigma_V (t)$, the transport identity \eqref{2PH-trans-thm0} simplifies to
\begin{equation}\label{2PH-trans-thm1}
\frac{d}{dt} \int_{\Sigma_V (t)} \!\! \phi^\Sigma \,do = \int_{\Sigma_V (t)}\!\!  \big( \partial_t^\Sigma \phi^\Sigma
- \phi^\Sigma \kappa_\Sigma V_\Sigma \big)\,do;
\end{equation}
cf.\ \cite{BPtripleline}.
Below, any control volume which intersects $\Sigma (t)$ is assumed to be of this type,
which is sufficient for deriving the local balance equations (in differential form).
From here on, the time dependence of $\Sigma_V (t)$ and $\partial \Sigma_V (t)$ in integration domains
is suppressed for better readability.

\begin{figure}
\includegraphics[width=7cm]{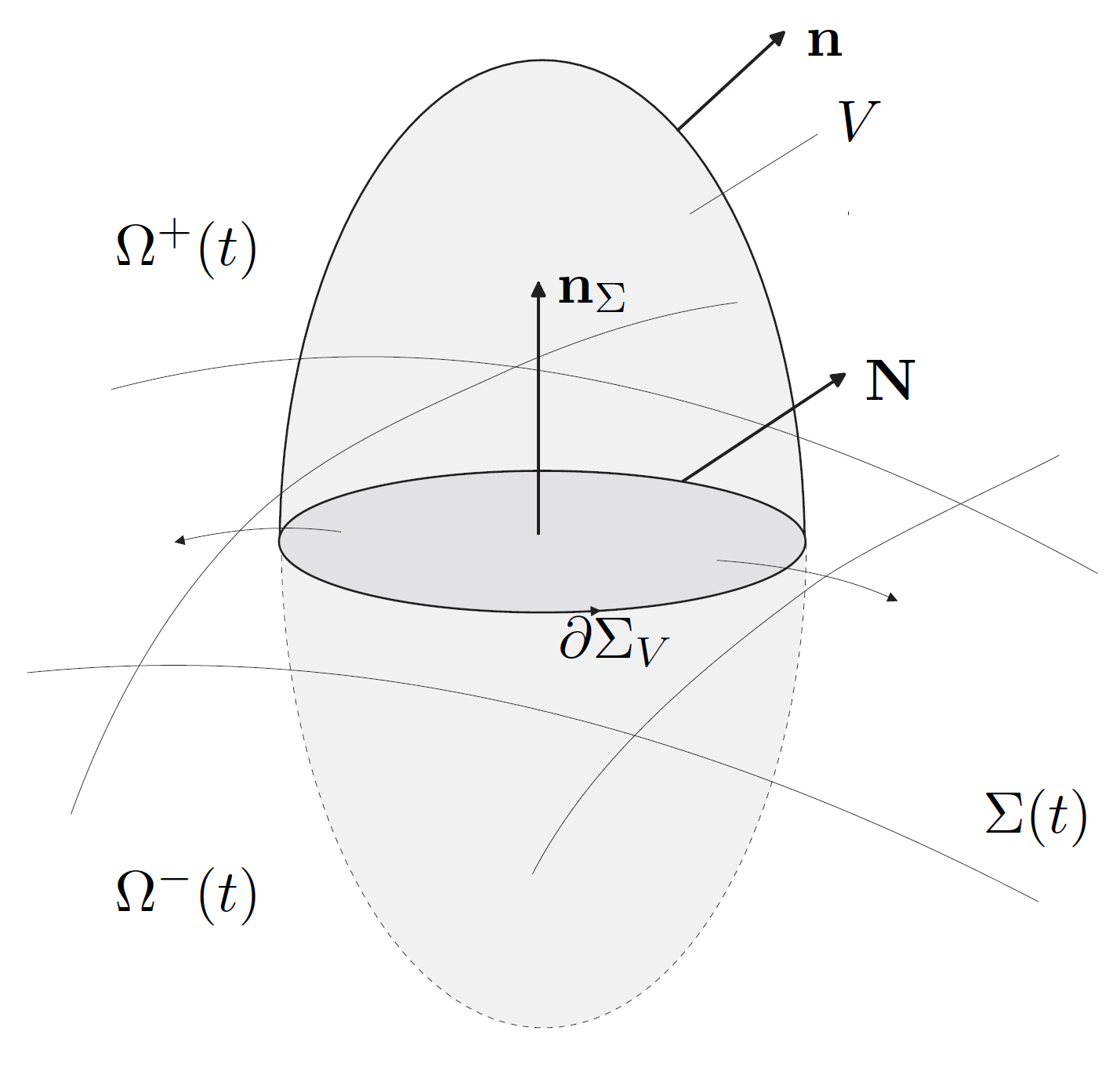}
\caption{Configuration of the phases and basic notations.}\label{CV-fig}
\end{figure}

To obtain a generic balance equation, we start with {\it integral balances}, since they are valid across the interface,
while a direct formulation of interfacial transmission and jump conditions is not obvious.
Let $\Phi$ denote an extensive quantity (i.e., mass additive such as mass, momentum, energy, etc.)
with volumetric density $\phi$ (i.e.\ $\phi^+$ and $\phi^-$ in the phase
$\Omega^+$ and $\Omega^-$, respectively) and interfacial, area specific density $\phi^\Sigma$.
Below, we consider a fixed control volume $V$, because the concept of a
material control volume is somewhat complicated here due to  possible differences between the
one-sided limits of the velocity fields of individual mixture components, the velocities of
absorbed forms of these species on the interface and the interfacial velocity (geometrical or
physical) itself.
Let $\mathbf{N}$ denote the outer
unit normal to $\partial \Sigma_V$ which is at the same time tangential to $\Sigma$; cf.\ Figure~\ref{CV-fig}.
With these notations, the generic integral balance for the amount of an extensive quantity $\Phi$ inside $V$ is given by
\begin{eqnarray}
\label{integral-master}
\frac{d}{dt} \big[ \int_{V} \phi\,dx + \int_{\Sigma_V} \phi^\Sigma \,do \big]
& = &
-\int_{\partial V} \mathbf{j}_{\rm tot}\cdot\mathbf{n}\,do +\int_{V} f\,dx\\[1ex]\nonumber
& & - \int_{\partial \Sigma_V} \mathbf{j}^{\Sigma}_{\rm tot}\cdot\mathbf{N}\,ds +\int_{\Sigma_V} f^{\Sigma}\,do,
\end{eqnarray}
where $\mathbf{j}_{\rm tot}$ is the flux of $\Phi$ inside the bulk phases, $\mathbf{j}^{\Sigma}_{\rm tot}$
the corresponding flux inside the interface, $f$ the source terms acting in the
bulk and, finally, $f^\Sigma$ the interfacial source term.
It is important to note that both $\mathbf{j}_{\rm tot}$ and $\mathbf{j}^\Sigma_{\rm tot}$ are total fluxes
(i.e.\ convective plus diffusive)
with respect to the reference frame in which the fixed control volume is given.
We require the interfacial fluxes to be tangential
to the interface. This is no restriction, since only $\mathbf{j}^{\Sigma}_{\rm tot}\cdot\mathbf{N}$ enters the balance equation,
but has to be accounted for in the constitutive modeling.
In order to obtain the local form of the generic balance equation, more notation and some mathematical tools are required.
We only list the relevant identities, more details can be found in \cite{Slattery-two-phase, Slattery-etal-Interfaces},
for proofs see \cite{Kimura.2008, PrSi15} or \cite{Garcke-Handbook}.
The two-phase transport theorem in the version for fixed control volumes $V$ states that
\begin{equation}\label{2PH-trans-thm-Euler}
\frac{d}{dt} \int_{V} \phi \,dx = \int_{V\setminus \Sigma} \partial_t \phi \, dx
- \int_{\Sigma_V} [\![ \phi ]\!] V_\Sigma \,do,
\end{equation}
where
\begin{equation}
\label{jump-notation}
[\![ \phi ]\!] (t,{\bf x}) := \lim_{h\to 0+} \big( \phi (t,{\bf x}+h \mathbf{n}_{\Sigma})
- \phi (t,{\bf x}-h \mathbf{n}_{\Sigma}) \big)
\end{equation}
denotes the jump of a quantity $\phi$ across the interface in a form which makes the jump conditions orientation invariant.
The two-phase divergence theorem provides the identity
\begin{equation}\label{2PH-div-thm}
\int_{\partial V} {\bf f}\cdot {\bf n} \,do = \int_{V\setminus \Sigma} \div {\bf f}\, dx
+ \int_{\Sigma_V} [\![ {\bf f}\cdot {\bf n}_\Sigma ]\!] \,do.
\end{equation}
We use $V\setminus \Sigma$ as the domain of integration to avoid confusion with distributional derivatives which would include a Dirac delta contribution supported on the interface; the latter would in fact correspond to the jump bracket term; cf.\ \cite{Kanwal}.
Finally, the surface divergence theorem states that
\begin{equation}\label{int-divM-f}
\int_{\partial \Sigma_V} \mathbf{f}^{\Sigma} \cdot {\bf N} \,ds
= \int_{\Sigma_V} {\rm div}_\Sigma \, \mathbf{f}^{\Sigma} \, do
\end{equation}
for vector fields $\mathbf{f}^{\Sigma}$ being tangential to $\Sigma$.

Applying the two-phase transport theorem to the first term in \eqref{integral-master},
the extended surface transport theorem in the form \eqref{2PH-trans-thm1}
to the second term, the two-phase divergence theorem to the third term and the surface divergence theorem
to the fifth term yields
\begin{eqnarray}
\label{integral-master3}
\nonumber
\int_{V\setminus \Sigma} ( \partial_t \phi + \div {\bf j}_{\rm tot} ) \, dx
+ \int_{\Sigma_V} (-[\![ \phi ]\!] V_\Sigma + [\![ \mathbf{j}_{\rm tot} \cdot \mathbf{n}_{\Sigma} ]\!] )\, do\\[1ex]
+ \int_{\Sigma_V} \big( \pa_t^\Sigma \phi^\Sigma + \divS (\phi^\Sigma V_\Sigma {\bf n}_\Sigma + {\bf j}^\Sigma_{\rm tot}) \big)\, do  = \int_{V} f \, dx  + \int_{\Sigma_V} f^{\Sigma}\,do.
\end{eqnarray}
Collecting bulk and interface terms, we obtain the differential form of the generic balance equation in an Eulerian form
by means of the usual localization procedure.
The result reads as\vspace{-0.05in}
\begin{eqnarray}
\label{master-bal-E1}
\partial_t \phi + \div {\bf j}_{\rm tot} =f & \mbox{ in } \Omega\setminus \Sigma,\\[1ex]
\pa_t^\Sigma \phi^\Sigma + \divS (\phi^\Sigma V_\Sigma {\bf n}_\Sigma + {\bf j}^\Sigma_{\rm tot})
 -[\![ \phi ]\!] V_\Sigma + [\![ \mathbf{j}_{\rm tot} \cdot \mathbf{n}_{\Sigma} ]\!] =f^\Sigma & \mbox{ on } \Sigma.
\end{eqnarray}
Recall that both $\mathbf{j}_{\rm tot}$ and $\mathbf{j}^\Sigma_{\rm tot}$ are total fluxes with respect to the
reference frame in which the chosen control volumes are fixed, and $\mathbf{j}^\Sigma_{\rm tot}$ is tangential to $\Sigma (t)$.
If these are decomposed into a convective and a diffusive (also called molecular) flux according to
\begin{equation}
\label{flux-decomposition}
\mathbf{j}_{\rm tot}= \phi \, {\bf v} +  \mathbf{j}, \qquad
\mathbf{j}^\Sigma_{\rm tot}=  \phi^\Sigma  \, {\bf v}^\Sigma_{||} +  \mathbf{j}^\Sigma,
\end{equation}
the generic local balance equation for an extensive quantity $\Phi$ with bulk density $\phi$ and interfacial
density $\phi^\Sigma$ reads as
\begin{eqnarray}
\label{master-bal-E2}
\partial_t \phi + \div \big( \phi \, {\bf v} +  \mathbf{j} \big) =f & \mbox{ in } \Omega\setminus \Sigma,\\[1ex]
\pa_t^\Sigma \phi^\Sigma + \divS (\phi^\Sigma  \, {\bf v}^\Sigma +  \mathbf{j}^\Sigma)
+ [\![ \phi ({\bf v} - {\bf v}^\Sigma )]\!] \cdot \mathbf{n}_{\Sigma}  + [\![ \mathbf{j} \cdot \mathbf{n}_{\Sigma}  ]\!] =f^\Sigma & \mbox{ on } \Sigma.
\end{eqnarray}
Concerning the jump bracket notation, we treat interfacial quantities as having one-sided bulk limits of the same value,
e.g.\ $[\![ {\bf j}\cdot \mathbf{n}_{\Sigma} ]\!]$ above is the same as $[\![ {\bf j} ]\!]\cdot \mathbf{n}_{\Sigma}$.
Hence, the generic interfacial balance can be written somewhat more compactly as
\begin{equation}
\label{master-bal-int}
\pa_t^\Sigma \phi^\Sigma + \divS (\phi^\Sigma  \, {\bf v}^\Sigma +  \mathbf{j}^\Sigma)
+ [\![ \phi ({\bf v} - {\bf v}^\Sigma ) + \mathbf{j} ]\!]  \cdot \mathbf{n}_{\Sigma}=f^\Sigma \;\; \mbox{ on } \Sigma.
\end{equation}
%
%
\section{Sharp-Interface balance equations}
We suppose that the fluid phases are composed of $N$ different constituents (chemical species) $A_1,\ldots ,A_N$.
Typically, all species will be present at least in one bulk phase and on the interface. We do not separately index the
constituents in the different phases; if a species is not present, its mass density simply vanishes.
We account for bulk chemical reactions between the $A_i$ according to
\begin{equation}\label{chem-react-bulks}
\alpha_1^a \, A_1 + \ldots + \alpha_N^a \, A_N
\rightleftharpoons
\beta_1^a \, A_1 + \ldots + \beta_N^a \, A_N
\quad \mbox{ for } a=1,\ldots ,N_R
\end{equation}
with stoichiometric coefficients $\alpha_i^a, \beta_i^a \in \N_0$.
Note that we allow for separate chemical reaction networks in the different bulk phases, i.e.\ in full detail
the stoichiometric coefficients read $\alpha_i^{+,a}, \beta_i^{+,a} \in \N_0$ for $a=1,\ldots ,N_R^+$ to describe
the chemistry in $\Omega^+(t)$ and
$\alpha_i^{-,a}, \beta_i^{-,a} \in \N_0$ for $a=1,\ldots ,N_R^-$ for the chemistry in $\Omega^-(t)$.
Since the aspect of chemical reactions is not the main point, here, we suppress the phase index for better readability
whenever this is possible without loss of understanding.
All reactions are considered with their forward and backward direction. The molar reaction rate
(i.e.\ the number of chemical conversions in multiples of the Avogadro number per time and volume)
of the $a^{\rm th}$ reaction
is denoted $R_a^f$ for the forward (from left to right in \eqref{chem-react-bulks}) and $R_a^b$ for the backward direction.
The total molar reaction rate of the $a^{\rm th}$ reaction hence is $R_a = R_a^f - R_a^b$
and the mass production rate of constituent $A_i$ due to the $a^{\rm th}$ reaction is $M_i r_i$ with
$M_i$ the molar mass of $A_i$ and
\begin{equation}
r_i = \sum\limits_{a=1}^{N_R} \nu_i^a R_a
\quad \mbox{ with } \nu_i^a:= \beta_i^a - \alpha_i^a.
\end{equation}
Because mass is conserved in every single reaction, it holds that
\begin{equation}
\sum\limits_{i=1}^N M_i \nu_i^a =0 \quad \mbox{ for all } a=1,\ldots ,N_R.
\end{equation}
We write $A_i^\Sigma$ for the adsorbed form of constituent $A_i$ in order to distinguish it from the latter,
i.e.\ from $A_i^\pm$ in the adjacent bulk phases.
We then account for interfacial chemical reactions between the $A_i^\Sigma$ according to
\begin{equation}
\alpha_1^{\Sigma, a} \, A_1^\Sigma + \ldots + \alpha_N^{\Sigma, a} \, A_N^\Sigma
\rightleftharpoons
\beta_1^{\Sigma, a} \, A_1^\Sigma + \ldots + \beta_N^{\Sigma, a} \, A_N^\Sigma
\quad \mbox{ for } a=1,\ldots ,N_R^\Sigma.
\end{equation}
In full analogy to the bulk case, we let
\begin{equation}
r_i^\Sigma = \sum\limits_{a=1}^{N_R^\Sigma} \nu_i^{\Sigma, a} R_a^\Sigma
\quad \mbox{ with } \nu_i^{\Sigma, a}:= \beta_i^{\Sigma, a} - \alpha_i^{\Sigma, a}
\end{equation}
to obtain $M_i r_i^\Sigma$ for the interfacial production rate of partial mass of $A_i^\Sigma$ and
have
\begin{equation}
\sum\limits_{i=1}^N M_i \nu_i^{\Sigma ,a} =0 \quad \mbox{ for all } a=1,\ldots ,N_R^\Sigma.
\end{equation}

We focus on the so-called class-I model, in which only the constituents' partial mass is balanced individually.
But we shall incorporate structural information from the class-II model, where the latter also includes partial momenta of the constituents and is given in the appendix.
Starting point for the class-I continuum thermodynamics of such systems therefore are the balance equations for partial and total mass, total momentum and total internal energy, both in the bulk phases and on the interface.

The partial mass balance of constituent $A_i$ (i.e.\ $A_i^\pm$ and $A_i^\Sigma$)
follows from \eqref{master-bal-E2}, \eqref{master-bal-int} with the choice of $\phi = \rho_i$ (partial mass density),
${\bf j}={\bf j}_i$ (diffusive mass flux), $f=M_i r_i$ (partial mass production rate due to chemical reactions) as well as the corresponding
interfacial quantities $\phi^\Sigma = \rho_i^\Sigma$, ${\bf j}^\Sigma={\bf j}_i^\Sigma$ and
$f^\Sigma=M_i r_i^\Sigma$ to the result
\begin{eqnarray}
\label{partial-mass-bal-bulk}
\partial_t \rho_i + \div (\rho_i {\bf v} + {\bf j}_i)=M_i r_i & \mbox{ in } \Omega\setminus \Sigma,\\[1ex]
\label{partial-mass-bal-int}
\partial^\Sigma_t  \rho_i^\Sigma  
+ \divS (\rho_i^\Sigma {\bf v}^\Sigma + {\bf j}_{i}^\Sigma)
 + [ \! [ \rho_i ({\bf v} - {\bf v}^\Sigma) + {\bf j}_i] \! ]\cdot {\bf n}_\Sigma =M_i r_i^{\Sigma} & \mbox{ on } \Sigma.
\end{eqnarray}
In addition, the total mass is conserved, i.e.\ the generic balance holds with $\phi=\rho$, ${\bf j}=0$ and $f=0$ as well as
$\phi^\Sigma=\rho^\Sigma$, ${\bf j}^\Sigma=0$ and $f^\Sigma=0$,
where the total mass density is related to the partial mass densities via
\begin{equation}\label{total-mass-density}
\rho = \sum_{i=1}^N \rho_i, \qquad
\rho^\Sigma = \sum_{i=1}^N \rho_i^\Sigma.
\end{equation}
The total mass balance therefore reads as
\begin{eqnarray}
\label{total-mass-bal-bulk}
\partial_t \rho + \div (\rho {\bf v})=0 & \mbox{ in } \Omega\setminus \Sigma,\\[1ex]
\label{total-mass-bal-int}
\partial^\Sigma_t  \rho^\Sigma  + \divS (\rho^\Sigma {\bf v}^\Sigma )
 + [ \! [ \rho ({\bf v} - {\bf v}^\Sigma)\cdot {\bf n}_\Sigma ] \! ]=0 & \mbox{ on } \Sigma.
\end{eqnarray}
Consistency between the total and the partial mass balances requires the constraints
\begin{equation}\label{flux-constraints}
\sum_{i=1}^N {\bf j}_i = 0, \qquad
\sum_{i=1}^N {\bf j}_i^\Sigma = 0,
\end{equation}
which play an important role in the closure for the diffusive fluxes; see below and cf.\ in particular \cite{BD-MCD}.
Let us note in passing that on the level of equations \eqref{partial-mass-bal-bulk} and \eqref{total-mass-bal-bulk}, say,
the condition $\div \big( \sum_{i=1}^N {\bf j}_i \big)= 0$ would be sufficient. But consistency with the class-II balance equations
actually requires \eqref{flux-constraints}, since ${\bf j}_i = \rho_i ({\bf v}_i - {\bf v})$ with ${\bf v}_i$ the continuum
mechanical velocity of constituent $A_i$ and ${\bf v}$ the barycentric velocity, the latter being consistent with
momentum conservation due to its definition via $\rho {\bf v}= \sum_{i=1}^N \rho_i {\bf v}_i$.

Since we focus on the class-I model in the main text,
there is only a single momentum balance. This concerns the total momentum and it follows from the generic balance by chosing
$\phi=\rho {\bf v}$ (momentum density with barycentric velocity), ${\bf j}=-{\bf S}$ (Cauchy stress), $f=\rho {\bf b}$ (body force) as well as
the corresponding interfacial quantities
$\phi^\Sigma=\rho^\Sigma {\bf v}^\Sigma$, ${\bf j}^\Sigma=-{\bf S}^\Sigma$ and $f^\Sigma=\rho^\Sigma {\bf b}^\Sigma$.
The resulting momentum balance reads
\begin{eqnarray}
\label{momentum-bal-bulk}
\partial_t ( \rho {\bf v}) + \div (\rho {\bf v}\otimes {\bf v} -\Sbf )  =\rho {\bf b},\\[1ex]
\label{momentum-bal-int}
\;\; \pa^\Sigma_t(\rho^\Sigma \vbf^\Sigma)+ \na_\Sigma \cdot(\rho^\Sigma \vbf^\Sigma \otimes \vbf^\Sigma - \Sbf^\Sigma)
 + [\![\rho \vbf \otimes (\vbf-\vbf^\Sigma)-\Sbf]\!] \cdot \nbf_\Sigma= \rho^\Sigma {\bf b}^\Sigma , \!\!\!\!
\end{eqnarray}
where \eqref{momentum-bal-bulk} holds in $\Omega\setminus \Sigma$ and \eqref{momentum-bal-int} holds on $\Sigma$.

We finally need the internal energy balance, where one usually starts with the balance of the total energy.
Here, it is helpful to bear in mind the partial momentum balances of a class-II model, since the individual constituents
may experience different body forces, for instance in an electrical field due to different charge numbers.
If $A_i$ is affected by the body force ${\bf b}_i$ with volume-specific density $\rho_i {\bf b}_i$, then
$\rho {\bf b}:= \sum_{i=1}^N \rho_i {\bf b}_i$ is the force density in \eqref{momentum-bal-bulk},
while $\sum_{i=1}^N {\bf v}_i \cdot \rho_i {\bf b}_i$ appears in the total energy balance.
The latter cannot be written solely by ${\bf v}$ and ${\bf b}$, but needs to be rewritten
in the class-I model using ${\bf v}_i = {\bf v} + {\bf j}_i / \rho_i$.
Consequently, the total energy balance results from the generic balance equation with the choice of
$\phi = \rho (e+{\bf v}^2 / 2)$ (density of internal plus kinetic energy), ${\bf j}={\bf q}- {\bf v}^{\sf T} {\bf S}$
(heat flux plus work due to stress inside the fluid), $f=\rho {\bf b} + \sum_{i=1}^N {\bf j}_i \cdot {\bf b}_i$ as well as the corresponding interfacial quantities
$\phi^\Sigma = \rho^\Sigma (e^\Sigma+ ({\bf v}^\Sigma)^2 / 2)$, ${\bf j}^\Sigma={\bf q}^\Sigma- ({\bf v}^\Sigma)^{\sf T} {\bf S}^\Sigma$ and ${\bf f}^\Sigma =\sum_{i=1}^N {\bf j}_i^\Sigma \cdot {\bf b}_i^\Sigma$.

Subtracting the balance of kinetic energy, derived from the momentum balance, a straightforward calculation yields
\begin{equation}\label{E74}
\pa_t(\rho e)+ \na \cdot (\rho e \vbf +\qbf)= \na\vbf:\Sbf+ \sum\limits^N_{i=1} \jbf_i \cdot \bbf_i \quad \mbox{in } \Omega\setminus \Sigma
\end{equation}
as the internal energy balance inside the bulk phases, and
\begin{equation}\label{E75}
\begin{split}
&\pa^\Sigma_t (\rho^\Sigma e^\Sigma)+\na_\Sigma \cdot(\rho^\Sigma e^\Sigma \vbf^\Sigma+ \qbf^\Sigma)
 + [\![\dot{m}\,\big( e+ \frac{(\vbf-\vbf^\Sigma)^2}{2}\big) ]\!]\\
& -[\![(\vbf-\vbf^\Sigma) \cdot (\Sbf \, \nbf_\Sigma)]\!]+ [\![\qbf \cdot \nbf_\Sigma]\!]= \na_\Sigma \vbf^\Sigma :\Sbf^\Sigma + \sum\limits^N_{i=1} \jbf^\Sigma_i \cdot \bbf^\Sigma_i \quad \mbox{on } \Sigma
\end{split}
\end{equation}
on the interface, where we employed the abbreviations
\begin{equation}\label{mdot}
\dot{m}^\pm := \rho^\pm ({\bf v} - {\bf v}^\Sigma )\cdot {\bf n}_\Sigma
\end{equation}
for the one-sided mass transfer fluxes.

For a better comparison with results from the literature, we also collect the non-conservative Lagrangian form
(w.r.\ to the barycentric interface velocity) of the interfacial
balances of mass, partial mass, momentum and internal energy. With the interfacial mass fractions $y_i^\Sigma := \rho_i^\Sigma /\rho^\Sigma$, these read as
\begin{align}\label{non-cons-balances1}
 \frac{D^\Sigma}{Dt} \rho^\Sigma  & + \rho^\Sigma \nabla_\Sigma \cdot \vbf^\Sigma + [\![\dot{m}  ]\!]= 0
 & \mbox{on } \Sigma,\\[0.5ex]
 \rho^\Sigma \frac{D^\Sigma}{Dt} y_i^\Sigma & + \nabla_\Sigma \cdot \jbf_i^\Sigma +[\![\dot{m}\,( y_i - y_i^\Sigma) + \jbf_i \cdot \nbf_\Sigma ]\!] = M_i r_i^\Sigma\label{non-cons-balances2}
  & \mbox{on } \Sigma,\\[0.5ex]
 \rho^\Sigma \frac{D^\Sigma}{Dt} \vbf^\Sigma & - \nabla_\Sigma \cdot \Sbf^\Sigma
+ [\![\dot{m}\,( \vbf - \vbf^\Sigma) - \Sbf \, \nbf_\Sigma ]\!] = \rho^\Sigma \bbf^\Sigma\label{non-cons-balances3}
& \mbox{on } \Sigma, \\[0.5ex]
 \rho^\Sigma \frac{D^\Sigma}{Dt} e^\Sigma & + \nabla_\Sigma \cdot \qbf^\Sigma  + [\![\dot{m}\,\big( e - e^\Sigma + \frac{(\vbf-\vbf^\Sigma)^2}{2}\big)]\!] \label{non-cons-balances4} & \\[-1.5ex]
& \hspace{-0.4in} -[\![(\vbf-\vbf^\Sigma) \cdot (\Sbf \, \nbf_\Sigma)]\!]+ [\![\qbf \cdot \nbf_\Sigma]\!]=
 \na_\Sigma \vbf^\Sigma :\Sbf^\Sigma + \sum\limits^N_{i=1} \jbf^\Sigma_i \cdot \bbf^\Sigma_i  & \mbox{on } \Sigma.\nonumber
\end{align}
The balance equations above (bulk and interface) coincide with the corresponding equations in the main references
\cite{Bedeaux} and \cite{Slattery-Interfaces}.
The balance equations contain the constitutive quantities $\jbf_i$, $r_i$, $\Sbf$, $\qbf$, $\jbf_i^\Sigma$, $r_i^\Sigma$, $\Sbf^\Sigma$ and $\qbf^\Sigma$, as well as all transfer rates, i.e.\ the one-sided values of the quantities in the jump brackets
in, say, equations \eqref{non-cons-balances1}--\eqref{non-cons-balances4}. The latter are not all independent, since the total
mass balance is the sum of the partial mass balances. Therefore, the primary constitutive quantities for transfer rates are
the partial mass transfer rates
\begin{align}\label{midot}
& \dot{m}_i := \big( \rho_i (\vbf - \vbf^\Sigma) + \jbf_i \big) \cdot \nbf_\Sigma \quad
\big( i.e.,\;\; \dot{m}_i^\pm := \big( \rho_i^\pm (\vbf^\pm - \vbf^\Sigma) + \jbf_i^\pm \big) \cdot \nbf_\Sigma \big).
\end{align}
Evidently, it holds that
\begin{align}\label{midot-mdot}
& \dot{m} = \sum_{i=1}^N \dot{m}_i, \qquad
 \dot{m}_i = y_i \, \dot{m} + \jbf_i \cdot \nbf_\Sigma.
\end{align}
In particular, the one-sided total mass transfer rates $\dot{m}^\pm$ are determined by the $\dot{m}_i^\pm$.
Next, knowing $\dot{m}$ and with $\vbf$ and $\vbf^\Sigma$ being independent variables in the balance equations, the jump term in \eqref{non-cons-balances3} contains the interfacial traces of the bulk stresses, i.e.
\begin{align}\label{stressbc}
 & \Sbf_{|\Sigma}\, \nbf_\Sigma \qquad \big(i.e., \;\; \big( \Sbf^\pm_{|\Sigma} \nbf_\Sigma \big)_{||}
 \mbox{ and } \nbf_\Sigma \cdot \Sbf^\pm_{|\Sigma} \nbf_\Sigma \big),
\end{align}
as constitutive quantities.
Finally, knowing $\dot{m}$ as well as $\Sbf_{|\Sigma} \nbf_\Sigma$
and with $\vbf$, $e$ and $\vbf^\Sigma$, $e^\Sigma$ being independent variables in the balance equations, the jump term in \eqref{non-cons-balances4}
is determined up to the remaining constitutive (one-sided) quantities
\begin{align}\label{heatfluxbc}
 & \qbf_{|\Sigma} \cdot \nbf_\Sigma \qquad \big(i.e.,\;\; \qbf^{\pm}_{|\Sigma} \cdot \nbf_\Sigma\big).
\end{align}
All constitutive quantities need to be related to the independent variables (and derivatives thereof)
by material dependent functions, the so-called closure relations.
These functions cannot be arbitrary, but need to satisfy certain principles, namely the principle of material frame indifference
and, most important for the present study, the second law of thermodynamics, i.e.\ the entropy principle including
the entropy inequality. We focus on the latter and only refer to \cite{M85, BD2015, Dreyer-Guhlke-M} and the references given there
concerning the principle of material frame indifference and other principles which might be imposed.
%
%
%
\section{Entropy Principle - The Second Law of Thermodynamics}
We employ the entropy principle in a sharpened version which has been introduced in \cite{BD2015}, but extend it from single-phase
to two-phase multicomponent fluid systems.
In order to state the entropy principle in an axiomatic form, we assign to every physical quantity an associated {\it parity},
called {\it positive} (+1) or {\it negative} (-1).
The original concept of parity for a time-dependent quantity characterizes its behavior under time reversal (see \cite{Meixner73}),
but in the continuum setting, the symmetry of the solutions under time reversal is lost, because the partial
differential equations (PDEs)
describe irreversible processes. However, for studying the {\em structure} of the (unclosed) balance equations,
the concept of reversibility does {\em not} play a role.
This observation allows to adapt and extend the notion of parity to the balance equations of continuum mechanics.
It turns out (see \cite{BD2015} for details) that
the parity of a physical quantity is simply obtained by assigning the factor -1 if the unit of time, ''second'', appears with an uneven power and +1 if it appears with even power. Here we rely on the SI base units m, kg, s, K, mol, A, cd. Note that the unit \textit{Ampere} (A) for the electric current, which does not occur in the present mixture theory\footnote{cf.\ \cite{Dreyer-Guhlke-M} for an extension to electrothermodynamics}, is also among the SI base units. If the latter unit is involved, a further factor -1 is assigned if the unit Ampere appears with uneven power. For instance, the combination A\hspace{1pt}s leads to even parity. The following example yields the parity of the densities of mass, momentum and internal energy according to
\begin{equation} \label{parity-examples}
[\rho]=\frac{\rm kg}{\rm m^3} \to+1,\qquad [\rho {\bf v}]=\frac{\rm kg}{\rm m^2 s}\to -1,\qquad [\rho e]=\frac{\rm kg}{\rm m\, s^2}\to+1.
\end{equation}
Evidently, time derivatives alter the parity, while spatial derivatives keep the parity unchanged.

Any solution of the above system of partial
differential equations, composed of the balance equations \eqref{partial-mass-bal-bulk}, \eqref{partial-mass-bal-int},
\eqref{momentum-bal-bulk}, \eqref{momentum-bal-int}, \eqref{E74} and \eqref{E75},
is called a {\it thermodynamic process}. Here, by a solution we
mean functions which satisfy the equations in a local sense, i.e.\ they do not form the solution
of a complete initial boundary value problem.
In particular, the value of a quantity and of its spatial derivative can
thus be chosen independently by adjusting the time derivative appropriately.
With this notion, the {\em second law of thermodynamics} consists
of the following axioms.\\[-2ex]
\begin{enumerate}
\item[(I)]
There are entropy/entropy-flux pairs $(\rho s, \Phi)$ and $(\rho^\Sigma s^\Sigma, \Phi^\Sigma)$ as material dependent quantities,
where $\rho s$, $\rho^\Sigma s^\Sigma$ are objective scalars and $\Phi$, $\Phi^\Sigma$ are objective vectors.\vspace{0.05in}
\item[(II)]
The pairs $(\rho s, \Phi)$, $(\rho^\Sigma s^\Sigma, \Phi^\Sigma)$ satisfy the balance equations
\begin{eqnarray}\label{E69}
\pa_t(\rho s)+ \na \cdot (\rho s \vbf+\Phibf)=\zeta & \mbox{in } \Omega\setminus \Sigma,\\[1ex]
\label{E70}
\qquad \pa^\Sigma_t(\rho^\Sigma s^\Sigma) + \na_\Sigma \cdot (\rho^\Sigma s^\Sigma \vbf^\Sigma+ \Phibf^\Sigma)+ [\![\rho s(\vbf-\vbf^\Sigma)+\Phibf]\!] \cdot \nbf_\Sigma=\zeta^\Sigma & \mbox{on } \Sigma,
\end{eqnarray}
where $\zeta$ and $\zeta^\Sigma$ are the entropy production rates in the bulk and on the interface, respectively.
The specific entropies $s,\, s^\Sigma$ have the physical dimension ${\rm J}\, {\rm kg}^{-1}\, {\rm K}^{-1}={\rm m}^2
{\rm s}^{-2} {\rm K}^{-1}$,
hence are of positive parity. The entropy fluxes and the entropy productions thus have negative parity.\vspace{0.05in}
\item[(III)]
Any admissible pair of entropy fluxes $\Phi$, $\Phi^\Sigma$ is such that\\[0.5ex]
    (i) the resulting entropy production rates $\zeta$ and $\zeta^\Sigma$ both consist of sums of binary products according to
    \begin{equation}\label{M15}
    \zeta = \sum_m \mathcal{N}_m \mathcal{P}_m, \qquad \zeta^\Sigma = \sum_{m^\Sigma} \mathcal{N}_{m^\Sigma}^\Sigma \mathcal{P}_{m^\Sigma}^\Sigma,
    \end{equation}
    where the $\mathcal{N}_m$, $\mathcal{N}_{m^\Sigma}^\Sigma$ denote objective quantities of negative parity, while $\mathcal{P}_m$, $\mathcal{P}_{m^\Sigma}^\Sigma$ refer to objective quantities of positive parity.\\[0.25ex]
    (ii) Each binary product corresponds to a dissipative mechanism, and the set of these must be fixed in advance. Then,
    \begin{equation}\label{strong-form}
    \mathcal{N}_m \, \mathcal{P}_m \geq 0 \;\mbox{ and }\; \mathcal{N}_{m^\Sigma}^\Sigma \, \mathcal{P}_{m^\Sigma}^\Sigma \geq 0
    \end{equation}
    for all $m$, ${m^\Sigma}$ and for every thermodynamic process.\\[1ex]
\hspace{-0.4in} In addition to these universal axioms, we impose two more specific ones
which refer to the most general constitutive models we are interested in. These are:\vspace{0.1in}
\item[(IV)]
For the class of fluid mixtures under consideration, we restrict the dependence of the entropy according to
\begin{equation}\label{E72}
\rho s=\rho s(\rho e, \rho_1, \dots, \rho_N) \quad\text{ and }\quad \rho^\Sigma s^\Sigma=\rho^\Sigma s^\Sigma(\rho^\Sigma e^\Sigma, \rho^\Sigma_1, \dots, \rho^\Sigma_N),
\end{equation}
where $\rho s$ and $\rho^\Sigma s^\Sigma$ are {\it strictly concave functions} which are strictly increasing in the first argument
and satisfy the principle of material frame indifference.
By means of these functions, we define the
{\it (absolute) temperature} $T$ and {\it chemical potentials} $\mu_i$ according to
\begin{equation}\label{E73}
\frac{1}{T}:= \frac{\pa \rho s}{\pa \rho e}, \quad - \frac{\mu_i}{T}:= \frac{\pa \rho s}{\pa \rho_i} \quad\text{and}\quad \frac{1}{T^\Sigma}:= \frac{\pa \rho^\Sigma s^\Sigma}{\pa \rho^\Sigma e^\Sigma}, \quad - \frac{\mu^\Sigma_i}{T^\Sigma}:= \frac{\pa \rho^\Sigma s^\Sigma}{\pa \rho^\Sigma_i}.
\end{equation}
\vspace{0.05in}
\item[(V)]
The following {\it dissipative mechanisms} occur in the fluid mixtures under consideration:
{\it mass diffusion}, {\it chemical reaction},
{\it viscous flow} (including bulk and shear viscosity) and {\it heat conduction} both in the bulk phases and in the interface
as well as {\it sliding friction}, {\it energy transfer} and {\it ad- and desorption} between the two bulk phases and the interface.
\end{enumerate}
At this point, several remarks about the above axiomatic entropy principle are in order.
A more comprehensive discussion of the entropy principle can be found in \cite{BD2015} and the references given there.\\[0.5ex]
\indent
1. Axiom (I) and Axiom (III-i) restrict the possible constitutive relations for the
entropy fluxes. These condition will not necessarily lead to a unique
choice of $\Phi$ and $\Phi^\Sigma$. More than this,
the full set of axioms constituting the above entropy principle does not
lead to a uniquely defined model, even for the same set of independent variables.
In fact, it is possible to modify the entropy flux such that the same entropy production is equivalently
rewritten with, say, an additional binary product, possibly accompanied by changes in other binary products.
In other words, a zero addition is then distributed to the different terms in the entropy production, changing
the entropy flux as well as a binary product and adding a new dissipative mechanism.
Two examples of this type can be found in \cite{BD2015}.\\[0.5ex]
\indent
2. In total, axiom (III) implies that {\it the entropy inequality} holds,
stating that $\zeta \geq 0$ and $\zeta^\Sigma \geq 0$ for all thermodynamic processes.
The entropy inequality is also called {\it the 2$^{nd}$ law of thermodynamics}.
A thermodynamic process for which $\zeta =0$ and $\zeta^\Sigma =0$ is said to be in {\it thermodynamic equilibrium}.
This statement is to be understood in a pointwise sense; in particular, this must not hold everywhere,
i.e.\ thermodynamic equilibrium can be attained locally.
A thermodynamic process is called {\it reversible} if $\zeta =0$ and $\zeta^\Sigma =0$ everywhere.\\[0.5ex]
\indent
3. Axiom (III-ii) is a strengthened form of the principal of detailed balance,
which essentially says that in equilibrium all individual dissipative mechanisms are themselves in equilibrium,
i.e.\ zero entropy production implies that all binary products vanish.
The latter property can be guaranteed if every binary product has non-negative entropy production for any thermodynamic
process, hence the content of axiom (III-ii).

The required non-negativity of each individual binary product has strong implications.
It required a block-diagonal closure, where the different blocks correspond to the dissipative mechanisms.
Consequently, cross-couplings have to be introduced by mixing between different binary products,
thus merging a certain number of binary products into a new single one. We named this modeling step
{\it entropy invariant mixing between dissipative mechanisms} in \cite{BD2015}.
The advantage of this approach, as compared to the classical closure with cross-couplings,
is that Onsager or Onsager-Casimir relations follow automatically.
In order for this to work properly, the concept of parity is crucial.
Note that if entropy invariant mixing is applied, the new combined mechanism is also
required to be represented by a binary product (of objective quantities of a certain tensorial rank)
with one factor being of positive and one of negative parity.
In other words, one can mix only terms within the same parity group.
For details we refer to sections~8 and 11 in \cite{BD2015}.\\[0.5ex]
\indent
4. There is no axiom (III-i) in the entropy principles of M\"uller in
\cite{M85} or Alt in \cite{Al09}. Instead, the versions of Müller and Alt rely on the
{\it principle of equipresence} whereupon all
constitutive quantities may depend on the same set of variables.
For example, if the velocity gradient appears in the constitutive
law of the stress tensor it should also appear in the constitutive
law of the entropy density. The entropy flux in their theories is completely determined by
the principle of equipresence.
The above version does not use this principle and, therefore, is much simpler to exploit.
In particular for advanced constitutive models the equipresence principle requires an enormous
modeling effort, cf.\ \cite{Liu}.
However, without equipresence, more preliminary knowledge is required.

Concerning model derivation via maximization of the entropy production, see \cite{Raj-Srini}.\\[0.5ex]
\indent
5. Axiom (IV) implicitly contains an assumption of local thermodynamical equilibrium in order
to formulate relations of the type \eqref{E72}. But note that this does not include a statement
of local thermodynamic equilibrium between the interface and the adjacent bulk positions.
For a discussion of the assumption of local thermodynamical equilibrium see \cite{Savin}.\\[0.5ex]
\indent
6. Axiom (V) requires a priori information on the underlying physics, which is to be modeled,
in terms of the dissipative mechanisms; cf.\ Remark~1 above.
At this point, knowledge on equilibrium thermodynamics is helpful,
identifying those quantities that vanish in thermodynamic equilibrium.
For example, we know by experience that in a single-phase multicomponent fluid at thermodynamic equilibrium
we must have
\begin{equation}\label{equi1}
R_a =0 , \quad {\bf j}_i =0 ,\quad \nabla {\bf v} + \nabla {\bf v}^{\sf T} = 0,
\quad  \nabla T=0.
\end{equation}
Correspondingly, these conditions motivate the constitutive axiom (V), saying that the (bulk) mechanisms that drive a fluid mixture to equilibrium are: chemical reactions, mass diffusion, viscous flow and heat
conduction.\\[1ex]
\noindent
In order to utilize the entropy principle, we need to calculate the structure of the entropy production rates
by means of the left-hand sides of the bulk and interface entropy balance equations.
For this purpose, we insert the entropy representations from \eqref{E72} into the entropy balances (\ref{E69}) and (\ref{E70}).
Then, using the chain rule and \eqref{E73}, the balance equations for mass, momentum and internal energy are employed to eliminate
all time derivatives. At this point, we make a general simplification by assuming the fluids to be non-polar in the sense
that the stresses are symmetric, i.e.\
\begin{equation}\label{sym-stress}
{\bf S}^{\sf T} = {\bf S}   \quad \mbox{ and } \quad \big({\bf S}^\Sigma \big)^{\sf T} = {\bf S}^\Sigma.
\end{equation}
We also split the stresses into an isotropic and a traceless part according to
\begin{equation}\label{spli-stress}
{\bf S}=- P {\bf I} + {\bf S}^\circ   \quad \mbox{ and } \quad {\bf S}^\Sigma = - P^\Sigma {\bf I}_\Sigma + {\bf S}^{\Sigma, \circ}
\end{equation}
with the {\em mechanical pressure} $P$ (not to be mistaken with the thermodynamic pressure $p$ introduced below)
and the {\em mechanical interfacial pressure} $P^\Sigma$ defined as
\begin{equation}\label{mech-pressure}
P=\, - \frac 1 3 {\rm trace} ( {\bf S}) \quad \mbox{ and } \quad
P^\Sigma = - \frac 1 2 {\rm trace} ({\bf S}^\Sigma ).
\end{equation}
Moreover, the ''identity'' ${\bf I}_\Sigma$ on the interface denotes the projection onto the local tangential plane, defined as
\begin{equation}\label{intidentity}
{\bf I}_\Sigma = {\bf I} - {\bf n}_\Sigma \otimes {\bf n}_\Sigma.
\end{equation}
After lengthy but straightforward calculations, this procedure yields
\begin{equation}\label{E76}
\begin{split}
\zeta= & \na \cdot(\Phibf- \frac{\qbf}{T}+ \sum\limits^N_{i=1} \frac{\mu_i \jbf_i}{T})- \frac{1}{T} (\rho e + P-\rho sT- \sum\limits_{i=1}^N \rho_i \mu_i) \na \cdot \vbf\\
& + \qbf \cdot \na \frac{1}{T}+ \frac{1}{T} \Sbf^{\circ} :{\bf D}^{\circ}- \sum\limits_{i=1}^N \jbf_i \cdot (\na \frac{\mu_i}{T}- \frac{\bbf_i}{T})- \frac{1}{T} \sum\limits_{a=1}^{N_R} R_a \mathcal{A}_a,
\end{split}
\end{equation}
where ${\bf D}=\frac 1 2 \big( \na \vbf + (\na \vbf)^{\sf T} \big)$ is the symmetric velocity gradient, ${\bf D}^{\circ}$
its traceless part and $\mathcal{A}_a$  are the affinities associated with the $a^{\rm th}$ chemical bulk reaction, defined as
\begin{equation}\label{E77}
\mathcal{A}_a = \sum\limits^N_{i=1} M_i \mu_i \nu^a_i.
\end{equation}
In between, we exploited the symmetry of ${\bf S}$ for rewriting the co-factor as the objective tensor ${\bf D}$ rather than
$\na \vbf$, which is not objective.

For the interfacial entropy production, we obtain
\begin{align}
\label{E78}
\zeta^\Sigma & = \na_\Sigma \cdot(\Phibf^\Sigma- \frac{\qbf^\Sigma}{T^\Sigma}+ \sum\limits^N_{i=1} \frac{\mu^\Sigma_i \jbf^\Sigma_i}{T^\Sigma})\\[-0.75ex]
 & - \frac{1}{T^\Sigma} \big(\rho^\Sigma e^\Sigma + P^\Sigma -\rho^\Sigma s^\Sigma T^\Sigma - \sum\limits^N_{i=1} \rho^\Sigma_i \mu^\Sigma_i \big) \na_\Sigma \cdot \vbf^\Sigma  + \qbf^\Sigma \cdot \na_\Sigma \frac{1}{T^\Sigma}\nonumber\\[-0.75ex]
&  + \frac{1}{T^\Sigma} \Sbf^{\Sigma, \circ} : {\bf D}^{\Sigma,\circ} 
- \sum\limits^N_{i=1} \jbf^\Sigma_i \cdot \big(\na_\Sigma \frac{\mu^\Sigma_i}{T^\Sigma}- \frac{\bbf^\Sigma_i}{T^\Sigma}\big)- \frac{1}{T^\Sigma} \sum\limits^{N_R^\Sigma}_{a=1} R^\Sigma_a \mathcal{A}^\Sigma_a  \nonumber\\[0.75ex]
	& +  \frac{1}{T^\Sigma}  [\![ (\vbf -\vbf^\Sigma) \cdot (\Sbf  \, \nbf_\Sigma ) ]\!] 
+ [\![ \Big(\frac{1}{T}- \frac{1}{T^\Sigma}\Big) \qbf \cdot \nbf_\Sigma ]\!] + [\![  \dot{m}\, s  ]\!]\nonumber\\[0.75ex]
& - \frac{1}{T^\Sigma} [\![ \dot{m} \,\big(  e + \frac{(\vbf -\vbf^\Sigma)^2}{2} \big) ]\!]
+ \sum\limits^N_{i=1} [\![ \dot{m}_i \frac{\mu^\Sigma_i}{T^\Sigma} ]\!]
- \sum\limits^N_{i=1} [\![ \frac{\mu_i}{T} \, {\bf j}_i \cdot \nbf_\Sigma ]\!]. \nonumber
\end{align}
In \eqref{E78}, we use the notation
\begin{equation}\label{rate-of-defo-int}
{\bf D}^\Sigma=\frac 1 2 {\bf I}_\Sigma \big( \na_\Sigma \vbf^\Sigma + (\na_\Sigma \vbf^\Sigma)^{\sf T} \big) {\bf I}_\Sigma
\end{equation}
for the interfacial rate of deformation tensor and ${\bf D}^{\Sigma,\circ}$ for its traceless part. We write
\begin{equation}\label{dotmi}
\dot{m}_i^\pm = \big( \rho_i^\pm ( {\bf v}- {\bf v}^\Sigma ) + {\bf j}_i \big) \cdot {\bf n}_\Sigma
\end{equation}
for the one-sided mass transfer rates of constituent $A_i$, describing the net mass exchange rate between an adjacent bulk phase and
the interface, i.e.\ adsorption minus desorption (or the other way around, depending on the orientation of ${\bf n}_\Sigma$).
Let us note that $\dot{m}_i^\pm = \rho_i^\pm ( {\bf v}_i- {\bf v}^\Sigma )\cdot {\bf n}_\Sigma$ holds
in the class-II model given in the appendix.

The decomposition of the entropy production rates into binary products can be done in different ways as it can be seen in
the literature and will be recalled in section~\ref{comparison}. Before we compare especially the interfacial entropy production rate and its structure with known representations,
we first discuss the closure procedure for the constitutive quantities. This also allows for a more complete review of the published results.
\section{Closure Relations}
According to the entropy principle above, the entropy fluxes are constitutive quantities which have to be modeled as objective vectors in such a way that the resulting entropy productions become sums of binary products with objective factors,
one for each dissipative mechanism.
Due to the structure of the first term in the respective production rate, we choose the constitutive relations
\begin{equation}\label{ef-bothe}
\Phibf= \frac{\qbf}{T}- \sum\limits^N_{i=1} \frac{\mu_i \jbf_i}{T} \quad\text{ and }\quad \Phibf^\Sigma= \frac{\qbf^\Sigma}{T^\Sigma}- \sum\limits^N_{i=1} \frac{\mu^\Sigma_i \jbf^\Sigma_i}{T^\Sigma}
\end{equation}
for the entropy fluxes. Hence the so-called {\em reduced entropy production rates} are
\begin{equation}\label{ep-bulk-reduced}
\begin{split}
\zeta= &  - \frac{1}{T} (\rho e + P-\rho sT- \sum\limits_{i=1}^N \rho_i \mu_i) \na \cdot \vbf  + \qbf \cdot \na \frac{1}{T}\\
& + \frac{1}{T} \Sbf^{\circ} :{\bf D}^{\circ}- \sum\limits_{i=1}^N \jbf_i \cdot (\na \frac{\mu_i}{T}- \frac{\bbf_i}{T})- \frac{1}{T} \sum\limits_{a=1}^{N_R} R_a \mathcal{A}_a
\end{split}
\end{equation}
for the bulk phases and
\begin{align}
\label{E78reduced}
 \zeta^\Sigma = &
 - \frac{1}{T^\Sigma}\big(\rho^\Sigma e^\Sigma + P^\Sigma -\rho^\Sigma s^\Sigma T^\Sigma - \sum\limits^N_{i=1} \rho^\Sigma_i \mu^\Sigma_i \big) \na_\Sigma \cdot \vbf^\Sigma  + \qbf^\Sigma \cdot \na_\Sigma \frac{1}{T^\Sigma} \\
& + \frac{1}{T^\Sigma} \Sbf^{\Sigma, \circ} : {\bf D}^{\Sigma,\circ} 
- \sum\limits^N_{i=1} \jbf^\Sigma_i \cdot \left(\na_\Sigma \frac{\mu^\Sigma_i}{T^\Sigma}- \frac{\bbf^\Sigma_i}{T^\Sigma}\right)- \frac{1}{T^\Sigma} \sum\limits^{N_R^\Sigma}_{a=1} R^\Sigma_a \mathcal{A}^\Sigma_a  \nonumber\\
	& +  \frac{1}{T^\Sigma}  [\![ (\vbf -\vbf^\Sigma) \cdot (\Sbf  \, \nbf_\Sigma ) ]\!] 
+ [\![ \Big(\frac{1}{T}- \frac{1}{T^\Sigma}\Big) \qbf \cdot \nbf_\Sigma ]\!] + [\![  \dot{m}\, s  ]\!]\nonumber\\
& - \frac{1}{T^\Sigma} [\![ \dot{m} \,\big(  e + \frac{(\vbf -\vbf^\Sigma)^2}{2} \big) ]\!]
+ \sum\limits^N_{i=1} [\![ \dot{m}_i \frac{\mu^\Sigma_i}{T^\Sigma} ]\!]
- \sum\limits^N_{i=1} [\![ \frac{\mu_i}{T} \, {\bf j}_i \cdot \nbf_\Sigma ]\!]. \nonumber
\end{align}
for the interface.\\[0.5ex]
\noindent
{\bf Closure relations for bulk constitutive quantities}.
While the focus lies on the interfacial dissipative mechanisms and closure procedure, we start with a few remarks on the
bulk closure.
For the first three binary products in \eqref{ep-bulk-reduced},
we employ the standard linear (in the co-factors) closure, i.e.\ we let
\begin{align}
\rho e - \rho s T + P - \sum_{i=1}^N \rho_i \mu_i  &  =\,   - \lambda \, \div {\bf v},\vspace{-0.1in}\label{closure-1}\\
{\bf q} & = \tilde\alpha \nabla \frac 1 T = - \alpha \nabla  T,\label{closure-2}\\[1.5ex]
\stress^\circ & = 2 \eta {\bf D}^\circ,\label{closure-3}
\end{align}
where the phenomenological coefficients $\lambda, \alpha, \eta \geq 0$ depend on the basic thermodynamic variables which are, after a Legendre transformation, $T,\rho_1, \ldots ,\rho_N$.

Since a closure without cross-effects renders every individual binary product non-negative for any thermodynamic process, it is clear that the co-factors of the constitutive quantities in \eqref{closure-1}, \eqref{closure-2} and \eqref{closure-3} must vanish in equilibrium, i.e.\
\begin{equation}\label{equilibrium}
\div {\bf v} =0, \quad \nabla T = 0, \quad {\bf D}^\circ = {\bf 0}.
\end{equation}
Therefore, the mechanical pressure $P$ at equilibrium satisfies
\begin{equation}\label{mech-press-equi}
P_{|equ} = - \rho e + \rho s T + \sum_{i=1}^N \rho_i \mu_i.
\end{equation}
Motivated by \eqref{mech-press-equi}, we define the \emph{thermodynamic pressure} as
\begin{equation}\label{pressure}
p = - \rho e + \rho s T + \sum_{i=1}^N \rho_i \mu_i.
\end{equation}
In order to interchange $\rho e$ with $T$ as independent variables, we introduce the free energy density as
\begin{equation}\label{free-energy}
\rho \psi = \rho e - \rho s T.
\end{equation}
It is straightforward to show that with $\psi = \psi (T,\rho, y_1, \ldots ,y_N)$, where the mass fractions
\begin{equation}\label{mass-fractions}
y_i = \frac{\rho_i}{\rho}
\end{equation}
satisfy the constraint $\sum_{i=1}^N y_i = 1$,
the thermodynamic pressure fulfills the relation
\begin{equation}\label{pressure-MR}
p = \rho^2 \frac{\partial \psi}{\partial \rho}.
\end{equation}
Conversely, if the thermodynamic pressure is defined according to \eqref{pressure-MR}, as is often done, than
\eqref{mech-press-equi} shows that $p$
necessarily satisfies the Euler (or, Gibbs-Duhem) relation, i.e.\
\begin{equation}\label{Gibbs-Duhem}
\rho e - \rho s T + p = \sum_{i=1}^N \rho_i \mu_i.\vspace{-0.1in}
\end{equation}
We now let\vspace{-0.2in}\\
\begin{equation}\label{dyn-pressure}
\pi = P-p
\end{equation}
denote the irreversible pressure part, also called frictional (or dynamic) pressure part.
Then the closure \eqref{closure-1} becomes
\[
\pi = - \lambda \, \div {\bf v}
\]
and models a viscous pressure contribution due to volume variations, where the material dependent parameter $\lambda$ is called
bulk viscosity.

The fourth product in \eqref{ep-bulk-reduced} corresponds to multicomponent diffusion in the bulk phases.
The standard linear closure requires first to incorporate the constraint \eqref{flux-constraints} on the diffusion fluxes,
say by eliminating $\jbf_N$. This leads to the mass diffusion fluxes
\begin{equation}\label{E31}
\jbf_i=- \sum\limits^{N-1}_{k=1} L_{ik} \cdot \left(\na \frac{\mu_k- \mu_N}{T}- \frac{\bbf_k-\bbf_N}{T}\right),
\end{equation}
where the matrix $[L_{ik}]$ of mobilities (so-called Onsager coefficients) is positive definite. It can also be shown that $[L_{ik}]$ is symmetric, i.e.\ the Onsager symmetries hold; see \cite{CT62} for the symmetry of the matrix $[L_{ik}]$ and
\cite{BD2015} for a rigorous theory of Onsager symmetries for transport coefficients. A difficulty with this Fick-Onsager form is the fact that the mobilities are complicated functions of the composition - this needs to be the case, since for constant mobilities the positivity of solutions would not be guaranteed. This is a rather involved topic and we
refer to \cite{BD-MCD} for a comprehensive investigation of the structure of continuum thermodynamical diffusion fluxes,
including a novel closure which combines the advantages of the (explicit) Fick-Onsager closure above and the
(implict) Maxwell-Stefan form of multicomponent diffusion modeling.

Next, for completeness, we reproduce the closure of the mass production rates according to \cite{BD2015},
where we recall that $R_a$ is given as forward minus backward rate according to
\[
R_a=R_a^f -R_a^b.
\]
Since chemical reactions are activated processes which often occur far from equilibrium,
a linear (in the affinities) closure for $R_a$ is not appropriate.
Instead, we use the nonlinear closure
\begin{equation}
\label{closure-chem-rates}
\ln \frac{R_a^f}{R_a^b}= - \gamma_a \frac{\mathcal{ A}_a}{RT}
\quad \mbox{ with } \gamma_a >0
\end{equation}
which implies
\[
\zeta_{\rm CHEM} = R \sum_{a=1}^{N_R} \frac{1}{\gamma_a}
(R_a^f - R_a^b) (\ln R_a^f - \ln {R_a^b}) \geq 0
\]
for the contribution by chemical reactions  to the entropy production in the bulk phases,
since the logarithm is monotone increasing. Note that still one of the rates --
either for the forward or the backward path -- needs to be
modeled, while the form of the other one then follows from \eqref{closure-chem-rates}.
This logarithmic closure not only allows to include standard mass action kinetics into this framework (see \cite{BD2015}),
but to provide thermodynamically consistent extensions;
cf.\ \cite{Dreyer-Guhlke-M}, where this is employed for a rigorous derivation and
an extension of the Butler-Volmer equation for fluid interfaces.
Because of the strict monotonicity of the logarithm, the reactive contribution to the entropy
production only vanishes if {\it all reaction are separately in
equilibrium}, i.e.\ all forward and corresponding backward rates
coincide. This is an instance of the {\it principle of detailed
balance}, called Wegscheider's condition in the context of
chemical reaction kinetics.\\[0.5ex]
\noindent
{\bf Closure relations for interfacial constitutive quantities}.
First of all, we rewrite the reduced interfacial entropy production rate, employing the Euler relation
\eqref{Gibbs-Duhem} to express the specific entropy as $s=(e+p/\rho -\sum_{i=1}^N y_i \mu_i)/T$. This yields
\begin{align}
\label{E78reduced2}
 \zeta^\Sigma & =
 - \frac{1}{T^\Sigma}\big(\rho^\Sigma e^\Sigma + P^\Sigma -\rho^\Sigma s^\Sigma T^\Sigma - \sum\limits^N_{i=1} \rho^\Sigma_i \mu^\Sigma_i \big) \na_\Sigma \cdot \vbf^\Sigma  + \qbf^\Sigma \cdot \na_\Sigma \frac{1}{T^\Sigma} \\
& + \frac{1}{T^\Sigma} \Sbf^{\Sigma, \circ} : {\bf D}^{\Sigma,\circ} 
- \sum\limits^N_{i=1} \jbf^\Sigma_i \cdot \left(\na_\Sigma \frac{\mu^\Sigma_i}{T^\Sigma}- \frac{\bbf^\Sigma_i}{T^\Sigma}\right)- \frac{1}{T^\Sigma} \sum\limits^{N_R^\Sigma}_{a=1} R^\Sigma_a \mathcal{A}^\Sigma_a  \nonumber\\
	& +  \frac{1}{T^\Sigma}  [\![ (\vbf -\vbf^\Sigma) \cdot (\Sbf  \, \nbf_\Sigma ) ]\!] 
+ [\![ \Big(\frac{1}{T}- \frac{1}{T^\Sigma}\Big) \big( \dot{m} \, e + \qbf \cdot \nbf_\Sigma \big) ]\!]\nonumber\\
& - \frac{1}{T^\Sigma} [\![ \dot{m} \, \frac{(\vbf -\vbf^\Sigma)^2}{2} ]\!] + [\![  \dot{m} \frac{p}{\rho T} ]\!]
+ \sum\limits^N_{i=1} [\![ \dot{m}_i \big( \frac{\mu^\Sigma_i}{T^\Sigma} - \frac{\mu_i}{T} \big) ]\!]. \nonumber
\end{align}
Since the first binary product with a jump term includes both tangential and normal parts, it is advisable to decompose it into
the corresponding components. From the much simpler case of a single-component two-phase fluid system without interfacial mass,
it is known that the specific enthalpy rather than the specific internal energy is the relevant quantity for energy transmission.
We therefore let
\begin{equation}\label{enthalpies}
h=e + \frac{p}{\rho} \quad \mbox{ and } \quad
h^\Sigma=e^\Sigma + \frac{p^\Sigma}{\rho^\Sigma}
\end{equation}
denote the specific bulk and interface enthalpies.
In order to arrive at an appropriate representation of $\zeta^\Sigma$, we employ the stress decompositon
\begin{equation}\label{irr-stress}
{\bf S} = -p\, {\bf I} + {\bf S}^{\rm irr} \quad \mbox{ and } \quad
{\bf S}^\Sigma = -p^\Sigma {\bf I} + {\bf S}^{\Sigma, \rm irr}
\end{equation}
which defines the irreversible part of the bulk and interface stresses.
With this decomposition, the interfacial entropy production rate can be rewritten as
\begin{align}
\label{E78reduced3}
 \zeta^\Sigma & =
 - \frac{1}{T^\Sigma}\big(\rho^\Sigma e^\Sigma + P^\Sigma -\rho^\Sigma s^\Sigma T^\Sigma - \sum\limits^N_{i=1} \rho^\Sigma_i \mu^\Sigma_i \big) \na_\Sigma \cdot \vbf^\Sigma  + \qbf^\Sigma \cdot \na_\Sigma \frac{1}{T^\Sigma} \\
& + \frac{1}{T^\Sigma} \Sbf^{\Sigma, \circ} : {\bf D}^{\Sigma,\circ} 
- \sum\limits^N_{i=1} \jbf^\Sigma_i \cdot \left(\na_\Sigma \frac{\mu^\Sigma_i}{T^\Sigma}- \frac{\bbf^\Sigma_i}{T^\Sigma}\right)- \frac{1}{T^\Sigma} \sum\limits^{N_R^\Sigma}_{a=1} R^\Sigma_a \mathcal{A}^\Sigma_a  \nonumber\\
	& +  \frac{1}{T^\Sigma}  [\![ (\vbf -\vbf^\Sigma)_{||} \cdot (\Sbf  \, \nbf_\Sigma )_{||} ]\!] 
+ [\![ \Big(\frac{1}{T}- \frac{1}{T^\Sigma}\Big) \big( \dot{m} \, h + \qbf \cdot \nbf_\Sigma \big) ]\!]\nonumber\\
& + \sum\limits^N_{i=1} [\![ \dot{m}_i \big( \frac{\mu^\Sigma_i}{T^\Sigma} - \frac{\mu_i}{T} \big) ]\!]
- \frac{1}{T^\Sigma} [\![ \dot{m} \,  \Big( \frac{(\vbf -\vbf^\Sigma)^2}{2}- \nbf_\Sigma \cdot \frac{{\bf S}^{\rm irr}}{\rho}  \cdot \nbf_\Sigma  \Big) ]\!]. \nonumber
\end{align}
At this point it is not clear, where (i.e., to which dissipative mechanism) the last term belongs to.
One important observation can be made by specializing to the case of continuous temperature, i.e.\ $T^\pm_{|\Sigma}=T^\Sigma$.
In this case the energy transmission does not produce entropy, while the last term stays unchanged, which indicates
that the last term in \eqref{E78reduced3} belongs to the mass transfer entropy production.
Indeed, this can be achieved by using the identity $\dot{m}=\sum_{i=1}^N \dot{m}_i$.
This conjecture is confirmed by inspecting the entropy production rate for the class-II model, which is derived in the appendix
in particular for this reason. In fact, the corresponding term in \eqref{red-CII-ep-int} reads as
\[
\sum\limits_{i=1}^N [\![ \dot{m}_i \Big(\frac{\mu^\Sigma_i}{T^\Sigma}- \frac{\mu_i}{T}
- \frac{1}{T^\Sigma} \Big( \frac{(\vbf_i-\vbf^\Sigma_i)^2}{2 T^\Sigma}
- \nbf_\Sigma \cdot \frac{\Sbf^{\rm irr}_i}{\rho_i T^\Sigma} \cdot \nbf_\Sigma \Big)\Big)]\!].
\]
Next, before the closure procedure can be applied, it is important to notice that in the transfer terms,
the binary products appear inside the jump terms and, hence, cannot be closed in this form.
In fact, the transfer terms always represent two one-sided bulk-interface (dissipative) transfer mechanisms, which are to be closed separately. In order to be able to do so, we rewrite these binary products using the identity
\begin{equation}\label{explicit-jump}
[\![ {\bf f} \cdot  \nbf_\Sigma ]\!] = \, - {\bf f}^+ \cdot {\bf n}^+ - {\bf f}^- \cdot {\bf n}^-,
\end{equation}
where ${\bf n}^\pm$ are the outer unit normal fields to $\Omega^\pm$.
The reduced interfacial entropy production rate then finally becomes
\begin{align}
\label{E29}
	& \zeta^\Sigma = - \frac{1}{T^\Sigma}\big(\rho^\Sigma e^\Sigma +P^\Sigma -\rho^\Sigma s^\Sigma T^\Sigma - \sum\limits^N_{i=1} \rho^\Sigma_i \mu^\Sigma_i \big) \na_\Sigma \cdot \vbf^\Sigma
 + \qbf^\Sigma \cdot \na_\Sigma \frac{1}{T^\Sigma}\\
	&+ \frac{1}{T^\Sigma} \Sbf^{\Sigma,\circ}: {\bf D}^{\Sigma,\circ}  - \sum\limits^N_{i=1} \jbf^\Sigma_i \cdot \left( \na_\Sigma \frac{\mu^\Sigma_i}{T^\Sigma}- \frac{\bbf^\Sigma_i}{T^\Sigma} \right)-
 \frac{1}{T^\Sigma} \sum\limits^{N_R^\Sigma}_{a=1} R^\Sigma_a \mathcal{A}^\Sigma_a\nonumber\\
	&- \frac{1}{T^\Sigma} (\vbf^+-\vbf^\Sigma)_{||} \cdot (\Sbf^+ \nbf^+)_{||}- \frac{1}{T^\Sigma} (\vbf^- -\vbf^\Sigma)_{||} \cdot (\Sbf^- \nbf^-)_{||}\nonumber\\
	&+ \left( \frac{1}{T^\Sigma}- \frac{1}{T^+}\right) \left(\dot{m}^{+,\Sigma}\, h^+ + \qbf^+ \cdot \nbf^+ \right)\nonumber
  + \left( \frac{1}{T^\Sigma}- \frac{1}{T^-}\right) \left( \dot{m}^{-,\Sigma}\, h^-+  \qbf^- \cdot \nbf^-\right)\nonumber\\
	&+ \sum\limits^N_{i=1} \dot{m}^{+,\Sigma}_i \left(\frac{\mu^+_i}{T^+}- \frac{\mu^\Sigma_i}{T^\Sigma}+ \frac{1}{T^\Sigma} \Big(\frac{(\vbf^+-\vbf^\Sigma)^2}{2}-\nbf^+ \cdot \frac{\Sbf^{+,\rm irr}}{\rho^+} \cdot \nbf^+\Big)\right)\nonumber\\
	&+ \sum\limits^N_{i=1} \dot{m}^{-,\Sigma}_i \left(\frac{\mu^-_i}{T^-}- \frac{\mu^\Sigma_i}{T^\Sigma}+ \frac{1}{T^\Sigma} \Big(\frac{(\vbf^--\vbf^\Sigma)^2}{2}-\nbf^- \cdot \frac{\Sbf^{-,\rm irr}}{\rho^-} \cdot \nbf^-\Big)\right).\nonumber
\end{align}
Here the symbols $\dot{m}^{\pm,\Sigma}$ stand for the mass transfer rate from the bulk phase $\Omega^\pm$ to the interface $\Sigma$,
i.e.\ with this fixed direction, independently of the orientation of $\nbf_\Sigma$, which enters into the definition
of $\dot{m}^\pm$.

The first five terms in \eqref{E29} are the direct interface analog of the corresponding bulk terms.
Therefore, we only collect the resulting closure relations except for the multi\-component diffusion mechanism, where we
briefly show how to obtain the Maxwell-Stefan form of the closure.
The first term leads to the definition of the thermodynamic interface pressure $p^\Sigma$ via the interfacial Euler
(or, Gibbs-Duhem) relation, i.e.\
\begin{equation}\label{E80}
\rho^\Sigma e^\Sigma + p^\Sigma-\rho^\Sigma s^\Sigma T^\Sigma= \sum\limits^N_{i=1} \rho^\Sigma_i \mu^\Sigma_i.
\end{equation}
Then the irreversible part $\pi^\Sigma$ of the interface pressure is defined via
\begin{equation}\label{irrev-int-pressure}
P^\Sigma = p^\Sigma + \pi^\Sigma
\end{equation}
and the corresponding closure relation reads as
\begin{equation}\label{irrev-int-p-closure}
\pi^\Sigma = - \lambda^\Sigma {\rm div}_\Sigma \vbf^\Sigma \quad \mbox{ with } \lambda^\Sigma \geq 0.
\end{equation}
Note that $\lambda^\Sigma$ is allowed to depend on $T^\Sigma$ and the $\rho^\Sigma_i$. The latter is also true for all further closure parameters which are introduced below without explicit mentioning.
It is also important to note that the interfacial tension, denoted as $\gamma$ in the physico-chemical interface sciences and as $\sigma$ in the hydrodynamical theory of capillary flows, is related to the interface pressure via
\[
\gamma \, = \,  \sigma \, = \, - p^\Sigma.
\]
Let us remark that in the physico-chemical interface sciences, the notion of interface (or surface) pressure often refers to
the difference $\gamma^{\rm clean} - \gamma^{\rm contam}$ between the interface tension of the clean interface and that of
the interface with surface active agents adsorbed to it.

The second binary product, which relates to heat conduction, leads to Fourier's law for the interfacial heat flux, i.e.\ \begin{equation}\label{E30}
\qbf^\Sigma= -\alpha^\Sigma \na_\Sigma T^\Sigma \quad\text{ with }\quad \alpha^\Sigma \geq 0.
\end{equation}
The interface temperature $T^\Sigma$ is, in general, different from the one-sided limits of the bulk temperature;
see the review paper \cite{Persad-Ward2016review} and the references given there.

The third binary product describes the entropy production due to viscous dissipation inside the interface,
a process related to intrinsic surface viscosities.
Linear (in the co-factor) closure here corresponds to a fluid interface with Newtonian rheology.
This leads to the Boussinesq-Scriven surface stress (\cite{Scriven}, cf.\ also \cite{BP2010}),
i.e.\ \begin{equation}\label{E20}
\Sbf^\Sigma=(-p^\Sigma+(\eta^\Sigma_d-\eta^\Sigma_s) \na_\Sigma \cdot \vbf^\Sigma)\, {\bf I}_\Sigma+ 2\eta^\Sigma_s \, \Dbf^\Sigma
\end{equation}
with the interface dilatational and shear viscosities $\eta^\Sigma_d \geq \eta^\Sigma_s \geq 0$.
For further information about interfacial rheology we refer to \cite{4, Sagis2011, Sagis-Fischer}.

The fourth product in (\ref{E29}) corresponds to multicomponent diffusion on the interface. To derive the Maxwell-Stefan form of interface diffusion, it should first be noted that the diffusion flux is related to a diffusion velocity $\ubf^\Sigma_i$ via
\begin{equation}\label{E32}
\jbf^\Sigma_i= \rho^\Sigma_i \ubf^\Sigma_i:= \rho^\Sigma_i(\vbf^\Sigma_i-\vbf^\Sigma),
\end{equation}
where $\vbf^\Sigma_i$ is the continuum mechanical velocity of the individual adsorbed component $A_i^\Sigma$;
cf.\ the class-II model derived in the appendix.
Then the starting point is the following reformulation of the relevant entropy production contribution. We have
\begin{equation}\label{E33}
\zeta^\Sigma_{\rm DIFF}=-\sum\limits^N_{i=1} \ubf^\Sigma_i \cdot \left(\rho^\Sigma_i \na_\Sigma \frac{\mu^\Sigma_i}{T^\Sigma}- \frac{\rho^\Sigma_i \bbf^\Sigma_i}{T^\Sigma}- \rho^\Sigma_i \Lambda\right)
\end{equation}
with a Lagrange parameter $\Lambda$ which can be arbitrary due to \eqref{flux-constraints}$_2$.
This parameter is chosen in such a way that the co-factors in (\ref{E33}) sum up to zero, hence
\begin{equation}\label{E34}
\Lambda= \frac{1}{\rho^\Sigma} \sum\limits_{i=1}^N \left(\rho_i^\Sigma \na_\Sigma \frac{\mu^\Sigma_i}{T^\Sigma}- \frac{\rho^\Sigma_i \bbf^\Sigma_i}{T^\Sigma}\right).
\end{equation}
Using (\ref{E80}) and the fundamental differential relation
\begin{equation}\label{E27}
d(\rho^\Sigma \psi^\Sigma)=-\rho^\Sigma s^\Sigma dT^\Sigma+ \sum\limits_{i=1}^N \mu^\Sigma_i d\rho^\Sigma_i,
\end{equation}
a straightforward computation shows that (\ref{E33}) becomes
\begin{equation}\label{E35}
\zeta^\Sigma_{\rm DIFF}=- \sum\limits^N_{i=1} \ubf^\Sigma_i \cdot \dbf^\Sigma_i
\end{equation}
with the generalized thermodynamic interface driving forces
\begin{equation}\label{E36}
\dbf^\Sigma_i=\rho^\Sigma_i \na_\Sigma \frac{\mu^\Sigma_i}{T^\Sigma}-\rho^\Sigma_i \frac{\bbf^\Sigma_i-\bbf^\Sigma}{T^\Sigma} -\frac{y^\Sigma_i}{T^\Sigma} \na_\Sigma p^\Sigma-y^\Sigma_i h^\Sigma \na_\Sigma \frac{1}{T^\Sigma},
\end{equation}
where $y^\Sigma_i=\rho^\Sigma_i/ \rho^\Sigma$ are the interfacial mass fractions, $\bbf^\Sigma= \sum_i y^\Sigma_i \bbf^\Sigma_i$ the total interface body force density and $h^\Sigma$ is the interface enthalpy density. From a kinetic theory, or from a more elaborate continuum theory employing partial momenta (see the appendix and \cite{BD2015}), the interface driving forces are
\begin{equation}\label{E37}
\dbf^\Sigma_i= \rho^\Sigma_i \na_\Sigma \frac{\mu^\Sigma_i}{T^\Sigma}- \rho^\Sigma_i \frac{\bbf^\Sigma_i -\bbf^\Sigma}{T^\Sigma} -\frac{y^\Sigma_i}{T^\Sigma} \na_\Sigma p^\Sigma- h^\Sigma_i \na_\Sigma \frac{1}{T^\Sigma}
\end{equation}
with the partial interface enthalpy densities $h^\Sigma_i$. Note that both agree in the isothermal case or if $h^\Sigma_i = y^\Sigma_i h^\Sigma$.
Now, equation (\ref{E35}) is used to derive constitutive equations for the $\dbf^\Sigma_i$, after $\dbf^\Sigma_N$, say, is eliminated by means of the constraint
\begin{equation}\label{E38}
\sum\limits^N_{i=1} \dbf^\Sigma_i=0.
\end{equation}
Linear closure yields
\begin{equation}\label{E39}
\dbf^\Sigma_i= - \sum\limits^{N-1}_{k=1} \tau_{ik} (\ubf^\Sigma_k- \ubf^\Sigma_N) \quad\text{ for }\quad i=1, \dots, N-1
\end{equation}
with a positive definite matrix $[\tau_{ik}]$ of interaction coefficients.
Extension of $[\tau_{ik}]$ to a positive semi-definite $N\times N$ matrix such that
\begin{equation}
\sum\limits^{N}_{k=1} \tau_{ik} \, = \, 0 \, = \, \sum\limits^{N}_{k=1} \tau_{ki} \quad\text{ for all }\quad i=1, \dots, N
\end{equation}
yields
\begin{equation}
\dbf^\Sigma_i= \sum\limits^{N}_{k=1} \tau_{ik} (\ubf^\Sigma_i- \ubf^\Sigma_k) \quad\text{ for }\quad i=1, \dots, N.
\end{equation}
By arguments fully analogous to those given in \cite{BD2015} for the bulk diffusion case, the assumption of binary interactions, i.e.\ $\tau_{ik}=\tau_{ik}(T^\Sigma, \rho_i^\Sigma, \rho_k^\Sigma) \to 0$ if $\rho^\Sigma_i$ or $\rho^\Sigma_k$ tend to $0$, leads to the Maxwell-Stefan form
\begin{equation}\label{E40}
\dbf^\Sigma_i= - \sum\limits^N_{k=1} f_{ik} \rho^\Sigma_i \rho^\Sigma_k (\ubf^\Sigma_i- \ubf^\Sigma_k) \quad\text{ for }\quad i=1, \dots, N
\end{equation}
with symmetric friction coefficients $f_{ik}$. An equivalent form for molar quantities is
\begin{equation}\label{E41}
-\sum\limits^N_{k=1} \frac{x^\Sigma_k \Jbf^\Sigma_i- x^\Sigma_i \Jbf^\Sigma_k}{D^\Sigma_{ik}}= \dbf^\Sigma_i \quad\text{ for }\quad i=1, \dots, N
\end{equation}
with the interfacial Maxwell-Stefan diffusivities $D^\Sigma_{ik}$. It turns out that inversion of the system (\ref{E41}) together with (\ref{flux-constraints})$_2$ leads to diffusion fluxes which guarantee positivity of the solutions of the final partial differential equations for the species concentrations; cf.\ \cite{11, BD-MCD}. The interface Maxwell-Stefan equations are also derived in \cite{12}, but under the assumption of quasi-stationary hydrodynamics which is not needed here.
Let us also note that if friction coefficients $f_{ik}$ are given such that $f_{ik}=f_{ki}>0$, then the resulting
matrix $[\tau_{ik}]$ is positive (semi-)definite.

The fifth term in (\ref{E29}) represents the entropy production due to interface chemical reactions, where $R^\Sigma_a$ is the rate of the $a^{\rm th}$ reaction and $\mathcal{A}^\Sigma_a$ the associated affinity. In analogy to the bulk closure, we employ the nonlinear closure
\begin{equation}
\label{closure-chem-rates-int}
\ln \frac{R_a^{\Sigma, f}}{R_a^{\Sigma ,b}}= - \gamma_a^\Sigma \frac{\mathcal{ A}_a^\Sigma}{RT^\Sigma}
\quad \mbox{ with } \gamma_a^\Sigma >0.
\end{equation}
As for the bulk reactions, one of the rates (forward or backward) needs to be modeled using for instance a microscopic theory.

The $6^{\rm th}$ and $7^{\rm th}$ binary products in (\ref{E29}) correspond to the dissipation due to momentum transfer
(friction) between the bulk phases and the interface. Linear closure yields the one-sided Navier slip conditions
\begin{equation}\label{E42}
\alpha^\pm (\vbf^\pm -\vbf^\Sigma)_{||}+ (\Sbf^\pm \cdot \nbf^\pm)_{||}=0 \quad\text{ with }\quad \alpha^\pm \geq 0.
\end{equation}
There are two common special choices: one is $\alpha^\pm=0$, which leads to the free (or, perfect) slip condition, i.e.\ $(\Sbf^\pm \cdot \nbf^\pm)_{||}=0$ at $\Sigma$. The other one corresponds to the limit $\alpha^\pm \to \infty$, which yields continuity of the tangential velocity at $\Sigma$,
i.e.\ $[\![ \vbf_{||} ]\!] =0$, and $\vbf^\Sigma_{||} = \vbf_{||}$.
In the general case of $\alpha^\pm \in (0, \infty )$, the relations \eqref{E42}
link the one-sided limits of the bulk velocity fields to the tangential interface velocity $\vbf^\Sigma_{||}$,
where the latter is determined from \eqref{momentum-bal-int} in case of $\rho^\Sigma >0$.
Let us note in passing that if $\rho^\Sigma =0$ is assumed, or if the time scales in a concrete process
are such that the storage of mass on the interface can be neglected (cf.\  \eqref{partial-mass-bal-int-nondim} below),
the interfacial momentum jump condition \eqref{momentum-bal-int} reduces to
\begin{equation}
\dot{m} [\![\vbf ]\!]  - [\![ \Sbf \cdot \nbf_\Sigma ]\!]= \na_\Sigma \cdot \Sbf^\Sigma
+ \rho^\Sigma {\bf b}^\Sigma.
\end{equation}
Combining this with \eqref{E42} allows to obtain the tangential interface velocity as
\begin{equation}
\vbf^\Sigma_{||} = \frac{1}{\alpha^+ + \alpha^-} \Big( \alpha^+ \vbf^+_{||} + \alpha^- \vbf^-_{||}
-  \dot{m} [\![\vbf_{||} ]\!]  + (\na_\Sigma \cdot \Sbf^\Sigma )_{||}
+ \rho^\Sigma {\bf b}^\Sigma_{||} \Big).
\end{equation}
This allows to eliminate $\vbf^\Sigma_{||}$ from the one-sided Navier slip conditions.
If, moreover,  ${\bf b}^\Sigma =0$ and interfacial viscosities are neglected, i.e.\ if the interfacial stress tensor in
\eqref{E20} is given by $\Sbf^\Sigma= \sigma {\bf I}_\Sigma$, the tangential interface velocity becomes
\begin{equation}
\vbf^\Sigma_{||} = \frac{\alpha^+ \vbf^+_{||} + \alpha^- \vbf^-_{||}
-  \dot{m} [\![\vbf_{||} ]\!]  + \na_\Sigma \sigma}{\alpha^+ + \alpha^-}.
\end{equation}

The $8^{\rm th}$ and $9^{\rm th}$ binary products in (\ref{E29}) refer to entropy production due to energy transmission from the bulk phases to the interface.
Linear closure yields the relations
\begin{equation}\label{E43}
\frac{1}{T^\Sigma}- \frac{1}{T^\pm}=\beta^\pm \big( \dot{m}(e^\pm+ \frac{p^\pm}{\rho^\pm})+ \qbf^\pm \cdot \nbf^\pm \big) \quad\text{ with }\quad \beta^\pm \geq 0.
\end{equation}
A common special choice is $\beta^\pm=0$, implying continuity of the temperature at $\Sigma$,
while $\beta^\pm \to \infty$ leads to energetically isolated (adiabatic) bulk phases.

The $10^{\rm th}$ and $11^{\rm th}$ binary products in (\ref{E29}) refer to entropy production due to one-sided mass transfer from the bulk phases to the interface.
If cross-effects between the different species transfer processes are ignored, linear closure gives
\begin{equation}\label{E44}
\dot{m}^{\pm,\Sigma}_i= \gamma_i^\pm \Big(\frac{\mu^\pm_i}{T^\pm}- \frac{\mu^\Sigma_i}{T^\Sigma}+ \frac{1}{T^\Sigma} \big( \frac{(\vbf^\pm-\vbf^\Sigma)^2}{2}-\nbf^\pm \cdot \frac{\Sbf^{\pm,\rm visc}}{\rho^\pm}\cdot \nbf^\pm \big)\Big)
\end{equation}
with $\gamma_i^\pm \geq 0$.
As already noted above in the context of chemical reactions,
one of the rates--either the ad- or the desorption rate--has to be modeled based on a micro-theory or experimental knowledge. Then the form of the other rate follows from (\ref{E44}).

In the limiting cases, as $\gamma_i^\pm \to\infty$, one obtains
\begin{equation}\label{E45}
\frac{\mu^\pm_i}{T^\pm}= \frac{\mu^\Sigma_i}{T^\Sigma}- \frac{1}{T^\Sigma} \big(\frac{(\vbf^\pm-\vbf^\Sigma)^2}{2}-\nbf^\pm \cdot \frac{\Sbf^{\pm,\rm visc}}{\rho^\pm}\cdot \nbf^\pm \big).
\end{equation}
These limiting cases correspond to vanishing (one-sided) interfacial resistance against mass transfer. If the temperature is assumed to be continuous, an often imposed assumption, and if the kinetic and viscous terms can be neglected compared to the chemical potentials, then (\ref{E45}) yields
\begin{equation}\label{E46}
\mu^+_i=\mu^-_i=\mu^\Sigma_i;
\end{equation}
cf.\ \cite{1} for an estimation of the strength of the different contributions in (\ref{E45}). Let us note that the first equation in (\ref{E46}), which is usually employed to describe the concentration jump at the interface, has to be evaluated with care, since not only the respective one-sided limits of bulk compositions enter the chemical potentials, but also the different pressures,
so that, in detail, it reads as
\begin{equation}\label{E47}
\mu^+_i(T, p^+, x^+_1, \dots, x^+_N)= \mu^-_i(T, p^-, x^-_1, \dots, x^-_N).
\end{equation}
Since the pressure also has a jump at $\Sigma$ and the height of this jump depends on the curvature, the pressure dependence of the chemical potentials introduces a curvature influence into (\ref{E47}), implying in particular that smaller bubbles display an increased solubility.
This is the Gibbs-Thomson effect, while in the context of thermally driven phase transfer, the analogous effect is described by the Kelvin equation; see \cite{Buff1956, Powles}.

Equation (\ref{E44}) shows that mass transfer is driven by chemical potential differences but, in addition, a kinetic term and viscous forces contribute to the driving force. In mass transfer applications, the systems are usually far from equilibrium in particular
concerning the dissipative mass transfer processes. Then, in analogy to the closure of chemical reaction rates,
a non-linear closure turns out to be more appropriate.
Actually, together with the conception of mass transfer as a series of two one-sided
bulk-interface transfer processes, we obtain a novel model for mass transfer which includes the local effects of adsorbed surface active agents.
These local effects are not only mediated via the (non-homogeneous) changes in surface tension with
resulting Marangoni effect, but also due to a direct hindrance of the local transfer rates,
where the latter might be viewed as a steric
effect because of a local blockage of the interface by surfactant molecules.
The next section is devoted to this enhanced mass transfer modeling.

Let us finally check that all constitutive quantities as listed above have been treated.
This is clear for all intrinsic bulk and interface quantities, hence we only need to check the bulk-interface transfer terms.
In this regard, we first obtain the $\dot{m}_i^\pm$ from \eqref{E44}, say, or a more refined closure for this term as given below.
Summing up the $\dot{m}_i^\pm$, this also yields $\dot{m}^\pm$. Next, the tangential components of $\Sbf^\pm \nbf_\Sigma$ follow from \eqref{E42}, unless
it is replaced by the condition $ \vbf_{||}^\pm  =\vbf^\Sigma_{||}$ in the limiting case of (one-sided)
no-slip between bulk and interface.
Finally, in the general case of interfacial temperature jumps,
$\qbf^\pm \cdot \nbf_\Sigma$ then follows from \eqref{E43}, since all other quantities therein
are either thermodynamic state variables or belong to the primitive variables, i.e.\ the balanced quantities.
\section{Mass transfer influenced by adsorbed species}

In order to understand the influence of surface coverage by surfactant to the mass transfer of another chemical species, we have to employ a general closure like (\ref{E44}). But, viewing the mass transfer from the bulk phases to the interface as ad- and desorption processes,  we prefer a non-linear, logarithmic closure in analogy to chemical reactions
since the system can be far away from mass transfer equilibrium.
The subsequent derivations have been first introduced in the conference proceedings paper \cite{Bothe-IBW7}.

We split the one-sided mass transfer terms into an adsorption (ad) and a desorption (de) term according to
\begin{equation}\label{E16}
\dot{m}^{+,\Sigma}_i=s^{ad,+}_i -s^{de,+}_i, \quad \dot{m}^{-,\Sigma}_i= s^{ad,-}_i -s^{de,-}_i.
\end{equation}
The crucial point is that under the assumption that the interface can carry also a transfer component $A_i$
with positive interfacial mass density $\rho_i^\Sigma$, the transfer
of $A_i$ from bulk phase $+$ to bulk phase $-$, say, is modeled as a series of two sorption processes,
one from phase $+$ onto the interface, followed by another sorption process from $\Sigma$ to bulk phase $-$.
The one-sided sorption rates need to be modeled using constitutive, i.e.\ material-dependent, relations.

For technical simplicity, we assume continuous temperature, i.e.\ $T^\pm_{|\Sigma} = T^\Sigma$,
and neglect the kinetic and viscous terms, i.e.\ we exploit the reduced binary products
\begin{equation}\label{E48}
\zeta^\pm_{\rm TRANS}= \frac{1}{T} \sum\limits_{i=1}^N \big(s^{ad,\pm}_i- s^{de,\pm}_i\big) \left(\mu^\pm_i- \mu^\Sigma_i\right)
\end{equation}
to derive the closure relations (ignoring direct cross-effects between different species)
\begin{equation}\label{E49}
\ln \frac{s^{ad,\pm}_i}{s^{de,\pm}_i}= \frac{a^\pm_i}{RT} \big(\mu^\pm_i-\mu^\Sigma_i \big) \quad\text{ with }\quad a^\pm_i \geq 0.
\end{equation}
As mentioned above in the context of chemical reactions, one of the rates--either the ad- or the desorption rate--has to be modeled based on a micro-theory or experimental knowledge. Then the other rate follows from (\ref{E49}). Below, we will let $a^\pm_i=1$ for simplicity which already suffices to obtain an interesting enriched mass transfer model.

We first consider the case of a soluble surfactant which, for simplicity, is only present in $\Omega^+$, say, and on $\Sigma$. Desorption is often more easy to model, where the simplest rate function is
\begin{equation}\label{E50}
s^{de}_i= k^{de}_i x^\Sigma_i
\end{equation}
with the interfacial molar fraction $x^\Sigma_i=c^\Sigma_i/c^\Sigma$, where $c^\Sigma_i = \rho^\Sigma_i /M_i$ and
$c^\Sigma= \sum_{i=1}^N c^\Sigma_i$.
According to (\ref{E49}) with $a^\pm_i=1$, the associated adsorption rate is
\begin{equation}\label{E51}
s^{ad}_i= k^{de}_i x^\Sigma_i \exp \big(\frac{\mu^+_i- \mu^\Sigma_i}{RT} \big).
\end{equation}
To achieve a concrete result, assume ideal mixtures both in the bulk and on the interface, i.e.\ (we include all $x_k$
as independent variables, where $\sum_{k=1}^N x_k =1$ is implicitly assumed)
\begin{equation}\label{E52}
\mu^\pm_i(T,p,x_1, \dots, x_{N})= g^\pm_i(T,p)+ RT \ln x^\pm_i
\end{equation}
with $g^\pm_i(T,p)$ denoting the bulk Gibbs free energy of component $A_i$ under the temperature and pressure of the mixture and
\begin{equation}\label{E53}
\mu^\Sigma_i(T,p^\Sigma,x^\Sigma_1, \dots, x^\Sigma_{N})= g^\Sigma_i(T,p^\Sigma)+ RT \ln x^\Sigma_i
\end{equation}
with $g^\Sigma_i(T,p^\Sigma)$ denoting the surface Gibbs free energy of component $A_i$ under the interface
temperature and interface pressure of the mixture. Insertion of the chemical potentials into (\ref{E51}) yields
\begin{equation}\label{E54}
s^{ad}_i= k^{de}_i \exp \big(\frac{g^+_i-g^\Sigma_i}{RT} \big) x^+_i=: k^{ad}_i x^+_i,
\end{equation}
where $k^{ad}_i$ depends in particular on the surface tension.
Together, \eqref{E50} and \eqref{E54} yield the simplest ad- and desorption rates, leading to the so-called Henry isotherm; see \cite{Kralchevsky} for more details on sorption isotherms of surfactants.

Next, we consider a transfer component $A_i$, like a dissolving gas, which does not accumulate at the interface as a surfactant does, but nevertheless needs to pass through the transmission zone between the bulk phases, mathematically represented by the sharp interface.
Rewriting the interfacial mass balance \eqref{partial-mass-bal-int} for $A_i$ in non-dimensional form, we obtain
\begin{equation}
\label{partial-mass-bal-int-nondim}
\frac{\rho_{\rm ref}^\Sigma }{l_{\rm ref} \rho_{\rm ref} } \,
\Big( \partial^\Sigma_{t^\ast}  \rho_i^{\Sigma, \ast}  
+ \nabla_\Sigma^\ast \cdot (\rho_i^{\Sigma, \ast} {\bf v}^{\Sigma, \ast} - {\bf j}_i^{\Sigma, \ast} )\Big)
 + [ \! [ \dot{m}_i^\ast ] \! ] =0 \mbox{ on } \Sigma.
\end{equation}
Here $l_{\rm ref}$ denotes a characteristic length scale on which $\rho_i^\Sigma$ is expected to display significant changes. For a non-accumulating species, the length scale $\delta := \rho_{\rm ref}^\Sigma / \rho_{\rm ref}$ will characterize
the typical thickness of a layer in the bulk being adjacent to the interface such that this layer contains a similar amount of $A_i$ than is sitting on the interface (i.e.\ inside the thin transition zone represented by the interface).
So, while $l_{\rm ref}$ will essentially be the bubble or droplet size, $\delta$ will rather be on the nanometer scale.
In this case, equation \eqref{partial-mass-bal-int-nondim} and, hence, \eqref{partial-mass-bal-int} can very accurately be approximated by
\begin{equation}\label{E55}
[\![\dot{m}_i]\!]=0 \quad \Leftrightarrow \quad \dot{m}^{+,\Sigma}_i+ \dot{m}^{-,\Sigma}_i=0.
\end{equation}
Consequently, employing (\ref{E16}) and (\ref{E49}) with $a^\pm_i=1$, we obtain
\begin{equation}\label{E56}
s^{de,+}_i \Big(\exp \big( \frac{\mu^+_i-\mu^\Sigma_i}{RT}\big)-1\Big)+ s^{de,-}_i \Big(\exp \big( \frac{\mu^-_i-\mu^\Sigma_i}{RT}\big)-1\Big)=0.
\end{equation}
Employing the same simplifying assumption of ideal mixtures and (\ref{E50}), this implies
\begin{equation}\label{E57}
k^{de,+}_i \Big(\exp \big( \frac{g^+_i-g^\Sigma_i}{RT}\big)x^+_i-x^\Sigma_i\Big)+ k^{de,-}_i \Big(\exp \big( \frac{g^-_i-g^\Sigma_i}{RT}\big)x^-_i-x^\Sigma_i\Big)=0,
\end{equation}
hence
\begin{equation}\label{E58}
\displaystyle
x^\Sigma_i= \frac{k^{de,+}_i \exp \big(\frac{g^+_i-g^\Sigma_i}{RT}\big)x^+_i
+ k^{de,-}_i \exp \big( \frac{g^-_i-g^\Sigma_i}{RT}\big)x^-_i}{k^{de,+}_i+ k^{de,-}_i}.
\end{equation}
Inserting this value into the first expression in (\ref{E57}), equation (\ref{E16}) yields
\begin{equation}\label{E59}
\dot{m}^{+,\Sigma}_i 
= \frac{k^{de,+}_ik^{de,-}_i}{k^{de,+}_i+ k^{de,-}_i} \exp \big(- \frac{g^\Sigma_i}{RT}\big) \Big( \exp \big( \frac{g^+_i}{RT}\big) x^+_i- \exp \big( \frac{g^-_i}{RT}\big)x^-_i\Big).
\end{equation}
Let us compare (\ref{E59}) with a closure which does not account for the interface concentrations.
In this case, (\ref{E55}) is directly build into the entropy production such that only a single binary product per species remains,
namely (cf.\ \cite{1})
\begin{equation}\label{E60}
\zeta^\Sigma_{\rm TRANS}=- \frac{1}{T} \sum\limits^N_{i=1} \dot{m}_i [\![\mu_i+ \frac{(\vbf-\vbf^\Sigma)^2}{2}- \nbf_\Sigma \cdot \frac{\Sbf^{\rm visc}}{\rho}\cdot \nbf_\Sigma]\!].
\end{equation}
Neglecting again the kinetic and viscous terms, the non-linear closure analogous to (\ref{E49}) yields
\begin{equation}\label{E61}
\dot{m}^{+,\Sigma}_i(=- \dot{m}^{-,\Sigma}_i)= k_i \exp \big( -\frac{g^-_i}{RT} \big) \Big(\exp \big(\frac{g^+_i}{RT}\big) x^+_i- \exp\big(\frac{g^-_i}{RT}\big)x^-_i\Big),
\end{equation}
where the ideal mixture assumption is used.
The most important difference to (\ref{E59}) is that there, the mass transfer rate is influenced by the surface tension via the surface Gibbs free energy. In contrast to (\ref{E61}), this allows to account for the effect of surfactants on the mass transfer of the considered transfer component according to the causal chain as illustrated in Figure~\ref{causal-chain} in the introduction.

Concrete forms of the mass transfer relation \eqref{E59} of course depend on the employed model for the free energies.
For example, following \cite{RIMS}, one possibility is to assume the surface equation of state to be
given as
\begin{equation}\label{surfpress}
p^\Sigma = RT \sum_{i=1}^{N-1} c^\Sigma_i \, + \, K^\Sigma \Big(  \frac{c^\Sigma_N}{c^\Sigma_{\rm ref}} -1 \Big).
\end{equation}
The idea behind \eqref{surfpress} is that the interface with its surface tension in the clean state is built by component $A_N$ (the solvent, say) as the phase boun\-dary between a liquid and a gaseous (vapor plus non-condensable gases) phase.
The phase boundary is modeled as a compressible interface phase with compressibility $K^\Sigma$.
Then a consistent interface free energy is given by
\begin{equation}
\rho^\Sigma \psi^\Sigma =  - p^\Sigma + (K^\Sigma + p^\Sigma) \ln \! \big( 1 + \frac{p^\Sigma}{K^\Sigma} \big)
+ RT \sum_{i=1}^N c_i^\Sigma \ln x_i^\Sigma \!.
\end{equation}
Under these assumptions, (\ref{E59}) yields
\begin{equation}
\dot{m}^{+,\Sigma}_i=  \frac{k_i}{1 + p^\Sigma /K^\Sigma}
\Big( \exp \big( \frac{g^+_i}{RT}\big) x^+_i- \exp \big( \frac{g^-_i}{RT}\big)x^-_i\Big).
\end{equation}
Taking the clean surface as the reference state, this yields
\begin{equation}\label{E65}
\dot{m}^{\rm contam}_i=\frac{1+ p^\Sigma_{\rm clean}/K^\Sigma}{1+ p^\Sigma_{\rm contam}/K^\Sigma} \, \dot{m}^{\rm clean}_i
=\frac{K^\Sigma - \sigma_{\rm clean}}{K^\Sigma - \sigma_{\rm contam}} \, \dot{m}^{\rm clean}_i
\end{equation}
for the mass transfer in a system contaminated by surfactant, given as a fraction of the corresponding
rate without surfactant, i.e.\ with $p^\Sigma = p^\Sigma_{\rm clean}$.
The specific relation (\ref{E65}) evidently results from strong
assumptions which are not realistic in particular for high surfactant concentrations.
To account for the (different) area demands of the adsorbed species, a possible surface equation of state is
\begin{equation}\label{surfpressb}
p^\Sigma = - K^\Sigma + \frac{RT}{1-\theta}  \sum_{i=1}^{N} \alpha_i c^\Sigma_i
\;\; \mbox{ with } \; \theta = \sum_{i=1}^{N} c^\Sigma_i / c_i^{\Sigma , \infty},
\end{equation}
where $\theta$ is the total coverage of the interface.
Then the surface analog of the construction of a consistent free energy as explained in
\S 15 in \cite{BD2015} yields the corresponding interface free energy as
\begin{equation}
\rho^\Sigma \psi^\Sigma =  - (1-\theta) p^\Sigma+
RT \sum_{i=1}^{N} \alpha_i c^\Sigma_i
\ln \! \big( 1+ \frac{p^\Sigma }{K^\Sigma} \big)
+ RT \sum_{i=1}^N c_i^\Sigma \ln x_i^\Sigma.
\end{equation}
The corresponding (molar based) interfacial chemical potentials then are
\begin{equation}
\mu_i^\Sigma =  g^\Sigma_i(T,p^\Sigma) + RT \ln x_i^\Sigma
\quad \mbox{ for } i=1,\ldots , N
\end{equation}
with
\begin{equation}\label{E62b}
g^\Sigma_i(T,p^\Sigma)= \frac{p^\Sigma }{c^{\Sigma,\infty}_i }
+ RT \alpha_i \ln \big( 1 + \frac{p^\Sigma}{K^\Sigma} \big) \quad \mbox{ for } i=1,\ldots , N.
\end{equation}
Insertion of (\ref{E62b}) into (\ref{E59}) implies the relation
\begin{equation}
\dot{m}^{+,\Sigma}_i=  \frac{k_i}{(1 + p^\Sigma /K^\Sigma)^{\alpha_i}}
\exp \big( \frac{- p^\Sigma}{c_i^{\Sigma,\infty} RT}\big)
\Big( \exp \big( \frac{g^+_i}{RT}\big) x^+_i- \exp \big( \frac{g^-_i}{RT}\big)x^-_i\Big).
\end{equation}
Taking again the clean surface as the reference state, this yields
\begin{equation}\label{E65b}
\dot{m}^{\rm contam}_i=\Big( \frac{K^\Sigma - \sigma_{\rm clean}}{K^\Sigma - \sigma_{\rm contam}}\Big)^{\alpha_i} \,
\exp \Big( - \frac{\sigma_{\rm clean} - \sigma_{\rm contam}}{c_i^{\Sigma,\infty} RT}\Big)
\dot{m}^{\rm clean}_i.
\end{equation}
The additional mass transfer reduction in (\ref{E65b}) as compared to (\ref{E65}) corresponds to an exponential
damping factor of Boltzmann type, i.e.\ a factor
of the form $\exp (-a \, \Delta p^\Sigma /RT)$, in accordance with the energy barrier model due to Langmuir; see \cite{Langmuir}, \cite{Ciani} and the references given there.
An intuitive explanation of the energy barrier model assumes that, in order to cross the interface, a molecule
first has to ''open a whole'' in the contaminated interface against the increased surface pressure.
Only a fraction of
the molecules carries sufficient energy in order to do this work, hence the frequency of transfer of molecules
across the interface is reduced by a corresponding Boltzmann factor.

Experimental data for the transfer of $\rm{CO_2}$ from Taylor bubbles under the influence of different surfactants in
\cite{Aoki2015, Aoki2017}
supports the fact that the surface pressure of the contaminated system - not, in the first place, the surfactant concentration - determines the mass transfer resistance.
\section{Discussion and Comparison with the Literature}\label{comparison}
Since the present paper investigates {\em multicomponent} two-phase fluid systems with {\em interfacial mass},
we focus on corresponding results in our comparison with the literature. Nevertheless, let us mention that
rigorous sharp interface continuum thermodynamics started with the seminal paper \cite{BAM},
but for single-component two-phase flows and without interfacial mass.\\[0.5ex]
\indent
(i) Based on \cite{BAM}, Jeffrey Kovac derived in \cite{Kovac} an entropy production rate for multicomponent two-phase systems with interfacial mass, but without interfacial chemical reactions as well as without individual body forces.
He employed the entropy fluxes according to \eqref{ef-bothe} and arrived at his representation (4.15).
Rewritten in the notation of the present paper and after a minor correction
(in the last line of (4.15), the symbol $g$, meaning $\rho \vbf$ in our notation,
should be replaced by $g^+$), it reads as
\begin{align}
	\zeta^\Sigma &= \qbf^\Sigma \cdot \na_\Sigma \frac{1}{T^\Sigma}
+ \frac{1}{T^\Sigma} \Sbf^{\Sigma, {\rm irr}} : \Dbf^\Sigma
- \sum\limits^N_{i=1} \jbf^\Sigma_i \cdot \na_\Sigma \frac{\mu^\Sigma_i}{T^\Sigma}\label{ep-kovac}\\
& + [\![ \Big(\frac{1}{T}- \frac{1}{T^\Sigma}\Big) \Big( \dot m \big( e + \vbf \cdot (\frac 1 2 \vbf - \vbf^\Sigma) \big) + \qbf \cdot \nbf_\Sigma \Big) ]\!]  \nonumber\\
& + \frac{1}{T^\Sigma} [\![ (\vbf - \vbf^\Sigma ) \cdot  \Sbf \nbf_\Sigma ]\!]
-  \sum_{i=1}^N  [\![  \Big (\frac{\mu_i}{T}- \frac{\mu^\Sigma_i}{T^\Sigma} \Big)
\rho_i (\vbf_i - \vbf_i^\Sigma)\cdot \nbf_\Sigma ]\!]  \nonumber\\
& - \sum_{i=1}^N [\![ \frac{\mu_i}{T} ]\!]  \, \big( \rho_i^+ + \rho_i^- \big) \big( \vbf_i^\Sigma - \vbf^\Sigma \big) \cdot \nbf_\Sigma
\, + \,
[\![ \vbf \cdot \nbf_\Sigma ]\!] \, \Big( \frac{p^+}{T^+} + \frac{p^-}{T^-} \Big)
\nonumber\\
&- [\![ \frac{1}{T} ]\!] \,
\Big( \big( \frac 1 2 \vbf^+ - \vbf^\Sigma \big) \cdot \vbf^+ \, \dot{m}^+
+ \big( \frac 1 2 \vbf^- - \vbf^\Sigma \big) \cdot \vbf^- \, \dot{m}^- \Big). \nonumber
\end{align}
In contrast to all later papers, the consistency condition (cf.\ \eqref{visigma-consistency} in the appendix)
\[
\vbf_i^\Sigma \cdot \nbf_\Sigma = V_\Sigma \quad \mbox{ for all } i=1,\ldots ,N
\]
is not imposed, hence persistence of a single common interface for all constituents cannot be guaranteed.
In addition, while assuming consistency would somewhat simplify \eqref{ep-kovac}, we cannot obtain agreement to \eqref{E29}.
Note also that certain co-factors within some of the binary products are not objective, hence this representation
of the entropy production is not appropriate for a physically sound closure.\\[0.5ex]
%
%
\indent
(ii)
The sharp-interface continuum thermodynamics of two-phase fluid systems with interfacial mass really started with the
fundamental work of Dick Bedeaux in \cite{Bedeaux}, where systems with external forces and chemical reactions,
both in the bulk and on the interface, have been considered. The entropy flux in \cite{Bedeaux}
coincides with \eqref{ef-bothe}. The bulk and interface Euler relations \eqref{Gibbs-Duhem} and \eqref{E80}, respectively,
are not derived but assumed to hold true.
In our notation, the interfacial entropy production rate, equation (3.2.10) in \cite{Bedeaux},
becomes (after correcting a typo in the second line, where $\mathbf{\Pi}$, meaning $-\Sbf$ in our notation,
should be replaced by $\mathbf{\Pi}^{\rm s}$)
\begin{align}
	\zeta^\Sigma &= \qbf^\Sigma \cdot \na_\Sigma \frac{1}{T^\Sigma}
+\frac{1}{T^\Sigma} \Sbf^{\Sigma, {\rm irr}} : \nabla_\Sigma \vbf^\Sigma
- \sum\limits^N_{i=1} \jbf^\Sigma_i \cdot \left(\na_\Sigma \frac{\mu^\Sigma_i}{T^\Sigma} - \frac{\bbf^\Sigma_i}{T^\Sigma}\right)\label{ep-bedeaux}\\
&- \frac{1}{T^\Sigma} \sum\limits^{N_R}_{a=1} R^\Sigma_a \mathcal{A}^\Sigma_a
+ [\![ \Big(\frac{1}{T}- \frac{1}{T^\Sigma}\Big) \big(\dot{m} sT+ \qbf \cdot \nbf_\Sigma \big) ]\!] \nonumber\\
& +  \frac{1}{T^\Sigma}  [\![ (\vbf -\vbf^\Sigma)_{||} \cdot (\Sbf  \, \nbf_\Sigma )_{||} ]\!]
- \frac{1}{T^\Sigma}
\sum\limits^N_{i=1} [\![ \dot{m}_i \big(\mu_i -\mu^\Sigma_i - \nbf_\Sigma \cdot \frac{\Sbf^{\rm irr}}{\rho} \cdot \nbf_\Sigma \big) ]\!]
\nonumber\\
&- \frac{1}{T^\Sigma} [\![ \dot{m} \left( -\frac 1 2 (\vbf_{||})^2 + \frac 1 2 (\vbf^\Sigma_{||})^2
+ \frac 1 2 \big( (\vbf - \vbf^\Sigma )\cdot \nbf_\Sigma \big)^2 + \vbf \cdot (\vbf - \vbf^\Sigma) \right) ]\!]. \nonumber
\end{align}
It is easy to check that replacing $sT$ in the second line by $h=e+p/\rho$ can be exactly compensated by
replacing
\[
\frac{1}{T^\Sigma}
\sum\limits^N_{i=1} [\![ \dot{m}_i \big(\mu_i -\mu^\Sigma_i \big) ]\!]
\quad \mbox{ with } \quad
\sum\limits^N_{i=1} [\![ \dot{m}_i \big(\frac{\mu_i}{T} -\frac{\mu^\Sigma_i}{T^\Sigma} \big) ]\!]
\]
in the third line. Therefore, with some algebra, we obtain
\begin{align}
\zeta^\Sigma &= \qbf^\Sigma \cdot \na_\Sigma \frac{1}{T^\Sigma}
+\frac{1}{T^\Sigma} \Sbf^{\Sigma, {\rm irr}} : \nabla_\Sigma \vbf^\Sigma
- \sum\limits^N_{i=1} \jbf^\Sigma_i \cdot \left(\na_\Sigma \frac{\mu^\Sigma_i}{T^\Sigma}- \frac{\bbf^\Sigma_i}{T^\Sigma}\right)- \frac{1}{T^\Sigma} \sum\limits^{N_R}_{a=1} R^\Sigma_a \mathcal{A}^\Sigma_a \nonumber \\
	&+  \frac{1}{T^\Sigma}  [\![ (\vbf -\vbf^\Sigma)_{||} \cdot (\Sbf  \, \nbf_\Sigma )_{||} ]\!] 
+ [\![ \Big(\frac{1}{T}- \frac{1}{T^\Sigma}\Big) \Big(\dot{m} \, h+ \qbf \cdot \nbf_\Sigma \Big) ]\!] \nonumber\\
	&- \sum\limits^N_{i=1} [\![ \dot{m}_i \Big(\frac{\mu_i}{T}- \frac{\mu^\Sigma_i}{T^\Sigma}+ \frac{1}{T^\Sigma}
\Big(\frac{(\vbf -\vbf^\Sigma)^2}{2}- \nbf_\Sigma \cdot \frac{\Sbf^{\rm irr}}{\rho} \cdot \nbf_\Sigma \Big)\Big) 
 - [\![\frac{\dot{m}^2}{ \rho T^\Sigma}  \vbf \cdot \nbf_\Sigma ]\!]. \nonumber
\end{align}
This representation of the entropy production rate almost equals \eqref{E29}, except for the last term which is not an objective
scalar and should not be present.

In \cite{Bedeaux}, a constitutive theory is also given, where all possible (which respect the Curie principle, saying that
only quantities of the same tensorial rank can be coupled) couplings are included in the closure relations.
In order to perform the closure of the transfer terms, the problem of the appearance of the binary products inside
the jump terms is resolved (following Waldmann \cite{Waldmann}) by rewriting such terms according to
\begin{align}\label{ab}
& [\![ a\, b ]\!] =  < a >   [\![ b ]\!]  +  [\![ a  ]\!]   < b >,
\end{align}
where, e.g., $<a>:=(a^+ + a^-)/2$ denotes the arithmetic mean of the one-sided limits of $a$.
If applied for the constitutive theory of the transfer terms, it results in closure relations for the
rate of transfer through the interface and the total rate of both bulk-to-interface transfer rates, respectively.
Later on, it has been realized that use of this decomposition of the transfer terms is not optimal, since
the arithmetic mean of the one-sided bulk limits of an intensive quantity has no direct physical meaning \cite{Dick-private}.\\[0.5ex]
%
\indent(iii)
In \cite{Slattery-Interfaces}, John C.\ Slattery considers multicomponent two-phase fluid systems with external forces and chemical reactions, both in the bulk and on the interface; see \cite{Slattery-etal-Interfaces} for a revised second edition with co-authors
Leonhard Sagis and Eun-Suok Oh. The entropy flux coincides with \eqref{ef-bothe}.
The bulk and interface Euler relations \eqref{Gibbs-Duhem} and \eqref{E80}, respectively,
are not derived but assumed to hold true.
For the interfacial entropy production rate, equation (8-8) in Chapter~5.8 of \cite{Slattery-etal-Interfaces}
is obtained. In our notation, it reads
\begin{align}
	\zeta^\Sigma &= (\qbf^\Sigma - \sum_{i=1}^N \mu^\Sigma_i \jbf^\Sigma_i ) \cdot \na_\Sigma \frac{1}{T^\Sigma}
+\frac{1}{T^\Sigma} \Sbf^{\Sigma, {\rm irr}} : {\bf D}^\Sigma\label{ep-slattery}\\
&- \frac{1}{T^\Sigma} \sum\limits^N_{i=1} \jbf^\Sigma_i \cdot \big( \na_\Sigma \mu^\Sigma_i - \bbf^\Sigma_i \big)
- \frac{1}{T^\Sigma} \sum\limits^{N_R}_{a=1} R^\Sigma_a \mathcal{A}^\Sigma_a \nonumber\\
& + [\![ \dot{m} s ]\!] - \frac{1}{T^\Sigma} [\![ \dot{m} \big( e + \frac 1 2 (\vbf -\vbf^\Sigma)^2 \big) ]\!]
+ \frac{1}{T^\Sigma} \sum\limits^N_{i=1} [\![ \dot{m}_i \mu^\Sigma_i ]\!] \nonumber\\
&+ [\![ \Big(\frac{1}{T}- \frac{1}{T^\Sigma}\Big)  \qbf \cdot \nbf_\Sigma ]\!]
- [\![ \sum\limits^N_{i=1} \frac{\mu_i}{T} \jbf_i \cdot \nbf_\Sigma ]\!] 
+  \frac{1}{T^\Sigma}  [\![ (\vbf -\vbf^\Sigma) \cdot \Sbf  \, \nbf_\Sigma ]\!]. \nonumber
\end{align}
Equation \eqref{ep-slattery} is equivalent to our entropy production rate from \eqref{E29},
but the terms inside the entropy production are differently grouped.
In particular, the entropy production due to mass transfer is not clearly visible in \eqref{ep-slattery}.

The entropy production rate is then further simplified, using the assumptions
\[
T_{|\Sigma}=T^\Sigma,\quad
\vbf_{||} = \vbf_{||}^\Sigma, \quad
\mu_{k|\Sigma} = \mu_k^\Sigma.
\]
The resulting constitutive theory is, hence, no longer relevant for this comparison.\\[0.5ex]
%
%
\indent(iv)
In \cite{Sagis}, Leonard Sagis considers multicomponent two-phase fluid systems with external forces but without chemical reactions.
His entropy flux coincides with \eqref{ef-bothe}.
While the bulk Euler relation \eqref{Gibbs-Duhem} is assumed to hold true, the interfacial version  \eqref{E80} is not mentioned.
Building on \cite{Slattery-etal-Interfaces}, the author arrives at the following representation of
the interfacial entropy production, given in equation (42) in \cite{Sagis}.
After correcting a minor mistake (the factor 1/2 is missing in the third binary product below), it reads as
\begin{align}
	\zeta^\Sigma &= (\qbf^\Sigma - \sum_{i=1}^N \mu^\Sigma_i \jbf^\Sigma_i ) \cdot \na_\Sigma \frac{1}{T^\Sigma}
+ \frac{1}{T^\Sigma} \Sbf^{\Sigma, \circ} : {\bf D}^{\Sigma,\circ}
+ \frac{1}{2\, T^\Sigma}  {\rm tr}\big( \Sbf^\Sigma \big)\, {\rm tr}\big( {\bf D}^\Sigma \big) \label{ep-sagis}\\
& - \frac{1}{T^\Sigma} \sum\limits^N_{i=1} \jbf^\Sigma_i \cdot \big( \na_\Sigma \mu^\Sigma_i - \bbf^\Sigma_i \big)
 + [\![ \big(\frac{1}{T}- \frac{1}{T^\Sigma}\big) \big( \dot{m} \, h + \qbf \cdot \nbf_\Sigma \big) ]\!]\nonumber\\
&- \sum\limits^N_{i=1} [\![ \dot{m}_i \Big( \frac{\tilde{\mu}_i}{T} - \frac{\tilde{\mu}_i^\Sigma}{T^\Sigma}\Big) ]\!]
+  \frac{1}{T^\Sigma}  [\![ (\vbf -\vbf^\Sigma) \cdot \Sbf^{\rm irr} \nbf_\Sigma ]\!] \nonumber\\
& - \frac{1}{T^\Sigma}  [\![ \dot{m} \frac{(\vbf -\vbf^\Sigma)^2}{2} ]\!]
+ [\![ \dot{m} \Big( \frac{(\vbf^\Sigma)^2}{2 T^\Sigma} - \frac{\vbf^2}{2 T} \Big) ]\!]. \nonumber
\end{align}
This representation employs so-called modified chemical potentials
\[
\tilde{\mu}_i = \mu_i - \frac 1 2 \vbf^2
\quad \mbox{ and } \quad
\tilde{\mu}_i^\Sigma = \mu_i^\Sigma - \frac 1 2 (\vbf^\Sigma)^2.
\]
Note that the latter are not objective scalars, so their subsequent use in a constitutive theory is arguable.
The closure relations are not directly related to the co-factors in the respective binary products, but the entropy
production rate is rather employed to first define the independent variables which are to be used for the respective
constitutive quantity. Then, developing these dependencies up to first order leads to the closure relations,
and insertion of these functions into \eqref{ep-sagis} yields sign conditions for the coefficients.
Full cross-couplings and Onsager relations are not discussed.\\[0.5ex]
%
%
\indent(v)
In \cite{Bothe-IBW7}, multicomponent two-phase fluid systems with external forces and chemical reactions, both in the bulk and on the interface, but without interfacial viscosities are considered.
The entropy flux in \cite{Bothe-IBW7} coincides with \eqref{ef-bothe} and, in the present notation, the interfacial entropy production reads as
\begin{align}
	\zeta^\Sigma &=
- \frac{1}{T^\Sigma}(\rho^\Sigma e^\Sigma + p^\Sigma -\rho^\Sigma s^\Sigma T^\Sigma - \sum\limits^N_{i=1} \rho^\Sigma_i \mu^\Sigma_i) \na_\Sigma \cdot \vbf^\Sigma \label{E29-IBW7}\\
& + \qbf^\Sigma \cdot \na_\Sigma \frac{1}{T^\Sigma}- \sum\limits^N_{i=1} \jbf^\Sigma_i \cdot \left(\na_\Sigma \frac{\mu^\Sigma_i}{T^\Sigma}- \frac{\bbf^\Sigma_i}{T^\Sigma}\right)
- \frac{1}{T^\Sigma} \sum\limits^{N_R}_{a=1} R^\Sigma_a \mathcal{A}^\Sigma_a \nonumber\\
	&+  \frac{1}{T^\Sigma}  [\![ (\vbf -\vbf^\Sigma)_{||} \cdot (\Sbf  \, \nbf_\Sigma )_{||} ]\!] 
+ [\![ \Big(\frac{1}{T}- \frac{1}{T^\Sigma}\Big) \Big(\dot{m} (e + \frac{p}{\rho})+ \qbf \cdot \nbf_\Sigma \Big) ]\!] \nonumber\\
	&- \sum\limits^N_{i=1} [\![ \dot{m}_i \left(\frac{\mu_i}{T}- \frac{\mu^\Sigma_i}{T^\Sigma}+ \frac{1}{T^\Sigma}
\left(\frac{(\vbf -\vbf^\Sigma)^2}{2}- \nbf_\Sigma \cdot \frac{\Sbf^{\rm irr}}{\rho} \cdot \nbf_\Sigma \right)\right) ]\!]. \nonumber
\end{align}
The constitutive theory employs the same splitting of the jump terms into two binary products for one-sided bulk-interface
transfer processes. The interfacial Euler relation is derived instead of merely assumed to hold. Onsager relations are not discussed.\\[0.5ex]
%
%
\indent(vi)
In \cite{Dreyer-Guhlke-M}, a revised and concise version of the Ph.D.\ thesis \cite{Diss-Guhlke},
Wolfgang Dreyer, Clemens Guhlke and R\"udiger M\"uller
consider multicomponent bulk-surface fluid systems comprised of magnetizable, polarizable, elastic, viscous, heat conducting and
chemically reactive constituents. The developed framework builds on the entropy principle in the axiomatic
form as introduced in \cite{BD2015}.
Specialized to the present setting, i.e.\ without electro-magnetic fields, and rewritten in our notation,
the entropy production from equation (6.14) reads as
\begin{align}
	\zeta^\Sigma = &
- \frac{1}{T^\Sigma}(\rho^\Sigma e^\Sigma + p^\Sigma -\rho^\Sigma s^\Sigma T^\Sigma
- \sum\limits^N_{i=1} \rho^\Sigma_i \mu^\Sigma_i) \na_\Sigma \cdot \vbf^\Sigma
+ \frac{1}{T^\Sigma} \Sbf^{\Sigma, {\rm irr}} : \Dbf^\Sigma\label{ep-WIAS}\\
& + \qbf^\Sigma \cdot \na_\Sigma \frac{1}{T^\Sigma}
- \sum\limits_{i=1}^{N-1} \jbf^\Sigma_i \cdot \Big(\na_\Sigma \frac{\mu^\Sigma_i -\mu^\Sigma_N}{T^\Sigma}
- \frac{\bbf^\Sigma_i -\bbf^\Sigma_N}{T^\Sigma}\Big) \nonumber\\
&  - \frac{1}{T^\Sigma} \sum\limits^{N_R}_{a=1} R^\Sigma_a \mathcal{A}^\Sigma_a
+ [\![ \big(\frac{1}{T}- \frac{1}{T^\Sigma}\big) \big(\dot{m} \, h+ \qbf \cdot \nbf_\Sigma \big) ]\!] \nonumber\\
& +  \frac{1}{T^\Sigma}  [\![ (\vbf -\vbf^\Sigma)_{||} \cdot (\Sbf  \, \nbf_\Sigma )_{||} ]\!]
- \sum\limits_{i=1}^{N-1} [\![ \dot{m}_i \Big(\frac{\mu_i -\mu_N}{T}- \frac{\mu^\Sigma_i -\mu^\Sigma_N}{T^\Sigma} \Big) ]\!]\nonumber\\
& - [\![ \dot{m} \Big(\frac{\mu_N}{T}- \frac{\mu^\Sigma_N}{T^\Sigma}+ \frac{1}{T^\Sigma}
\big(\frac{(\vbf -\vbf^\Sigma)^2}{2}- \nbf_\Sigma \cdot \frac{\Sbf^{\rm irr}}{\rho} \cdot \nbf_\Sigma \big)\Big) ]\!]. \nonumber
\end{align}
While this representation is equivalent to \eqref{E29} and \eqref{flux-constraints}$_2$,
the arrangement of terms differs considerably.
The kinetic and viscous contributions to the co-factors in the binary products of the mass transfer entropy production in \eqref{E29}
have here been allotted to a single term.
For this purpose, the partial mass transfer flux $\dot{m}_N$ is eliminated by means of $\sum_{i=1}^N \dot{m}_i = \dot m$.
While this simplifies the co-factores of the remaining $\dot{m}_i$, it breaks the symmetry w.r.\ to the constituents.
This enforces the need for full cross-coupling and, moreover, is not consistent to the class-II entropy production rate
as derived in the appendix.
Note also that, in contrast to the closure for the diffusion fluxes, the relation $\sum_{i=1}^N \dot{m}_i = \dot m$ is not primarily a constraint, but yields $\dot m$, once the $\dot{m}_i$ are fixed.

The entropy principle in \cite{Dreyer-Guhlke-M} differs from the one in \cite{BD2015} mainly in two ways: the concept of parity is
not included in the set of axioms and the entropy inequality is not implemented in the strengthened form of \cite{BD2015}.
As a consequence, the Onsager relations for cross-coupling coefficients cannot be derived but have to be postulated.
However,  \cite{Dreyer-Guhlke-M} does not contain information on Onsager symmetries.\\[1ex]

\noindent
This comparison with the existing results shows that even if two representations of the entropy production rate are equivalent,
the resulting constitutive relations can only be identical if the co-factors in the binary products are objective quantities and
all possible cross-couplings are accounted for in the closure process.
This leads to a final model with plenty phenomenological coefficients, all of which need to be known in their dependence on
the thermodynamic state variables in order to complete the model.
Indeed, as noted in \cite{Bedeaux}, the sharp-interface model for $N$ constituents and $N_R$ chemical reactions (bulk and interface),
requires $(N+N_R+3)(N+N_R+4)/2$ phenomenological coefficients.
Therefore, in order to obtain a pragmatic model with the minimum number of coefficients, we employed additional information
from the class-II model (see appendix) to obtain the representation \eqref{E29} of the entropy production rate, for which a block-diagonal closure turns out to be appropriate. To arrive at this specific form, we have also been guided by the idea of a fundamental relevance of the entropy derivatives, i.e.\ the terms $1/T$, $1/T^\Sigma$ and $-\mu/T$, $-\mu^\Sigma / T^\Sigma$, as the amplification factors
connecting changes in the distribution of the conserved quantities with the rate of entropy production.
For example, we preferred the representation in which
mass transfer is driven by differences between $-\mu^\pm/T^\pm$ and $-\mu^\Sigma / T^\Sigma$ rather than differences between
$-\mu^\pm$ and $-\mu^\Sigma$. In the end, experiments need to be consulted to decide between different model variants.\\[1ex]
\noindent
{\bf Acknowledgment.}
This research has been financially supported
by the Deutsche Forschungsgemeinschaft (DFG, German Research Foundation) - Project-ID 265191195 - SFB~1194
''Interaction between Transport and Wetting Processes''.

The author would like to express his heartfelt thanks to Prof.\ Akio Tomiyama (Kobe University) for pointing out
the problem of mass transfer hindrance due to the presence of surfactants for the dissolution of CO$_2$ bubbles.

%
%
%
%
%
%
%
%
%
%
\appendix
\newpage
\section{Class-II sharp interface balance equations}
\noindent
We provide a concise derivation of a more general sharp-interface model for multicomponent
two-phase fluid systems in which not only partial mass densities for all constituents in the bulk phases and
on the interfaces are included, but also partial momenta of all species are individually balanced.
Hence, in this so-called class-II sharp-interface model for fluid systems with interfacial mass, every constituent
has its individual velocity, with independent bulk and interface fields.
We set up all required balance equations and derive the entropy production rates for the bulk phases and the interface.
The helps to understand the most natural arrangement of the binary products such that their
correspondence to dissipative mechanisms become evident. This class-II interfacial entropy production rate,
given in \eqref{red-CII-ep-int}, is new.

Many more details about the
general strategy of deriving the class-II continuum thermodynamics
(in the single-phase case without interface) can be found in \cite{BD2015}; cf.\ also \cite{Hutter-book}.
Let us also note that  mixture models with individual velocity fields for different constituents require appropriate boundary conditions for all species.
This is an intrinsic difficulty since, while one might know the total traction or the velocity of the mixture at the boundary, one typically does not know the individual contributions. This issue is discussed in some detail in the book on mixture theory by Rajagopal and Tao \cite{Raj-Tao}; cf.\ also \cite{bothe2013}.

Below, all integral balance equations are formulated for fixed control volumes.
The notations are the same as in the main text.\\[1ex]
{\bf Integral balance of partial mass}\\
\begin{align}
	& \frac{d}{dt} \Big[\int\limits_V \rho_i + \int\limits_{\Sigma_V} \rho^\Sigma_i \, \Big]=- \int\limits_{\pa V} \rho_i \vbf_i \cdot \nbf + \int\limits_V r_i- \int\limits_{\pa\Sigma_V} \rho^\Sigma_i \vbf^\Sigma_i \cdot {\bf N} + \int\limits_{\Sigma_V} r^\Sigma_i,
\end{align}
where we omitted the differentials $dx$, $do$ in the integrals in order to save some space.
Application of the bulk and surface transport theorems, the two-phase and the surface divergence theorem, followed by the usual localization procedure yields the\\[1ex]
{\bf Local form of partial mass balance}
\begin{align}\label{partial-mass-bulk}
& \pa_t \rho_i + \div (\rho_i \vbf_i)= r_i & \mbox{ in } \Omega \setminus \Sigma\\
& \pa^\Sigma_t \rho^\Sigma_i + \divS(\rho^\Sigma_i \vbf^\Sigma_i)+ [\![\rho_i(\vbf_i-\vbf^\Sigma_i) \cdot \nbf_\Sigma]\!] = r^\Sigma_i & \mbox{ on } \Sigma.\label{partial-mass-int}
\end{align}
Below, we also skip the domains for the balance equations to save space, since the time derivative determines whether the balance
is a bulk or an interface balance.
In order to maintain a single common interface between the two bulk phases, the individual interface velocities
$\vbf^\Sigma_i$ need to fulfil the {\em consistency requirement} (cf.\ \cite{Bo-2PH-ODE})
\begin{align}\label{visigma-consistency}
\vbf^\Sigma_i \cdot \nbf_\Sigma = V_\Sigma \; \mbox{ for all } i=1, \ldots ,N.
\end{align}
In other words, the surfaces of discontinuity of the individual mass densities move with the same normal speed.
But note that the individual interfacial velocities will in general have different tangential components.

The jump term in \eqref{partial-mass-int} represents two one-sided bulk-interface mass exchange terms.
In the physico-chemical sciences, these bulk-interface mass (species) exchange processes are named {\rm sorption},
with {\em adsorption} being the bulk-to-interface mass exchange and {\em desorption} for the reverse direction.
We therefore let
\begin{align}
\dot{m}^{\pm,\Sigma}_i := \rho^\pm_i(\vbf^\pm_i- \vbf^\Sigma_i) \cdot \nbf^\pm
\end{align}
with $\nbf^\pm$ the outer normals to $\Omega^\pm$
denote these one-sided bulk-to-interface exchange rates of partial mass.
Then the {\em sorption formulation} of the partial mass balances reads as
\begin{align}
		& \pa_t \rho_i + \div (\rho_i \vbf_i)= r_i, \\
    &  \pa^\Sigma_t \rho^\Sigma_i + \divS(\rho^\Sigma_i \vbf^\Sigma_i)
 = \dot{m}^{+,\Sigma}_i + \dot{m}^{-,\Sigma}_i + r^\Sigma_i.
\end{align}
We assume {\em conservation of total mass} to hold true, which means that
\begin{align}\label{mass-conservation}
	& \sum\limits_{i=1}^N r_i =0, \qquad \sum\limits_{i=1}^N r^\Sigma_i =0.
\end{align}
Summation of the partial mass balances \eqref{partial-mass-bulk} and \eqref{partial-mass-int}, together with \eqref{mass-conservation},
yields\\[1ex]
{\bf Local balance of total mass}
\begin{align}
	& \pa_t \rho + \div(\rho \vbf)=0, \\
    & \pa^\Sigma_t \rho^\Sigma + \div_\Sigma (\rho^\Sigma \vbf^\Sigma)
- [\![ \rho(\vbf -\vbf^\Sigma) \cdot \nbf_\Sigma]\!] =0,\hspace{2in}
\end{align}
where\vspace{-0.1in}
\begin{align}
	& \rho:= \sum\limits_{i=1}^N \rho_i, \quad \rho \vbf:= \sum\limits_{i=1}^N \rho_i \vbf_i, \quad \rho^\Sigma:= \sum\limits_{i=1}^N \rho^\Sigma_i, \quad \rho^\Sigma \vbf^\Sigma:= \sum\limits_{i=1}^N \rho^\Sigma_i \vbf^\Sigma_i.\vspace{0.2in}
\end{align}
We continue with the partial and total momenta.\\[1ex]
{\bf Integral balance of partial momentum}
\begin{align}
	& \frac{d}{dt} \Big[\int\limits_V \rho_i \vbf_i + \int\limits_{\Sigma_V} \rho^\Sigma_i \vbf^\Sigma_i \, \Big] = - \int\limits_{\pa V} \rho_i \vbf_i (\vbf_i \cdot \nbf) + \int\limits_{\pa V} \Sbf_i \nbf + \int\limits_V \rho_i \bbf_i \\
	& - \int\limits_{\pa\Sigma_V} \rho^\Sigma_i \vbf^\Sigma_i (\vbf^\Sigma_i \cdot {\bf N}) + \int\limits_{\pa\Sigma_V} \Sbf^\Sigma_i \cdot {\bf N} + \int\limits_{\Sigma_V} \rho^\Sigma_i\bbf^\Sigma_i 
\underbrace{+ \int\limits_V \fbf_i + \int\limits_{\Sigma_V} \fbf^\Sigma_i.}_{\text{ momentum exchange}}\nonumber
\end{align}
Application of the bulk and surface transport theorems, the two-phase and the surface divergence theorem, followed by the usual localization procedure yields the\\[1ex]
{\bf Local form of partial momentum balance}
	\begin{align}\label{partial-mom-bulk}
		& \pa_t(\rho_i \vbf_i) + \div(\rho_i \vbf_i \otimes \vbf_i-\Sbf_i)= \rho_i \bbf_i + {\bf f}_i, \\
		&	\pa^\Sigma_t (\rho^\Sigma_i \vbf^\Sigma_i) + \divS(\rho^\Sigma_i \vbf^\Sigma_i \otimes \vbf^\Sigma_i -\Sbf^\Sigma_i)+ [\![\dot{m}_i \vbf_i -\Sbf_i \nbf_\Sigma ]\!]= \rho^\Sigma_i \bbf^\Sigma_i+ \fbf^\Sigma_i\label{partial-mom-int}
\end{align}
with ${\bf f}_i$ and ${\bf f}_i^\Sigma$ the rate of momentum exchange between the species and the abbreviation
	\begin{align}
\dot{m}_i := \rho_i (\vbf_i-\vbf^\Sigma_i) \cdot \nbf_\Sigma = \rho_i (\vbf_i-\vbf^\Sigma) \cdot \nbf_\Sigma,
\end{align}
where the second equality comes from \eqref{visigma-consistency}. Note that this definition of
$\dot{m}_i$ is consistent with the class-I notion from the main text.
We assume {\em conservation of total momentum} to be valid, hence
\begin{align}
	& \sum\limits_{i=1}^N \fbf_i = 0, \qquad \sum\limits_{i=1}^N \fbf^\Sigma_i =0.
\end{align}
Employing the abbreviation
\begin{align}
  & \dot{m}:=\sum_{i=1}^N \dot{m}_i = \rho (\vbf-\vbf^\Sigma) \cdot \nbf_\Sigma,
\end{align}
this is applied to the sum of the partial momentum balances \eqref{partial-mom-bulk}, \eqref{partial-mom-int} to obtain the
\\[1ex]
{\bf Local balance of total momentum}
\begin{align}
	& \pa_t (\rho \vbf) + \div(\rho \vbf \otimes \vbf -\Sbf)= \rho \bbf,\\
	& \pa^\Sigma_t(\rho^\Sigma \vbf^\Sigma)+ \divS(\rho^\Sigma \vbf^\Sigma \otimes \vbf^\Sigma -\Sbf^\Sigma) + [\![\dot{m} \vbf -\Sbf \nbf_\Sigma ]\!]= \rho^\Sigma \bbf^\Sigma
\end{align}
with the total bulk and interface stresses
\begin{align}
	& \Sbf:=\sum\limits_{i=1}^N (\Sbf_i - \rho_i \ubf_i \otimes \ubf_i), \qquad
	& \Sbf^\Sigma:= \sum\limits_{i=1}^N (\Sbf^\Sigma_i -\rho^\Sigma_i \ubf^\Sigma_i \otimes \ubf^\Sigma_i),
\end{align}
containing the bulk and interface diffusion velocities
\begin{align}
	& \ubf_i:= \vbf_i-\vbf, \qquad  \ubf^\Sigma_i := \vbf^\Sigma_i-\vbf^\Sigma,
\end{align}
and the bulk and interface body force densities
\begin{align}
\rho \bbf:= \sum\limits_{i=1}^N \rho_i \bbf_i,\qquad \rho^\Sigma \bbf^\Sigma:= \sum\limits_{i=1}^N \rho^\Sigma_i \bbf^\Sigma_i.
\end{align}
Below, we employ the stress decompositions
\begin{align}
	& \Sbf_i = -P_i {\bf I} + \Sbf_i^\circ = - p_i {\bf I} + \Sbf_i^{\rm irr}, \qquad
	& \Sbf_i^\Sigma = -P_i^\Sigma {\bf I}_\Sigma + \Sbf_i^{\Sigma, \circ} = - p_i^\Sigma {\bf I}_\Sigma + \Sbf_i^{\Sigma, \rm irr}
\end{align}
with the mechanical partial pressures $P_i$ and $P_i^\Sigma$ defined as
\begin{align}
	& P_i := - \frac 1 3 {\rm trace} (\Sbf_i), \qquad
	& P_i^\Sigma := - \frac 1 2 {\rm trace} (\Sbf_i^\Sigma ),
\end{align}
which reduce to the thermodynamical partial pressure $p_i$ and $p_i^\Sigma$, respectively,
in equilibrium (i.e., the latter are to be given by thermal equations of state).
Moreover, let
\begin{align}
	& \pi_i := P_i - p_i, \qquad
	& \pi_i^\Sigma := P_i^\Sigma - p_i^\Sigma
\end{align}
denote the irreversible part of the bulk and interface pressure, respectively. Then
\begin{align}
	& \Sbf_i^{\rm irr} = -\pi_i {\bf I} + \Sbf_i^\circ, \qquad
	& \Sbf_i^{\Sigma, \rm irr} = -\pi_i^\Sigma {\bf I}_\Sigma + \Sbf_i^{\Sigma, \circ}.
\end{align}
We continue with the balance of energy, where we include partial energy balances since they provide
structural information even if they are not part of the final class-II model.\\[1ex]
{\bf Integral balance of partial energy}
\begin{align}
	& \frac{d}{dt} \Big[\int\limits_V \rho_i \big(e_i + \frac{\vbf^2_i}{2}\big)+ \int\limits_{\Sigma_V} \rho^\Sigma_i \big(e^\Sigma_i+ \frac{(\vbf^\Sigma_i)^2}{2}\big) \Big] =\\
	& -\int\limits_{\pa V} \rho_i \big(e_i + \frac{\vbf^2_i}{2}\big) \vbf_i \cdot \nbf - \int\limits_{\pa V} \qbf_i \cdot \nbf + \int\limits_{\pa V} \vbf_i \cdot \Sbf_i \nbf + \int\limits_V \vbf_i \cdot \rho_i \bbf_i + \int\limits_V k_i +\int\limits_{\Sigma_V} k^\Sigma_i\nonumber \\ \nonumber
	& -\int\limits_{\pa\Sigma_V} \rho^\Sigma_i \big(e^\Sigma_i + \frac{(\vbf^\Sigma_i)^2}{2}\big) \vbf^\Sigma_i \cdot \Nbf - \int\limits_{\pa\Sigma_V} \qbf^\Sigma_i \cdot \Nbf + \int\limits_{\pa\Sigma_V} \vbf^\Sigma_i \cdot \Sbf^\Sigma_i \Nbf + \int\limits_{\Sigma_V} \vbf^\Sigma_i \cdot \rho^\Sigma_i \bbf^\Sigma_i, \nonumber
\end{align}
where $k_i$ and $k_i^\Sigma$ denote the rate of energy exchange between the different constituents;
for simplicity of notation we do not include external energy supply by radiation.
Application of the surface transport theorem, the two-phase and the surface divergence theorem, followed by the usual localization procedure yields the\\[1ex]
{\bf Local form of partial energy balance}
\begin{align}
	& \pa_t \big(\rho_i \big(e_i+ \frac{\vbf_i^2}{2}\big)\big)+ \div \big(\rho_i \big(e_i+\frac{\vbf_i^2}{2}\big) \vbf_i
+ \qbf_i - \vbf_i \cdot \Sbf_i \big) = \rho_i \vbf_i \cdot \bbf_i + k_i,\\
	& \pa^\Sigma_t \big(\rho^\Sigma_i \big(e^\Sigma_i+ \frac{(\vbf^\Sigma_i)^2}{2}\big)\big) + \div_\Sigma \big(\rho^\Sigma_i \big(e^\Sigma_i + \frac{(\vbf^\Sigma_i)^2}{2}\big) \vbf^\Sigma_i +\qbf^\Sigma_i - \vbf^\Sigma_i\cdot \Sbf^\Sigma_i \big) \\
	& \quad + [\![\rho_i \big(e_i + \frac{\vbf_i^2}{2}\big) (\vbf_i-\vbf^\Sigma_i) \cdot \nbf_\Sigma ]\!] + [\![\qbf_i \cdot \nbf_\Sigma ]\!] - [\![\vbf_i \cdot \Sbf_i \nbf_\Sigma]\!] = \rho^\Sigma_i \vbf^\Sigma_i \cdot\bbf^\Sigma_i + k^\Sigma_i. \nonumber
\end{align}
Subtraction of the local balance of partial kinetic energy (as obtained from the local balance of partial mass and momentum) yields
the\\[1ex]
{\bf Local balance of partial internal energy}
\begin{align}\label{partial-intenergy-bulk}
	& \pa_t(\rho_i e_i) + \div(\rho_i e_i \vbf_i+ \qbf_i)= \na \vbf_i : \Sbf_i -\vbf_i \cdot \fbf_i+ \frac{1}{2} r_i \vbf^2_i+ k_i,\\
	& \pa^\Sigma_t(\rho^\Sigma_i e^\Sigma_i)+ \divS (\rho^\Sigma_i e^\Sigma_i \vbf^\Sigma_i + \qbf^\Sigma_i) + [\![\dot{m}_i (e_i+ \frac{(\vbf_i-\vbf^\Sigma_i)^2}{2})]\!]+ [\![\qbf_i \cdot \nbf_\Sigma]\!]\label{partial-intenergy-int} \\
	& - [\![(\vbf_i-\vbf^\Sigma_i) \cdot \Sbf_i \nbf_\Sigma]\!]
= \na_\Sigma \vbf^\Sigma_i : \Sbf^\Sigma_i- \vbf^\Sigma_i \cdot \fbf^\Sigma_i+ \frac{1}{2} r^\Sigma_i(\vbf^\Sigma_i)^2 + k^\Sigma_i. \nonumber
\end{align}
We assume {\em conservation of total energy} to be valid, i.e.\
\begin{align}
	& \sum\limits_{i=1}^N k_i = 0, \qquad \sum\limits_{i=1}^N k^\Sigma_i =0.
\end{align}
We next consider the full mixture internal energy, where the latter has to be carefully defined because there are different options.
In general, the total internal energy is defined to be the remainder if the total kinetic energy is subtracted from the total energy.
In a class-I context, this means to subtract $\rho \vbf^2 /2$ from the total energy density, while all partial kinetic energy
densities $\rho_i \vbf^2_i /2$ should be subtracted in the class-II context; cf.\ \cite{BD2015}. We
congruously employ the latter concept, i.e.\ we define
\begin{align}\label{intenergydensity}
  \rho e = \sum_{i=1}^N \rho_i e_i , \qquad   \rho^\Sigma e^\Sigma = \sum_{i=1}^N \rho_i^\Sigma e_i^\Sigma
\end{align}
to be the densities of the total internal energy in the bulk and on the interface.
This is applied to the sum of equations \eqref{partial-intenergy-bulk} and \eqref{partial-intenergy-int} to obtain the\\[1ex]
{\bf Local balance of total internal energy}
\begin{align}
	& \pa_t(\rho e) + \div (\rho e \vbf + \tilde{\qbf} ) = \sum\limits_{i=1}^N \Sbf_i : \na \vbf_i - \sum\limits_{i=1}^N \ubf_i \cdot \big(\fbf_i - r_i \vbf_i + \frac{r_i}{2} \ubf_i\big),\\
	& \pa^\Sigma_t(\rho^\Sigma e^\Sigma) + \div_\Sigma (\rho^\Sigma e^\Sigma \vbf^\Sigma + \tilde{\qbf}^\Sigma)
+ [\![ \dot{m}\, e  ]\!]  + [\![ \tilde{\qbf} \cdot \nbf_\Sigma ]\!]\label{local-total-intenergy} \\
	&  + [\![  \sum\limits_{i=1}^N \dot{m}_i \big( \frac{(\vbf_i -\vbf^\Sigma_i)^2}{2} - \nbf_\Sigma \cdot \frac{\Sbf_i}{\rho_i} \cdot \nbf_\Sigma \big) ]\!]
- [\![ \sum\limits_{i=1}^N (\vbf_i -\vbf^\Sigma_i)_{||} \cdot (\Sbf_i \nbf_\Sigma)_{||} ]\!]\nonumber \\
    & = \sum\limits_{i=1}^N \Sbf^\Sigma_i : \na_\Sigma \vbf^\Sigma_i- \sum\limits_{i=1}^N \ubf^\Sigma_i \cdot \big(\fbf^\Sigma_i - r^\Sigma_i \big(\vbf^\Sigma_i - \frac{\ubf^\Sigma_i}{2}\big)\big) \nonumber
\end{align}
$\mbox{ }$\\[-4ex]
with\vspace{-0.2in}
\begin{align}
	& \tilde{\qbf}:= \sum\limits_{i=1}^N (\qbf_i + \rho_i e_i \ubf_i),
 \quad \tilde{\qbf}^\Sigma = \sum\limits_{i=1}^N \big(\qbf^\Sigma_i + \rho^\Sigma_i e^\Sigma_i \ubf^\Sigma_i\big).
\end{align}
We will later rewrite the latter balances, employing the final form of the heat fluxes which will only become
clear after the entropy production rates have been formulated in an appropriate form.\\[1ex]
The form of the entropy balance is identical to the one for class-I, i.e.\ we have the\\[1ex]
{\bf Local balance of entropy}
\begin{align}
	& \pa_t(\rho s) + \div(\rho s \vbf + \Phibf)=\zeta,\\
	& \pa^\Sigma_t (\rho^\Sigma s^\Sigma) + \divS(\rho^\Sigma s^\Sigma \vbf^\Sigma + \Phibf^\Sigma) + [\![\dot{m} s + \Phibf \cdot\nbf_\Sigma]\!]= \zeta^\Sigma.
\end{align}
We first consider the bulk entropy production.
We insert $\rho s = \rho s (\rho e,\rho_1, \ldots ,\rho_N)$
into the first two terms  and carry out the differentiations. Then
we eliminate the resulting time derivatives by means of the balance equations \eqref{partial-mass-bulk} and \eqref{local-total-intenergy}, introducing temperature and chemical potentials (defined as in \eqref{E73})
via the chain rule.
After straightforward computations, a first representation of the bulk entropy production is
\begin{align}
\label{entropy-production1}
\zeta =
\div \big( \Phi - \frac{{\bf q}}{T}+ \sum_i \frac{\rho_i \mu_i {\bf u}_i}{T} \big)
- \frac 1 T \big(  \rho e + p - T \rho s - \sum_i \rho_i \mu_i  \big)\, \div {\bf v}\nonumber \\
+ \frac 1 T \sum_i \stress_i^\circ : {\bf D}_i^\circ
- \frac 1 T \sum_i  \pi_i \, \div {\bf v}_i
+ {\bf q} \cdot \nabla \frac 1 T - \frac 1 T \sum_{a=1}^{N_R} R_a \mathcal{A}_a\\
- \sum_i {\bf u}_i \cdot \Big( \rho_i \nabla \frac{\mu_i}{T} +
\frac 1 T  \big( {\bf f}_i - r_i ( {\bf v}_i  - \frac{1}{2} {\bf u}_i) - \nabla p_i \big) \Big), \nonumber
\end{align}
where we have introduced the thermodynamic pressure $p$ and the heat flux $\qbf$ as
\begin{align}\label{p-heatflux}
& p:=\sum_{i=1}^N p_i, \qquad {\bf q} := \tilde{\bf q} + \sum_{i=1}^N p_i {\bf u}_i = \sum\limits_{i=1}^N \big( \qbf_i + (\rho_i e_i + p_i) \ubf_i \big).
\end{align}
Guided by the form of the first term in \eqref{entropy-production1}, we use the constitutive equation
\begin{align}
	& \Phibf:= \frac{\qbf}{T}- \sum\limits_{i=1}^N \frac{\rho_i \mu_i \ubf_i}{T}
\end{align}
for the bulk entropy flux. The second term is a binary product, where the first factor is a function of the thermodynamic state
variables, i.e.\ independent of irreversible process parts. Since the co-factor $\div {\bf v}$ has no sign, the entropy principle
requires the relation
\begin{align}
& \rho e + p - \rho s T = \sum\limits_{i=1}^N \rho_i \mu_i,
\end{align}
i.e.\ the same Euler relation as in the class-I model holds true.
Note that, here, it did not result from a closure
relation, since the constitutive quantities are the $\pi_i$,
which have already been separated and appear in the $N$ corresponding binary products.
The resulting reduced entropy production rate reads as
\begin{align}
\label{entropy-production2}
\zeta =
 \frac 1 T \sum\limits_{i=1}^N \stress_i^\circ : {\bf D}_i^\circ
- \frac 1 T \sum\limits_{i=1}^N  \pi_i \, \div {\bf v}_i
+ {\bf q} \cdot \nabla \frac 1 T - \frac 1 T \sum_{a=1}^{N_R} R_a \mathcal{A}_a\\
- \sum\limits_{i=1}^N {\bf u}_i \cdot \Big( \rho_i \nabla \frac{\mu_i}{T} +
\frac 1 T  \big( {\bf f}_i - r_i ( {\bf v}_i  - \frac{1}{2} {\bf u}_i) - \nabla p_i \big) \Big). \nonumber
\end{align}
Similarly, a lengthy calculation yields the interfacial entropy production rate as
\begin{align}
	\zeta^\Sigma  = & \nabla_\Sigma \cdot \big( \Phibf^\Sigma -\frac{\qbf^\Sigma}{T^\Sigma} + \sum\limits_{i=1}^N \frac{\rho^\Sigma_i \mu^\Sigma_i \ubf^\Sigma_i}{T^\Sigma} \big)
+ \big(  \rho^\Sigma e^\Sigma + p^\Sigma - \rho^\Sigma s^\Sigma T^\Sigma - \sum\limits_{i=1}^N \rho^\Sigma_i \mu^\Sigma_i \big) \nabla_\Sigma \cdot \vbf^\Sigma\nonumber \\
& + \frac{1}{T^\Sigma} \sum\limits_{i=1}^N \na_\Sigma \Sbf^{\Sigma, \circ}_i : {\bf D}^{\Sigma, \circ}_i
+ \frac{1}{T^\Sigma} \sum\limits_{i=1}^N \pi_i^\Sigma \nabla_\Sigma \cdot \vbf_i^\Sigma
+ \qbf^\Sigma \cdot \na_\Sigma \frac{1}{T^\Sigma}
- \frac{1}{T^\Sigma} \sum\limits_a R^\Sigma_a \mathcal{A}^\Sigma_a\nonumber \\
	& - \sum\limits_{i=1}^N \ubf^\Sigma_i \cdot \Big(\rho^\Sigma_i \na_\Sigma \frac{\mu^\Sigma_i}{T^\Sigma} + \frac{1}{T^\Sigma} \big(\fbf^\Sigma_i -r^\Sigma_i (\vbf^\Sigma_i-\frac{\ubf^\Sigma_i}{2})- \na_\Sigma p^\Sigma_i \big)\Big) \\
	& + \frac{1}{T^\Sigma} \sum\limits_{i=1}^N [\![(\vbf_i - \vbf^\Sigma_i)_{||} \cdot (\Sbf^{\rm irr}_i \nbf_\Sigma)_{||}]\!]
 + [\![ \big(\frac{1}{T} - \frac{1}{T^\Sigma}\big) (\dot{m}\, h + \qbf \cdot \nbf_\Sigma)]\!]\nonumber \\
	& + \sum\limits_{i=1}^N [\![  \dot{m}_i \big(\frac{\mu^\Sigma_i}{T^\Sigma}- \frac{\mu_i}{T}
-  \frac{1}{T^\Sigma}\big( \frac{(\vbf_i-\vbf^\Sigma_i)^2}{2} - \nbf_\Sigma \cdot \frac{\Sbf^{\rm irr}_i}{\rho_i} \cdot \nbf_\Sigma \big) \big)]\!], \nonumber
\end{align}
where we used the interfacial heat flux defined as
\begin{align}\label{CII-heatflux-int}
& \qbf^\Sigma= \sum\limits_{i=1}^N \Big(\qbf^\Sigma_i + \big(\rho^\Sigma_i e^\Sigma_i + p^\Sigma_i \big) \ubf^\Sigma_i \Big).
\end{align}
In analogy to the bulk closure for the entropy flux we employ the constitutive relation
\begin{align}
	& \Phibf^\Sigma:= \frac{\qbf^\Sigma}{T^\Sigma}- \sum\limits_{i=1}^N \frac{\rho^\Sigma_i \mu^\Sigma_i \ubf^\Sigma_i}{T^\Sigma}
\end{align}
and also obtain the interfacial Euler relation, i.e.\
\begin{align}
\quad \rho^\Sigma e^\Sigma + p^\Sigma - \rho^\Sigma s^\Sigma T^\Sigma = \sum\limits_{i=1}^N \rho^\Sigma_i \mu^\Sigma_i.
\end{align}
The resulting reduced interfacial entropy production rate reads as
\begin{align}
	\zeta^\Sigma  =
& \frac{1}{T^\Sigma} \sum\limits_{i=1}^N \na_\Sigma \Sbf^{\Sigma, \circ}_i : {\bf D}^{\Sigma, \circ}_i
+ \frac{1}{T^\Sigma} \sum\limits_{i=1}^N \pi_i^\Sigma \nabla_\Sigma \cdot \vbf_i^\Sigma
+ \qbf^\Sigma \cdot \na_\Sigma \frac{1}{T^\Sigma}
- \frac{1}{T^\Sigma} \sum\limits_a R^\Sigma_a \mathcal{A}^\Sigma_a\nonumber \\
	& - \sum\limits_{i=1}^N \ubf^\Sigma_i \cdot \Big(\rho^\Sigma_i \na_\Sigma \frac{\mu^\Sigma_i}{T^\Sigma} + \frac{1}{T^\Sigma} \big(\fbf^\Sigma_i -r^\Sigma_i (\vbf^\Sigma_i-\frac{\ubf^\Sigma_i}{2})- \na_\Sigma p^\Sigma_i \big)\Big)\nonumber \\
	& - \frac{1}{T^\Sigma} \sum\limits_{i=1}^N (\vbf_i^+ - \vbf^\Sigma_i)_{||} \cdot (\Sbf^{+,\rm irr}_i \nbf^+)_{||}
- \frac{1}{T^\Sigma} \sum\limits_{i=1}^N (\vbf_i^- - \vbf^\Sigma_i)_{||} \cdot (\Sbf^{-,\rm irr}_i \nbf^-)_{||}\label{red-CII-ep-int}\\
& - \Big(\frac{1}{T^+} - \frac{1}{T^\Sigma}\Big) (\dot{m}^{+, \Sigma}\, h^+ + \qbf^+ \cdot \nbf^+ )
- \Big(\frac{1}{T^-} - \frac{1}{T^\Sigma}\Big) (\dot{m}^{-, \Sigma}\, h^- + \qbf^- \cdot \nbf^- ) \nonumber \\
 	& + \sum\limits_{i=1}^N \dot{m}_i^{+, \Sigma} \Big(\frac{\mu_i^+}{T^+} - \frac{\mu^\Sigma_i}{T^\Sigma}
+  \frac{1}{T^\Sigma}\Big( \frac{(\vbf_i^+-\vbf^\Sigma_i)^2}{2} - \nbf^+ \cdot \frac{\Sbf^{+, \rm irr}_i}{\rho_i^+} \cdot \nbf^+ \Big) \Big) \nonumber \\
 	& + \sum\limits_{i=1}^N \dot{m}_i^{-, \Sigma} \Big(\frac{\mu_i^-}{T^-} - \frac{\mu^\Sigma_i}{T^\Sigma}
+  \frac{1}{T^\Sigma}\Big( \frac{(\vbf_i^- -\vbf^\Sigma_i)^2}{2} - \nbf^- \cdot \frac{\Sbf^{-, \rm irr}_i}{\rho_i^-} \cdot \nbf^- \Big) \Big), \nonumber
\end{align}
where we expanded the jump brackets into the one-sided terms. Equation \eqref{red-CII-ep-int} can now be exploited
for deriving constitutive equations, but we refrain from collecting the resulting closure relations,
just noting that a system of $N$ coupled Navier-Stokes-type bulk equations and $N$ coupled interfacial
Navier-Stokes-Boussinesq-Scriven equations would result; see \cite{BD2015} for the resulting bulk model.

We close this appendix with the alternative form of the total internal energy balance, which
employs the heat fluxes $\qbf$, $\qbf^\Sigma$ from \eqref{p-heatflux} and \eqref{CII-heatflux-int}, respectively.
It reads as
\begin{align*}
	& \pa_t(\rho e) + \div (\rho e \vbf +\qbf) = \sum\limits_{i=1}^N \Sbf^{\rm irr}_i : \na \vbf_i - p \,\div \vbf - \sum\limits_{i=1}^N \ubf_i \cdot \big(\fbf_i -r_i \vbf_i + \frac{r_i}{2} \ubf_i -\na p_i \big),
\end{align*}
\begin{align*}
	& \pa^\Sigma_t (\rho^\Sigma e^\Sigma) + \div_\Sigma (\rho^\Sigma e^\Sigma \vbf^\Sigma + \qbf^\Sigma) + [\![ \dot{m} \, h + \sum\limits_{i=1}^N y_i \big(\frac{(\vbf_i- \vbf^\Sigma_i)^2}{2} - \nbf_\Sigma \cdot \Sbf^{\rm irr}_i \cdot \nbf_\Sigma \big)\Big) ]\!] \\
	& + [\![ \sum\limits_{i=1}^N \jbf_i \cdot \nbf_\Sigma \Big(\frac{(\vbf_i -\vbf^\Sigma_i)^2}{2}- \nbf_\Sigma \cdot \Sbf^{\rm irr}_i \cdot \nbf_\Sigma \Big) ]\!]
	 + [\![ \qbf \cdot \nbf_\Sigma ]\!] - [\![ \sum\limits_{i=1}^N (\vbf_i -\vbf^\Sigma_i)_{||} \cdot (\Sbf^{\rm irr} \nbf_\Sigma)_{||} ]\!]\\
 & = \sum\limits_{i=1}^N \Sbf^{\Sigma,{\rm irr}}_i: \na_\Sigma \vbf^\Sigma_i -p^\Sigma \div_\Sigma\, \vbf^\Sigma
- \sum\limits_{i=1}^N \ubf^\Sigma_i \cdot \Big(\fbf^\Sigma_i- r^\Sigma_i \big(\vbf^\Sigma_i- \frac{\ubf^\Sigma_i}{2}\big) -\na_\Sigma\, p^\Sigma_i \Big).
\end{align*}
\end{document}